\documentclass[fleqn,usenatbib]{mnras}

\usepackage{newtxtext,newtxmath}

\usepackage[T1]{fontenc}
\usepackage{ae,aecompl}

\usepackage{graphicx}	
\usepackage{amsmath}	
\usepackage{amssymb}	
\usepackage[dvipsnames]{xcolor}
\usepackage{lscape}

\title[Scaling relations for GCS]{Scaling relations for globular cluster systems in early-type galaxies}
\author[J. P. Caso et al.]{Juan P. Caso$^{~1,2}$\thanks{E-mail:
jpcaso@fcaglp.unlp.edu.ar (JPC)}, Bruno J. De B\'ortoli$^{~1,2}$, Ana I. Ennis$^{~1,2}$ and
Lilia P. Bassino$^{~1,2}$\\
$^{1}$Facultad de Ciencias Astron\'omicas y Geof\'isicas de la Universidad Nacional de La Plata,    
and \\ Instituto de Astrof\'isica de La Plata (CCT La Plata -- CONICET, UNLP), Paseo del Bosque S/N,  
B1900FWA La Plata, Argentina\\   
$^{2}$Consejo Nacional de Investigaciones Cient\'ificas y T\'ecnicas, Godoy Cruz 2290, C1425FQB,  
Ciudad Aut\'onoma de Buenos Aires, Argentina}

\date{Released 2002 Xxxxx XX}

\pagerange{\pageref{firstpage}--\pageref{lastpage}} \pubyear{2002}

\begin{document}

\label{firstpage}

\maketitle

\begin{abstract}
The formation and growth of globular cluster systems (GCSs) is closely 
related to the evolutionary processes experienced by their host 
galaxies. In particular, their radial distributions scale with several
properties of the galaxies and their halos. We performed a photometric
study, by means of HST/ACS archival data of several intermediate 
luminosity galaxies located in low density environments. It was supplemented
with available photometric data of GCSs from the Virgo and Fornax clusters,
resulting in a sample of almost 30 GCSs for which we fitted their radial
profiles. The resulting overall properties agree with those from previous
studies, as we found that the effective radius, extension and 
concentration of the GCS radial profiles correlate with the stellar mass,
effective radius and number of globular clusters, presenting in some
cases a bilinear relation. The extension also correlates with the central 
velocity dispersion for central galaxies, but not for satellites. From
a statistical comparison with numerical simulations we obtained good 
agreement between the effective radius and extension of the GCS scale
with the effective and virial radius of the halos, respectively.
Finally, we analysed these results in the literature context.
\end{abstract}


\section{Introduction} 

Radial profiles of globular cluster systems (GCSs) have been extensively 
analysed in literature as part of the general context of the globular clusters 
(GCs) in individual galaxies \citep[e.g.][]{ush13,dur14,cas19}. 
GCSs span a broad range of properties, with giant ellipticals usually 
presenting several thousand members and large  systems up to several
tens of kiloparsecs from the galaxy centre \citep[e.g.][]{dir03a,bas06a}. 
On the other hand, dwarf galaxies usually have a few members 
\citep[e.g.][]{pen08,geo09}.

The connection of the GCS with both the stellar and halo (i.e., stellar 
plus dark) mass distributions of the host galaxy is well known, with metal-rich 
GCs tracing the surface-brightness profile of 
the host galaxy and metal-poor ones presenting a more extended 
distribution \citep{bas08,fos11,bas17}, similar to the X-ray emission 
of the hot gas \citep{for12,esc15}. 
These properties of GCSs in luminous early-type galaxies have been 
interpreted as evidence of the presence of two GC sub-populations with different origins.
In the current paradigm where GCs were formed during major starburst 
episodes \citep{kru14}, the origin of metal-poor GCs is connected 
with the primordial building blocks of the galaxies, while metal-rich ones have
their origin in later major mergers \citep{mur10,li14,cho18}. The accretion
of satellite galaxies was also relevant in the build up of the metal-poor
population \citep{ton13}. Observations support this scenario of two-phases, 
with the mean colours of the GCs presenting a colour gradient, 
getting more metal-poor
towards larger galactocentric distances \citep{for11,cas17,for18b}.
In this sense, \citet{pen06} found that the colour range spanned by a 
GCS correlates with the galaxy stellar mass, in a way that the bluer colours 
present in the outer part of GCSs agree with the main role assumed by 
accretion.

Some galaxies that have suffered late mergers present evidence 
of the existence of intermediate age GCs \citep{cas15a,ses16,str04b}, deviating from the 
typical scenario of two sub-populations.

The size-mass relation for galaxies is well documented in the literature
\citep[][and references therein]{spi17}, so that it is natural 
to wonder whether a similar relation exists for GCSs. A first attempt
was made by \citet{rho07}, comparing the extent of the GCS with the
stellar mass of their host galaxy for a small sample. \citet{kar14} analysed
the behaviour of the effective radius of the GCS with the stellar mass
and the effective radius of the galaxy. More recently, \citet{for17}
compared the extension of the GCS in early-type galaxies  
with the host galaxy size and its virial radius. In these studies a correlation
seems to exist between the parameters of the GCS and the host galaxy.

GCs have also proven to be 
useful tracers of the galaxy dynamics \citep{ric13,ric15,ala17,was18} up
to large galactocentric distances \citep{sch10,sch12}, implying their
dynamics is ruled by the total mass distribution. \citet{spi08} redefined 
the T-parameter from \citet{zep93}, considering the halo mass instead of 
just the stellar mass, and suggested that 
GCs were formed in direct proportion to the halo mass of the
host galaxy. \citet{spi09} found that the total mass of the GCS 
scales with the halo mass, later confirmed by \citet{hud14}. 
All these results point to a connection between the properties of 
the GCS and the mass distribution of the host galaxy, which has already  
been addressed by \citet{hud18} and \citet{for18} with samples of 
different characteristics. 

\medskip
We aim to extend the study of the scaling relations for GCSs 
to less massive early-type galaxies, taking advantage of a galaxy 
sample observed with the same instrument and similar photometric 
depth. In this context, we analyse relations found in literature 
for massive galaxies, looking for any possible changes when we 
move to less massive ones and poorer GCSs.

The paper is organized as follows. The observations and reduction 
procedures are described in Section\,2, and the observational and numerical 
catalogues are indicated in Section\,3. In Section\,4 we present the results, 
and Section\,5 is devoted to the discussion. Finally,
in Section\,6 we make a brief summary of the results.

\section{Observational data and reduction}

The observational dataset consists of observations 
centred on nearby early-type galaxies, taken
from the Mikulski Archive for Space Telescopes (MAST)\footnote{ Based on 
observations made with the NASA/ESA Hubble Space Telescope, obtained 
from the data archive at the Space Telescope Science Institute. STScI 
is operated by the Association of Universities for Research in Astronomy, 
Inc. under NASA contract NAS 5-26555.}. All the observations had been carried
out with the HST/ACS Wide Field Camera (WFC), with filters
commonly used in GCs studies. The analysed galaxies are listed in 
Table\,\ref{gal}, in descending order of $B$-band luminosity, 
together with a log of the observations.

\begin{table}   
\begin{minipage}{85mm}   
\begin{center}   
\caption{Basic data from observations. The extinction values are
those from the recalibration by \citet{sch11}.}    
\label{gal}   
\begin{tabular}{@{}l@{}c@{}c@{}c@{}c@{}c@{}}   
\hline   
\multicolumn{1}{@{}c}{Name}&\multicolumn{1}{c}{Filter}&\multicolumn{1}{c}{Obs.\,date}&\multicolumn{1}{c}{Exp.\,time}&\multicolumn{1}{c}{Prog.\,ID}&\multicolumn{1}{c}{$A_{\lambda}$}\\   
 & &\multicolumn{1}{c}{dd\,mm\,yyyy}&\multicolumn{1}{c}{sec.}& & \\   
\hline   
NGC\,3818\, & $F475W$  & $01\,01\,2006$ & $1380$ & $10554$ & 0.12\\ 	 
          & $F850LP$ & $01\,01\,2006$ & $2987$ & $10554$ & 0.05\\
NGC\,1340  & $F475W$  & $20\,09\,2004$ & $760$  & $10217$ & 0.06\\
          & $F850LP$ & $20\,09\,2004$ & $1130$ & $10217$ & 0.02\\
NGC\,4621 & $F475W$  & $19\,07\,2003$ & $750$  & $9401$ & 0.11\\
          & $F850LP$ & $19\,07\,2003$ & $1120$ & $9401$ & 0.04\\
NGC\,7173 & $F475W$  & $16\,05\,2006$ & $1375$ & $10554$ & 0.09\\ 
          & $F850LP$ & $16\,05\,2006$ & $3075$ & $10554$ & 0.03\\  
NGC\,1439 & $F475W$  & $21\,08\,2006$ &	$1375$ & $10554$ & 0.10\\ 
          & $F850LP$ & $21\,08\,2006$ &	$3023$ & $10554$ & 0.04\\ 
NGC\,1426 & $F475W$  & $21\,08\,2006$ & $1375$ & $10554$ & 0.05\\ 
          & $F850LP$ & $21\,08\,2006$ & $3023$ & $10554$ & 0.02\\ 
NGC\,3377 & $F475W$  & $13\,01\,2006$ & $1380$ & $10554$ & 0.11\\ 
          & $F850LP$ & $13\,01\,2006$ & $3005$ & $10554$ & 0.04\\  
NGC\,4033 & $F475W$  & $04\,01\,2006$ &	$1380$ & $10554$ & 0.16\\ 
          & $F850LP$ & $04\,01\,2006$ &	$3017$ & $10554$ & 0.06\\  
NGC\,1172 & $F475W$  & $17\,08\,2006$ & $1380$ & $10554$ & 0.22\\
          & $F850LP$ & $17\,08\,2006$ & $3005$ & $10554$ & 0.09\\ 
\hline
\end{tabular}    
\end{center}    
\end{minipage}   
\end{table}   

\subsection{Photometry and source selection}   
The surface-brightness profile of each galaxy was obtained in 
both filters (see Table\,\ref{gal}), using the task ELLIPSE within
{\sc iraf}. The ellipse parameters, e.g. ellipticity, centre 
coordinates and position angle, were fitted for the inner region of 
the galaxies, depending on the FOV and the galaxy surface brightness, 
typically up to $\approx 30$\,arcsec. For larger galactocentric 
distances these parameters were fixed.

Then, a synthetic model of the galaxies was generated and subtracted 
from the original image, to facilitate the detection of GC candidates. 
A first catalogue of sources was made with SE{\sc xtractor} \citep{ber96}, 
considering every 
detection of at least three connected pixels above a threshold of 
$3\sigma$ from the sky level as a positive identification. As shown 
in the literature, GC-like objects at distances similar to the ones 
corresponding to the galaxies in our sample might be marginally 
resolved \citep[e.g.][]{jor04,cas14,bas17}, and they usually present 
low eccentricities \citep[e.g.][]{har09a,chib11}. Then, in order to 
discard extended sources we selected
those with elongation smaller than 2 and full width at half-maximum 
(FWHM) smaller than 5\,px. Similar criteria have been previously used 
for identifying GCs on ACS images \citep[e.g.][]{jor04,jor07}.

Aperture photometry was performed in both filters with an aperture radius
of 5\,px. In order to calculate aperture corrections, we analysed the
change in the correction value with the effective radius of the sources
(${\rm r_{eff}}$). As a first 
step, we carried out PSF photometry on images of 47\,Tuc observed in dates 
close to the ones from the observations in our sample, and with the same 
filters. In each case, approximately 40 to 50 relatively isolated bright 
stars from the 47\,Tuc images were used to obtain the PSF. The derived PSF 
typically had FWHM $\approx0.08$\,arcsec. Then, the software ISHAPE 
\citep[][]{lar99} was used to calculate structural parameters for 
the sources in our photometric catalogue. We assumed a King profile 
\citep[][]{kin62,kin66} with a concentration parameter, i.e. the ratio 
of tidal over core radius, {\rm $c=30$}, used in previous studies of GCs. 
The mode of the distribution of ${\rm r_{eff}}$ for the galaxies in 
our sample spans $0.015-0.035$\,arcsec, with a tail towards more extended 
objects.

Those objects with signal-to-noise ratio larger than 50 (a condition 
required for an accurate calculus of the structural parameters, 
according to \citealt{lar99}) were split in samples according to 
their ${\rm r_{eff}}$, and aperture corrections were calculated
from each of them. This procedure allowed us to determine the 
variation in aperture correction with ${\rm r_{eff}}$, which was 
typically $\approx 0.06$\,mag for objects in the range ${\rm 0.01 
< r_{eff} < 0.08}$\,arcsec, which represents more than $80$ per cent of 
the GC candidates. Despite these variations, we applied mean 
corrections to GC candidates. These were calculated for 
candidates with ${\rm r_{eff}}$ around the mode of the 
distribution. Although the changes in 
the aperture corrections as a function of ${\rm r_{eff}}$ might 
be large, the present study is focused on 
analysing the radial distribution of GCs, and our simplified 
treatment does not lead to significant uncertainties in our 
results. 

\subsection{Photometric calibration and extinction corrections}
The instrumental magnitudes ($F475$, $F850$) were calibrated using
the relation:
\begin{equation}
m_{std} = m_{inst} + ZP
\end{equation}

\noindent with $m_{std}$ and $m_{inst}$ the standard and instrumental
magnitudes, respectively. The zero-points were taken from \citet{sir05},
$ZP_{F475}= 26.068$ and $ZP_{F850}=24.862$, and the resulting magnitudes
correspond to $g$ and $z$ bands, respectively.

Then, we applied corrections due to Galactic extinction from \citet{sch11},
listed in the last column of Table\,\ref{gal}.

Finally, we selected as GC candidates those sources with colours in the
range $0.6 < (g-z)_0 < 1.7$, in agreement with previous studies in
the same bands \citep[e.g.][]{jor05,cho12}.

\subsection{Completeness analysis}
\label{compl}   
\begin{figure}    
\includegraphics[width=80mm]{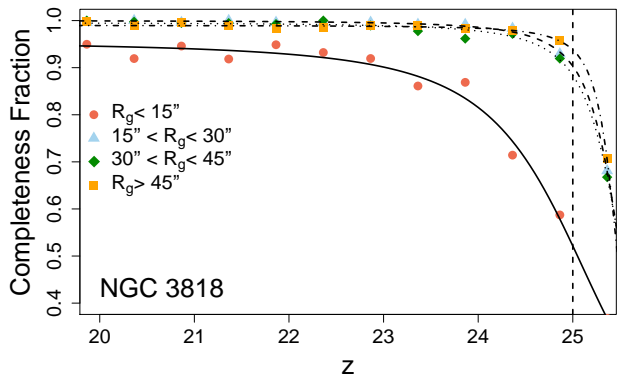}    
\caption{Completeness as a function of $z$ magnitude for NGC\,3818,
obtained from 60\,000 artificial stars. The completeness curves were
calculated in for different galactocentric ranges (${\rm R_g}$), and
the fits correspond to Equation\,2. The
dashed vertical line at $z=25$\,mag indicates the assumed magnitude limit.
Analogue analysis was performed for the other ellipticals in 
low-density environments.}    
\label{compl1}    
\end{figure}

The photometric completeness for each galaxy was obtained by adding
artificial stars to the images in both bands. We added 50 artificial 
stars per image using the PSF previously obtained from the 47\,Tuc
exposures. These artificial sources span the colour range of GCs 
and $21 < z_0 < 26$. We repeated the process 
$1\,200$ times to achieve a final sample of $60\,000$ artificial 
stars. The photometry was developed in the same manner as for the 
science fields, and the resulting catalogues were used to calculate 
the completeness curves in four different galactocentric ranges 
(Fig.\,\ref{compl1}). Typical completeness limits are selected at the 
$z$ magnitude for which completeness levels fall below 90 per cent. In 
order to apply completeness corrections in our analysis, we fitted 
an analytic function of the form:

\begin{equation}
f(m) = \beta \left( 1 - \frac{\alpha(m-m_0)}{\sqrt{1+\alpha^2(m-m_0)^2}}\right) 
\end{equation}

\noindent similar to that used by \citet{har09c}, with $\beta$, 
$\alpha$ and $m_0$ free parameters (curves are shown in Fig.\,\ref{compl1}).

\begin{figure}    
\includegraphics[width=80mm]{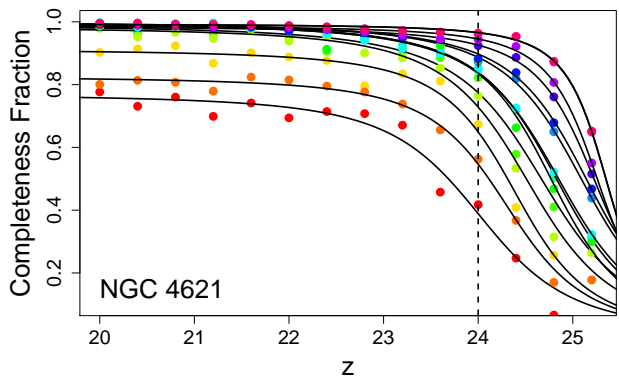}    
\caption{Completeness as a function of $z$ magnitude for NGC\,4621.
A detailed analysis of the completeness behaviour at different
radii was carried out from 250\,000 artificial stars, in order to 
model the completeness for the rest of the galaxies in the Virgo 
cluster. Different colours identify completeness curves for 
different galactocentric radii, i.e. different surface-brightness 
levels, ranging from $17.4$ to $21.9$\,mag\,arcsec$^{-2}$ in the 
$z$ filter. 
The dashed vertical line at $z=24$\,mag indicates a typical 
completeness limit.}    
\label{compl2}    
\end{figure}

The exception to this procedure were NGC\,4621 and NGC\,1340. In these 
cases a total of $250\,000$ artificial stars were added to the images 
in both bands, in order to obtain a more detailed evolution of the
completeness curves as a function of the galactocentric radii
(i.e. the surface-brightness level, see Fig.\,\ref{compl2}). From
these we calculated the completeness
corrections to be applied to the rest of the galaxies in their 
respective photometries.

\section{Data sources from literature}

\subsection{Catalogues of GCs from Virgo and Fornax galaxies}

We also fitted GCs radial profiles for a sample of ellipticals 
from the Virgo and Fornax clusters. We selected those galaxies which
presented an intermediate luminosity, and a number of GCs large enough 
to allow their radial profile fitting. 
We used the available photometry from \citet{jor09} and \citet{jor15}. 
In order to calculate the background level we used point sources in 
the ACS fields of several dwarf ellipticals, which present few GCs
(\citealt{pen08} for Virgo galaxies and \citealt{vil10} for Fornax 
ones). In order to apply completeness corrections to these photometries,
we calculated the mean surface-brightness in the $z$ band
($\mu_{mean,z}$) for different radial ranges, taking into 
account the profiles derived by \citet{fer06b} for galaxies in the 
Virgo cluster and profiles fitted by us for those belonging to Fornax. 
Then, we calculated the corresponding completeness corrections from
the completeness curve that matches the $\mu_{mean,z}$ from the analysis 
described in the previous Section for NGC\,4621 and NGC\,1340.

\begin{figure*}
\raggedright
\includegraphics[width=64mm]{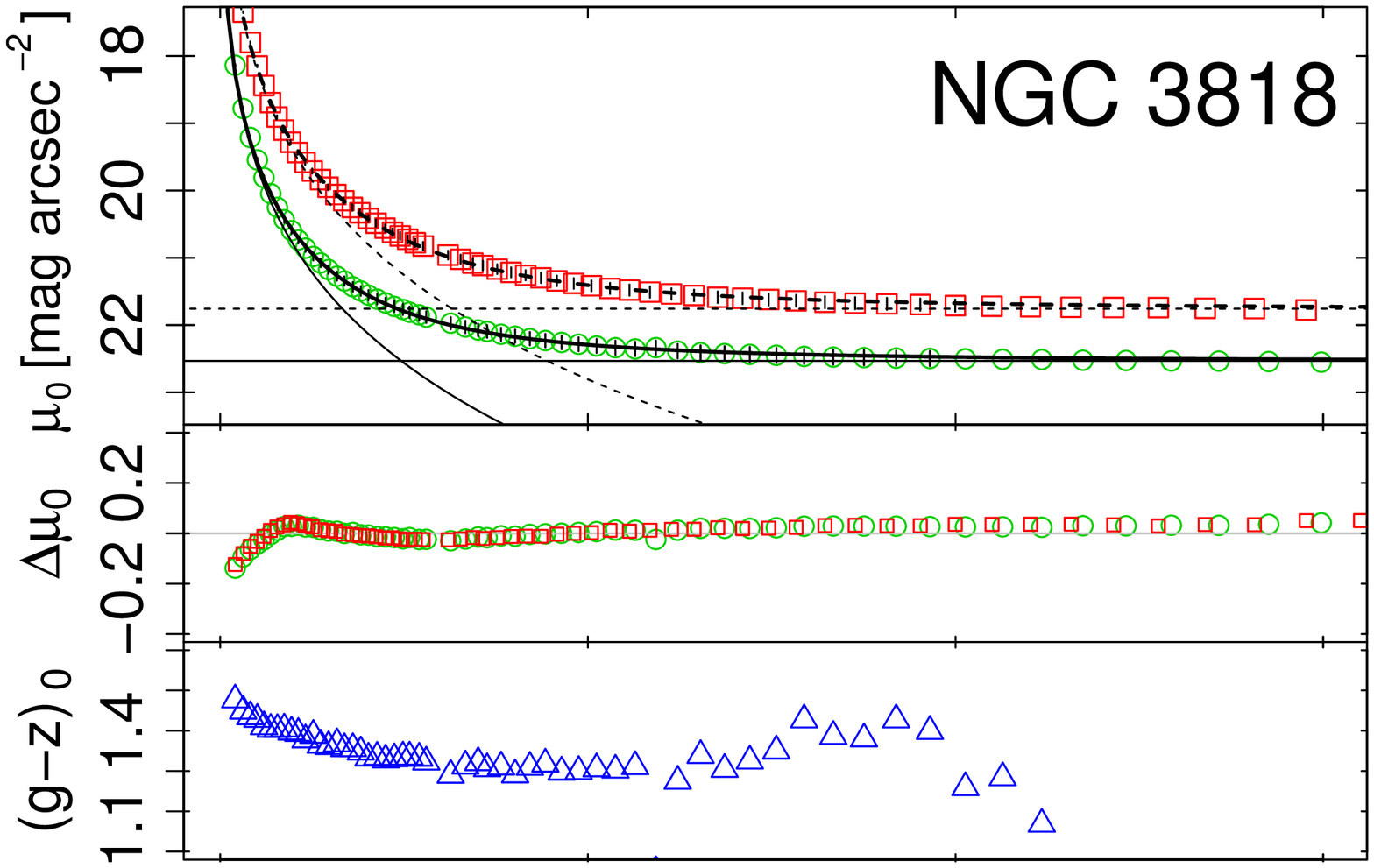}\\    
\includegraphics[width=64mm]{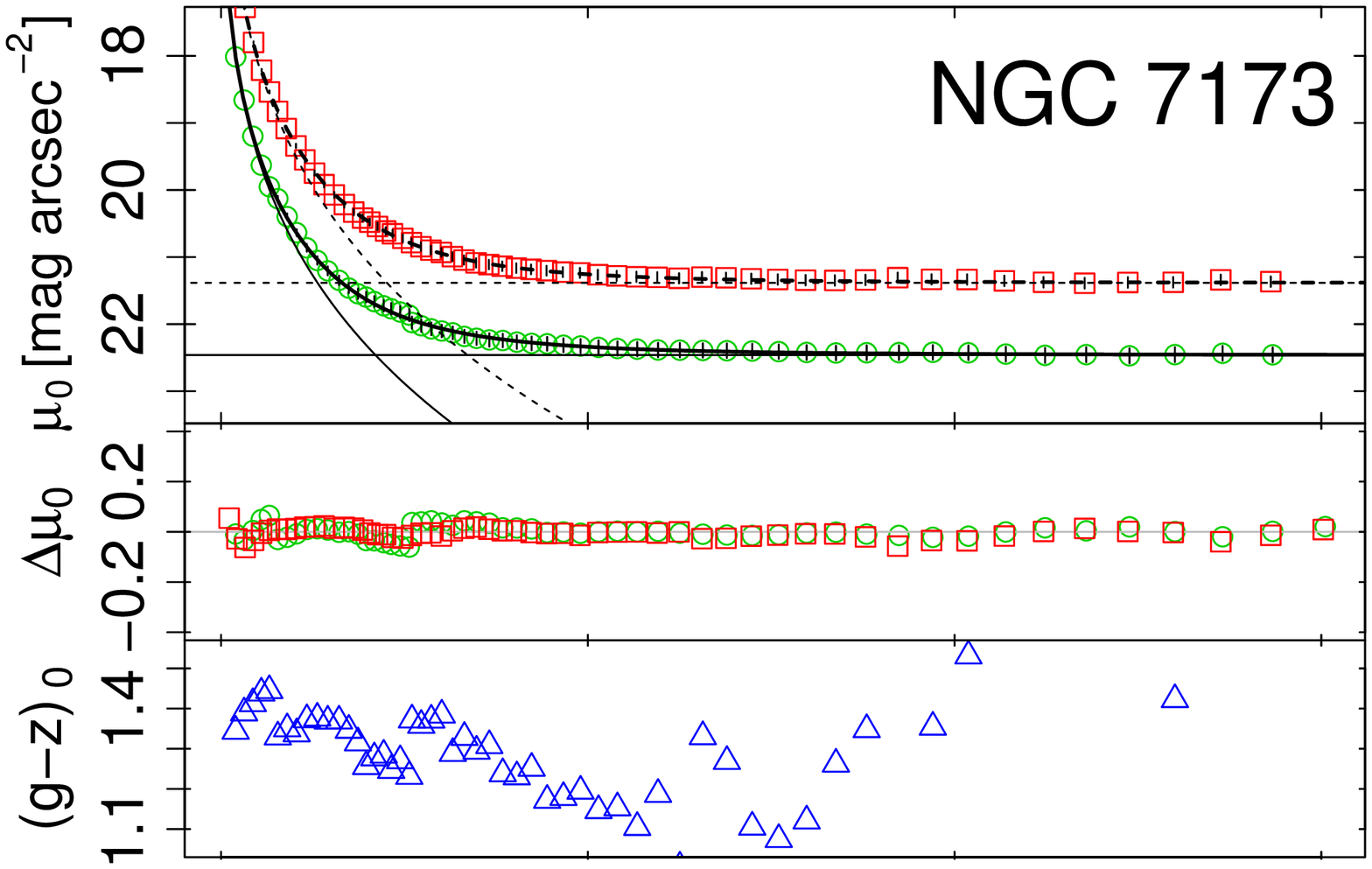}    
\includegraphics[width=55.5mm]{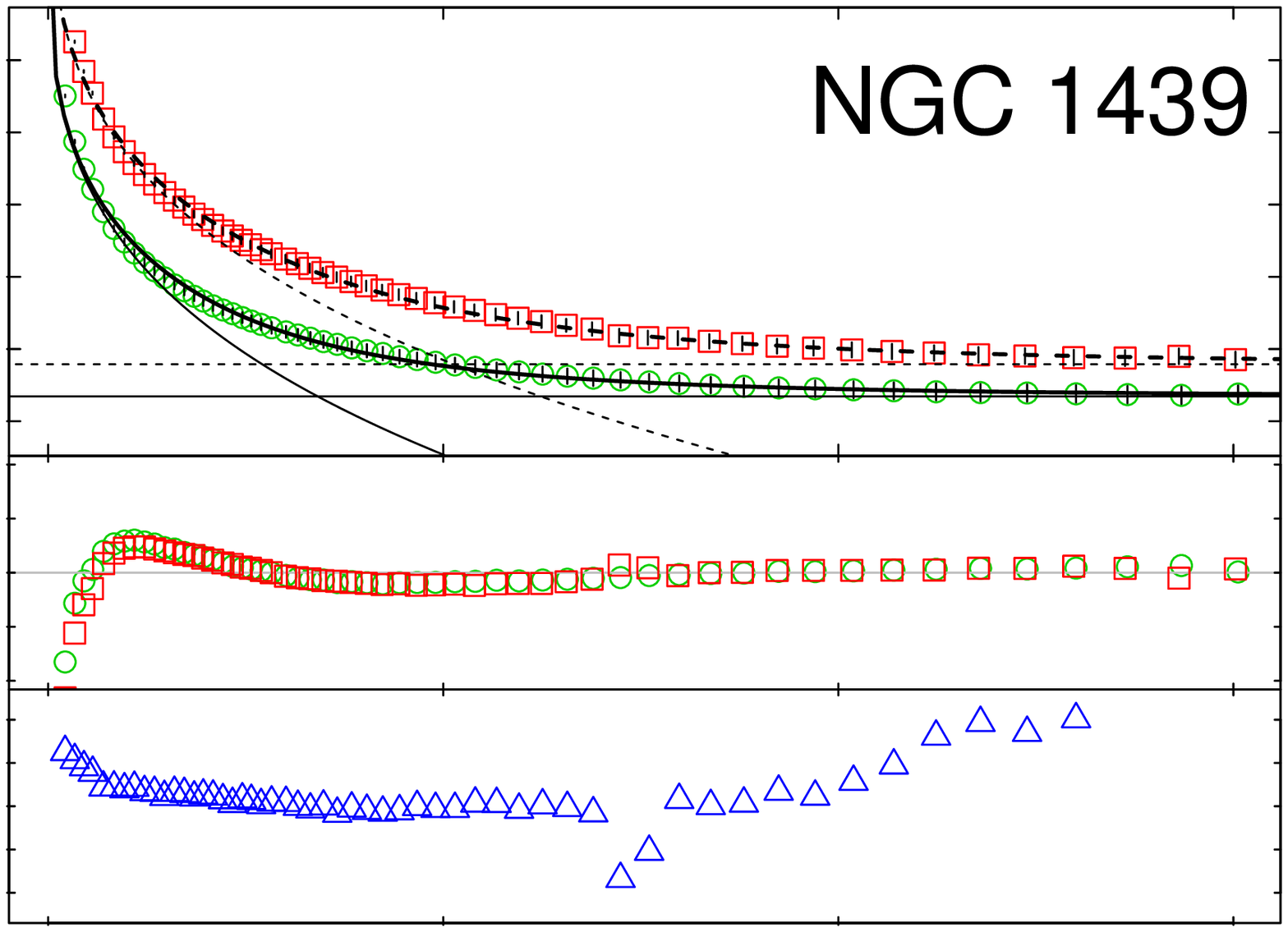}    
\includegraphics[width=55.5mm]{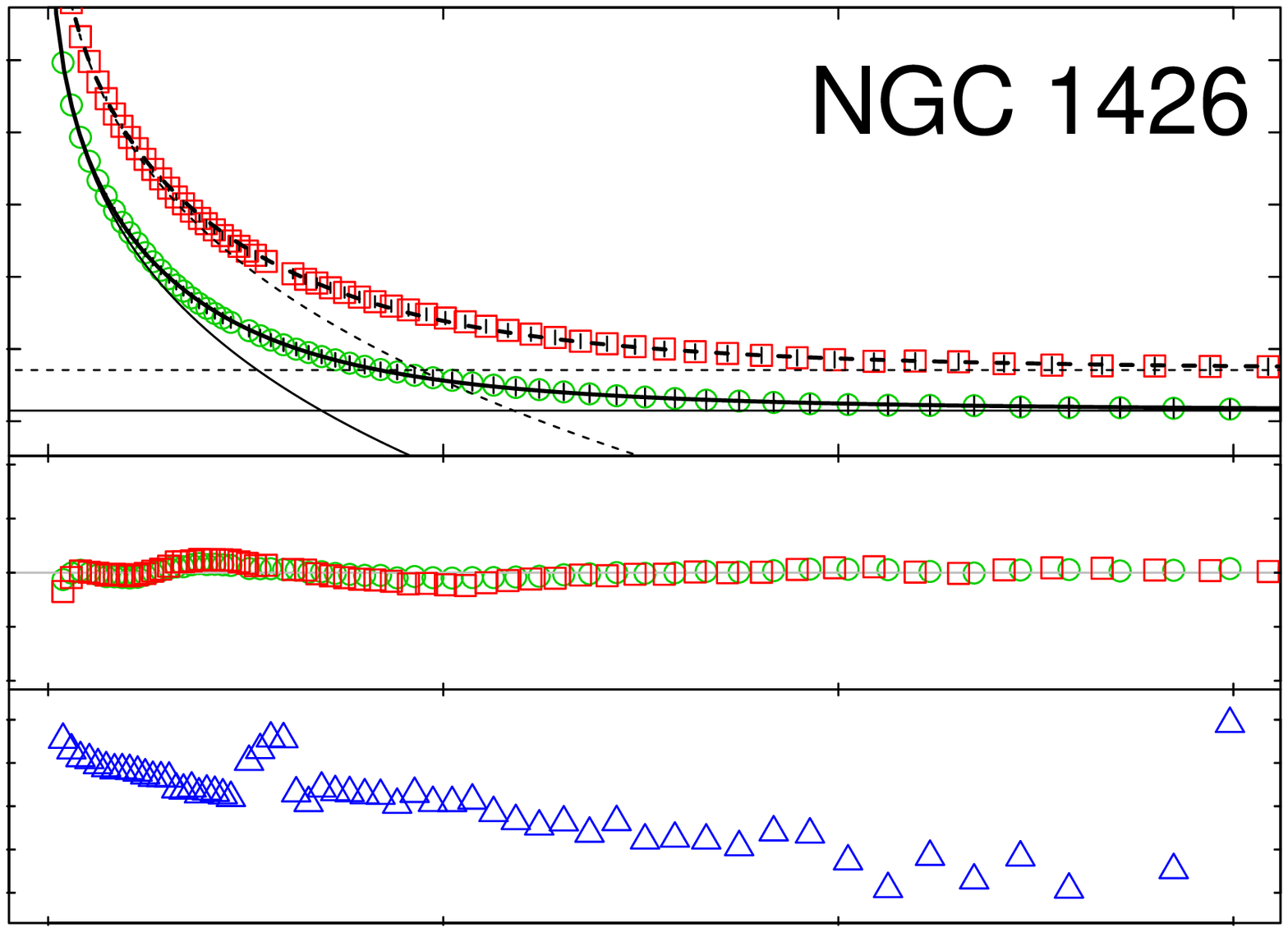}\\    
\includegraphics[width=64mm]{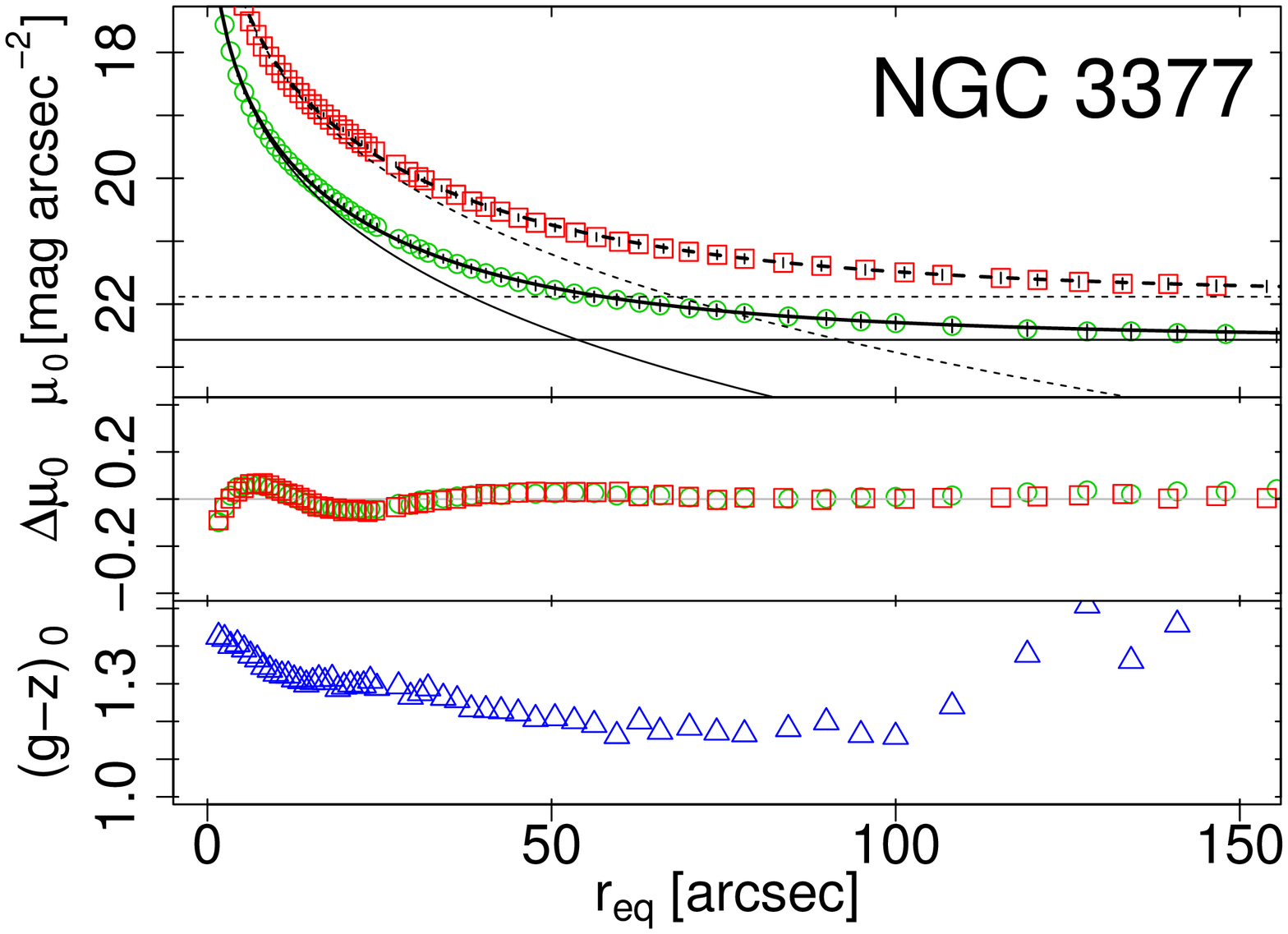}    
\includegraphics[width=55.5mm]{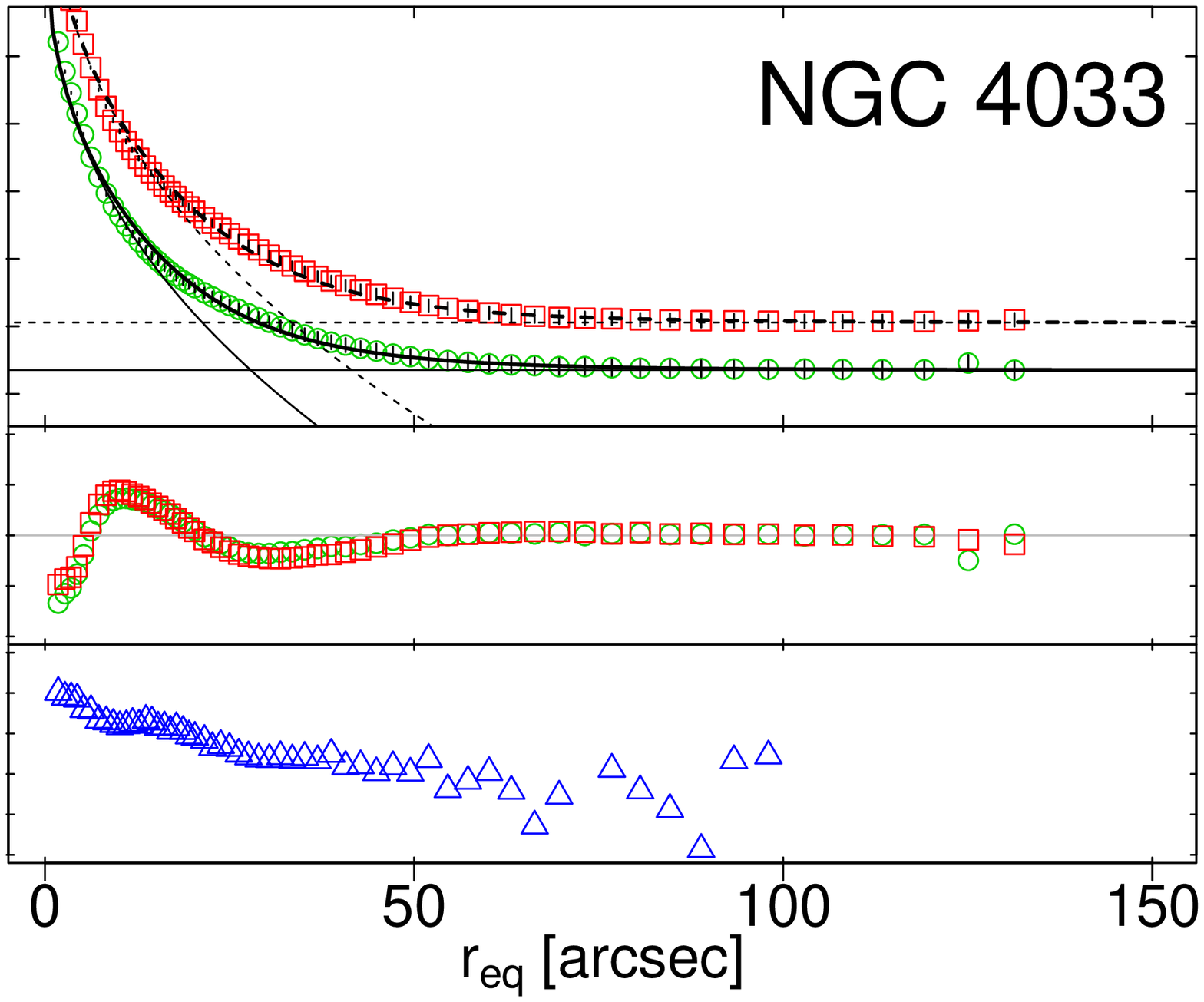}    
\includegraphics[width=55.5mm]{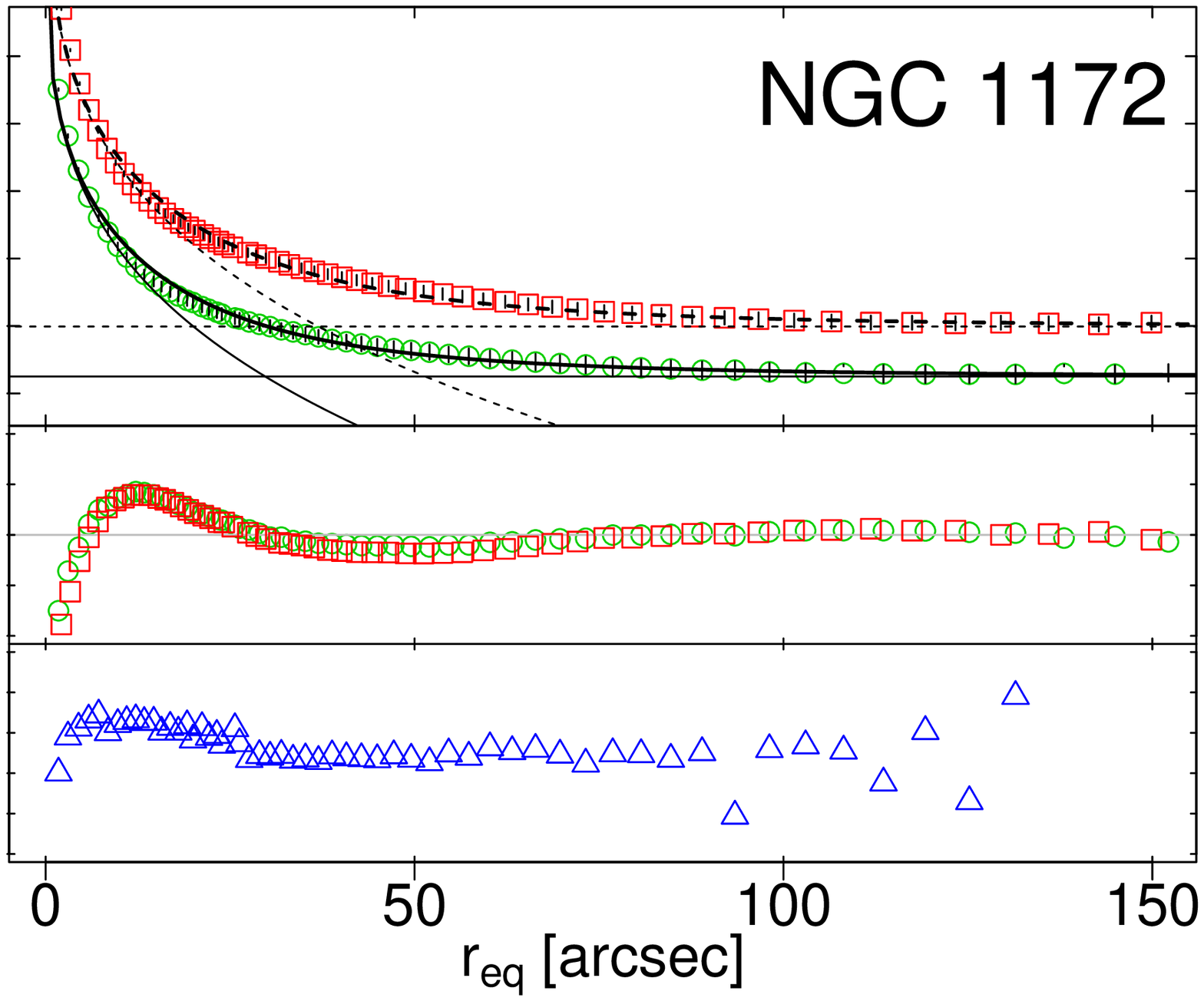}\\  
\caption{The upper panels show the surface-brightness profiles in $g$ (green 
circles) and $z$ (red squares) bands. The solid and dashed horizontal lines
show the background level fitted in each case, the thin curves correspond
to the S\'ersic profile, and the thick ones to the contribution of the galaxy plus 
background.
The fitting procedure was repeated iteratively. The 
middle panels represent the fit residuals, using the same symbols for each
band. The lower panels correspond to the colour profile in $(g-z)_0$.}    
\label{perfilgal}    
\end{figure*}    

\subsection{Dark matter simulation}

We analysed the SMDPL cosmological dark matter simulation, which is 
part of the Multidark project \citep{kly16}, 
and is publicly available through the official database of the project
\footnote{https://www.cosmosim.org/}.
This simulation consists 
of a periodic cubic volume of $400\,h^{-1}\textrm{Mpc}$ of size length, 
filled with $3840^3$
particles with a mass of $9.63\times10^7\,h^{-1}\textrm{M}_\odot$ and
it considers the cosmological parameters of the \citet{pla13}. The
dark matter haloes of the simulation detected with \textsc{Rockstar}
halo finder were analysed, specifically the 
catalogue corresponding to the local Universe ($z=0$).

This catalogue includes the main host haloes found over the background 
density and the satellite haloes (or subhaloes) lying within another halo. 
We consider each of these structures as the host of a unique galaxy, so
the main ones correspond to the central galaxies of each
system, and the satellite haloes, to the satellite galaxies.
For each halo we extracted from the catalogue its position, mass,
host/satellite relationships and the parameters of the mass profile.

In addition to the properties obtained from the catalogue, 
we assigned to each halo a luminosity in the $K$ band by using 
a simple implementation of a halo occupation distribution method 
\citep[HOD][]{val06, con06}, which appoints each luminosity 
in a non-parametric way. 
We assume a monotonic relation of the form 
\begin{equation}
   n_\textrm{g}( > L) = n_\textrm{h}(> M),
\end{equation}
\noindent where $n_\textrm{g}$ and $n_\textrm{h}$ are the number density of galaxies
and haloes, respectively. No distinction between main host and satellites was
made. The number density of galaxies must preserve the parametric luminosity 
function (LF) derived by \citet{sch76}, with the parameters for the $K$ band 
measured by \citet{koc01} from the 2MASS survey. 
Expressing the Schechter LF in terms of the magnitude and starting 
from the bright end of the distribution, rest frame $M_K$ magnitudes were 
assigned to all the haloes using a precision of 0.01 mag.
The most massive main haloes in SMDPL present
virial masses of $\approx 10^{15}\,M_{\odot}$, which are similar to 
the typical total mass derived for the Coma cluster 
\citep[e.g.][]{gel99,lok03,kub07}. Hence, we assumed it as an observational 
analogue to these massive main haloes. Because of this we chose the 
luminosity of 
NGC\,4889, its central galaxy, as the typical luminosity of a central galaxy
belonging to these massive main haloes. Considering for NGC\,4889 an apparent
magnitude of $K= 8.4$\,mag 
\citep{gav96} and a distance of 94\,Mpc, obtained from NED\footnote{This 
research has made use of the NASA/IPAC Extragalactic Database (NED) which is 
operated by the Jet Propulsion Laboratory, California Institute of Technology, under 
contract with the National Aeronautics and Space Administration.}, its absolute
magnitude in the $K$ filter is $M_K= -26.5$. This value was used as the upper 
limit luminosity for the HOD method.

\begin{table}   
\begin{minipage}{85mm}   
\begin{center}   
\caption{S\'ersic profiles and background levels fitted for each galaxy,
listed in decreasing $B$-band luminosity.}
\label{sersic}   
\scalebox{0.9}{
\begin{tabular}{@{}r@{}c@{}ccc@{}c@{}}   
\hline   
\multicolumn{1}{@{}c}{Name}&\multicolumn{1}{c}{$\mu_{\rm backg,0}$}&\multicolumn{1}{c}{$\mu_{\rm eff,0}$}&\multicolumn{1}{c}{${\rm r_{eff,gal}}$}&\multicolumn{1}{c}{n}&\multicolumn{1}{c}{$(g-z)_{\rm 0,gal}$}\\
\multicolumn{1}{@{}c}{}&\multicolumn{2}{c}{${\rm mag\,arcsec^{-2}}$}&\multicolumn{1}{c}{${\rm arcsec}$}& &\multicolumn{1}{c}{mag}\\
\hline   
NGC\,3818\,$g$ & $22.53$ & $23.0\pm0.2$ & $31.4\pm2.8$ & $6.7\pm0.4$ & 1.31\\ 
          $z$ & $21.76$ & $21.2\pm0.1$ & $23.9\pm1.3$ & $6.8\pm0.4$ & \\ 
NGC\,7173\,$g$ & $22.46$ & $21.0\pm0.1$ & $10.9\pm0.6$ & $4.4\pm0.2$ & 1.28\\ 
          $z$ & $21.37$ & $19.1\pm0.1$ &  $8.6\pm0.3$ & $4.1\pm0.1$ & \\ 
NGC\,1439\,$g$ & $22.66$ & $22.9\pm0.1$ & $38.1\pm1.6$ & $3.5\pm0.2$ & 1.32\\
          $z$ & $22.21$ & $21.6\pm0.1$ & $38.8\pm1.2$ & $3.5\pm0.2$ & \\
NGC\,1426\,$g$ & $22.84$ & $22.2\pm0.1$ & $25.6\pm0.8$ & $4.5\pm0.2$ & 1.33\\ 
          $z$ & $22.29$ & $20.7\pm0.1$ & $23.1\pm0.7$ & $4.5\pm0.2$ & \\ 
NGC\,3377\,$g$ & $22.57$ & $22.8\pm0.2$ & $59.4\pm4.6$ & $5.9\pm0.3$ & 1.18\\ 
          $z$ & $21.88$ & $21.1\pm0.1$ & $47.6\pm3.1$ & $6.0\pm0.3$ & \\ 
NGC\,4033\,$g$ & $22.65$ & $20.8\pm0.1$ & $12.8\pm0.4$ & $2.5\pm0.1$ & 1.30\\
          $z$ & $21.94$ & $19.2\pm0.1$ & $11.5\pm0.4$ & $2.4\pm0.1$ & \\
NGC\,1172\,$g$ & $22.75$ & $23.1\pm0.2$ & $35.4\pm2.3$ & $3.7\pm0.3$ & 1.22\\
          $z$ & $22.01$ & $21.7\pm0.1$ & $32.4\pm1.8$ & $3.9\pm0.4$ & \\
\hline
\end{tabular}}  
\end{center}    
\end{minipage}   
\end{table}

\section{Results}

\subsection{Galaxy surface-brightness profiles}

Because a comprehensive analysis has been carried out for Virgo and Fornax 
galaxies by \citet{fer06b},  \citet{cot07} and \citet{gla11}, we focused
on those located in low-density environments (HST programme ID 10554). The 
upper panels for each galaxy in Figure\,\ref{perfilgal} show the
surface-brightness profiles in $g$ (green circles) and $z$ (red squares)
bands 
as a function of the equivalent radius ${\rm r_{eq}}$. For each band we 
fitted S\'ersic profiles \citep{ser68} of the form:

\begin{equation}
\mu({\rm r_{eq}}) = \mu_{\rm eff} + 1.0857*{\rm b_n}\left[ \left( \frac{{\rm r_{eq}}}{\rm r_{ eff,gal}} \right)^{\frac{1}{\rm n}}-1 \right]
\end{equation}

\noindent where ${\rm r_{eq}}$ and $r_{\rm eff,gal}$ are in arcsec, the 
latter one corresponding to the effective radius, $\mu({\rm r_{eq}})$ 
and $\mu_{\rm eff}$ are in units of mag\,arcsec$^{-2}$, n is the S\'ersic 
shape index,and $b_n$ is calculated using the expression from
\citet{cio91}. 
We considered a single component for the profiles 
corresponding to the field-of-view  (FOV) of the ACS camera, 
achieving acceptable fits in all cases (see residuals in the
middle panels in Fig.\,\ref{perfilgal}). Due to the reduced FOV, an accurate 
measurement of the background level was not possible.
Hence, it was handled as a free parameter and it was fitted
from the counts level for galactocentric distances larger than
100\,arcsec. This value and the S\'ersic profile were fitted
iteratively, subtracting their corresponding contributions in 
each step. The procedure was repeated until the parameters 
converged and the residuals for the measurements with galactocentric
distances larger than 100\,arcsec achieved typically $10^{-2}$.
For each galaxy, the upper panels in Fig.\,\ref{perfilgal} 
show the S\'ersic profile fitted in the $g$ and $z$ bands with
solid and dashed thin curves, respectively. The background
levels are indicated with horizontal lines, and the thick
curves correspond to the contributions of galaxy plus background.
In Table\,\ref{sersic} we listed the S\'ersic parameters for the 
galaxies fitted in this work, and the corresponding background
levels ($\mu_{\rm backg,0}$) in units of mag\,arcsec$^{-2}$. 
Regarding this latter parameter, they do not present significant
differences from values expected from the ACS Exposure Time 
Calculator\footnote{http://etc.stsci.edu/etc/input/acs/imaging/}
in units of electrons per second for similar positions, filters, 
exposures and dates to the observations. The fitted values
for $\mu_{\rm backg,0}$ are also similar to
those obtained by \citet{jor04} for galaxies in the Virgo cluster,
with similar instrumental configuration. The last column shows
the integrated colours $(g-z)_{\rm 0,gal}$, obtained from the 
integration of the S\'ersic profiles.  These are $\approx 0.1$\,mag 
bluer than in galaxies with similar luminosities from the Virgo cluster 
\citep{smi13}, in agreement with studies from the literature that 
also measured bluer colours for elliptical galaxies in low-density 
environments \citep[e.g.][]{lac16}.
The lower panels for each galaxy present the colour profiles in 
$(g-z)_0$. In some galaxies the colours are missing at large radii, 
this is due their surface brightness profiles fall quickly to the background 
level, resulting in a large noise in the colour measurement. A negative 
colour gradient is evident in most of the galaxies.

\subsection{Effective radii of GCs}

Although ${\rm r_{eff}}$ of GCs in some galaxies of our sample have
already been measured \citep{jor05,mas10}, 
those hosts presenting intermediate luminosities and located in low-density 
environments (programme ID 10554) lack this analysis. Moreover,
the papers cited above point to the dependence of the ${\rm r_{eff}}$ 
of GCs with a mixture of properties of the host galaxy and the GCs
themselves. Hence, it is interesting to corroborate whether these 
galaxies in our sample follow similar relations. In Table\,\ref{t.reff}
the galaxies from programme\,10554 are listed, together with the 
absolute $B$ magnitude, the mean ${\rm r_{eff}}$ in the $z$-band of GCs
and their mean $(g-z)_0$ colour.
We did not find a clear gradient in the mean ${\rm r_{eff}}$ of 
the GCs (${\rm \overline{r_{eff}}}$) in terms of neither the host galaxy 
luminosity nor its colour listed in Table\,\ref{sersic}, but 
galaxies in this sample span a limited range for these properties.
However, the results listed in Table\,\ref{t.reff} are in agreement
with those in \citet{jor05} for similar galaxies. There seems to exist
a trend between ${\rm \overline{r_{eff}}}$ and mean colour for a GCS,
$(g-z)_{\rm 0,GCS}$, which is expected from the difference in typical 
${\rm r_{eff}}$ for blue and red GCs \citep[e.g.][]{jor05,mas10}.

\begin{figure}    
\includegraphics[width=80mm]{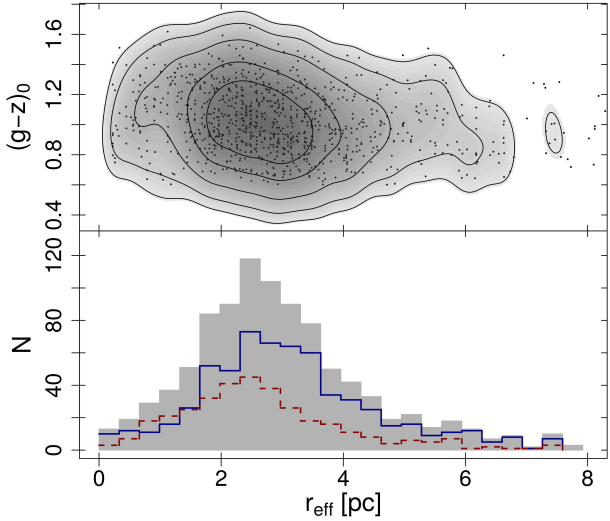}    
\caption{{\bf Upper panel:} Smoothed distribution of $(g-z)_0$ colour of GC
candidates from the joint sample as a function of their ${\rm r_{eff}}$. 
Candidates with ${\rm r_{eff}}$ in the range $10-20$\,pc, typical of
extended clusters, were excluded from the figure. 
{\bf Lower panel:} distribution of ${\rm r_{eff}}$ for all the GC candidates 
(filled histogram), the blue subpopulation (solid line histogram) and the
red one (dashed line histograms).}    
\label{reff1}    
\end{figure}

The upper panel of Figure\,\ref{reff1} shows the $(g-z)_0$ colours of the
joint sample of GCs belonging to these galaxies, as a
function of their ${\rm r_{eff}}$. The smoothed distribution
suggests that bluer GCs tend to present a distribution of ${\rm r_{eff}}$
that reaches larger values. The lower panel presents the 
distribution of ${\rm r_{eff}}$ of the entire sample of GCs (filled
histogram), and the blue (solid line histogram) and
red GCs (dashed line histogram), assuming $(g-z)_0=1.1$ as the colour
limit between both subpopulations. As in previous studies, the blue GCs 
present a larger ${\rm \overline{r_{eff}}}$ than the red ones, $2.96\pm0.1$\,pc
and $2.55\pm0.1$\,pc, respectively. These values imply that red GCs
are $\approx 15$ per cent smaller than their blue counterparts, which is in
good agreement with results from \citet{jor05} for GCs in the Virgo 
cluster. The mean ${\rm r_{eff}}$ for the entire sample is $2.81\pm0.07$\,pc,
similar to results from other systems \citep[e.g.][]{har09a,mas10}.
There is a small sample of $\approx20$ candidates which might be classified
as extended clusters \citep[e.g.][]{bro11,bru12}, with ${\rm r_{eff}}$ in
the range $10-20$\,pc, and typical blue colours. 

\begin{table}   
\begin{minipage}{85mm}   
\begin{center}   
\caption{Luminosity in the $B$ filter for galaxies located in 
low-density environments studied
in this paper with mean properties of their GCSs. The absolute 
magnitudes were obtained from the $B$ magnitudes
and the distance moduli $(m-M)$ listed in Table\,\ref{hubpar}, 
${\rm \overline{r_{eff}}}$ is the mean effective radius for all the GC
candidates, and $(g-z)_{\rm 0,GCS}$ their mean colours.}    
\label{t.reff}   
\begin{tabular}{@{}lccc@{}}   
\hline   
\multicolumn{1}{@{}c}{Name}&\multicolumn{1}{c}{$M_B$}&\multicolumn{1}{c}{${\rm \overline{r_{\rm eff}}}$}&\multicolumn{1}{c}{$(g-z)_{\rm 0,GCS}$}\\   
&\multicolumn{1}{c}{mag}&\multicolumn{1}{c}{pc}&\multicolumn{1}{c}{mag}\\   
\hline   
NGC\,3818 & $-20.33$  & $3.14\pm0.14$ & $0.97\pm0.02$\\ 	 
NGC\,7173 & $-19.96$  & $2.69\pm0.10$ & $1.13\pm0.02$\\ 
NGC\,1439 & $-19.95$  & $3.38\pm0.15$ & $0.93\pm0.03$\\ 
NGC\,1426 & $-19.65$  & $3.09\pm0.11$ & $0.98\pm0.02$\\ 
NGC\,3377 & $-19.18$  & $2.62\pm0.12$ & $1.12\pm0.02$\\ 
NGC\,4033 & $-19.11$  & $3.35\pm0.18$ & $0.96\pm0.03$\\ 
NGC\,1172 & $-18.93$  & $2.80\pm0.09$ & $1.08\pm0.01$\\
\hline
\end{tabular}    
\end{center}    
\end{minipage}   
\end{table}   

\subsection{Radial profiles}

Radial profiles of GCs are usually fitted by different 
mathematical expressions. Power-laws \citep[e.g.][]{esc15,sal15} and 
de Vaucouleurs laws \citep[e.g.][]{fai11} have been commonly used in
the past, but recent papers also applied S\'ersic \citep[e.g.][]{ush13,kar14}
and modified Hubble profiles \citep{bin87,bas17}. In this paper we 
adopt the latter option: 

\begin{equation}
n(r) = a \left( 1+ \left(\frac{r}{r_0} \right)^2 \right)^{-b}
\end{equation}

\noindent which provided accurate fits for the vast majority of GCS. This
profile
behaves as a power-law with an exponent $2\times b$ for large galactocentric 
distances and it presents a central flattening, 
ruled by the core radius $r_0$. These properties allow us to perform a direct 
comparison with a large sample from previous studies, and to analyse, 
for our galaxy sample, the paucity of GCs in the inner 
regions of GCSs, and its possible connection with GC erosion 
processes. Using the Hubble profile, we can also calculate the $r_{\rm eff,GCS}$ 
of the radial distribution, and compare our results with recent studies 
based on S\'ersic profiles \citep[e.g.][]{for17}.

Radial profiles were fitted to projected density distributions corrected by 
completeness and contamination. For those galaxies belonging to programme 
10554 (i.e., those in low-density environments), completeness functions
were derived individually (Fig\,\ref{compl1}), and a typical 
contamination density of ${\rm n_b} = 1$\,arcmin$^{-2}$ was assumed. 
This value was calculated by
\citet{cho12} for the same sample of galaxies from blank fields with
similar galactic coordinates. The colour-magnitude diagrams obtained
for our samples of GC candidates and those presented in their Figure\,3
agree, which indicates that both photometries are comparable and no bias 
was introduced by assuming their contamination level.
As mentioned above, a detailed completeness analysis was carried 
out for a single galaxy with intermediate luminosity belonging to the
Virgo and Fornax galaxy clusters, NGC\,4621 and NGC\,1340 respectively. The 
completeness corrections for the rest of the galaxies were derived from 
these analysis, taking into account their mean surface-brightness in the
$z$-band for several radial ranges. For the galaxies in Virgo we adopted
the S\'ersic profiles derived by \citet{fer06b}. For galaxies
in Fornax we derived them from ACS observations obtained from MAST.

In all cases the radial binning is constant on a logarithmic scale, with a 
typical size of $log_{10} \Delta r\,{\rm [arcsec]} = 8$, but varying from
galaxy to galaxy depending on the size of the sample of GCs. In each case
the bin breaks were slightly shifted around ten times to take into account
noise uncertainties, and the final parameters resulted from weighted means
of the parameters fitted in the individual runs (see Table\,\ref{hubpar}).
Figure\,\ref{hubprof} shows the projected radial distribution for GCSs 
fitted in this paper, corrected by completeness and contamination. The 
variation of the fitted functions due to changes in the bin breaks is
represented by the grey regions. The red solid curve corresponds to the
Hubble modified profile with parameters obtained from the weighted means.
The contamination level $n_b$ was calculated from fields centred on 
dwarf galaxies with few GC candidates, according with \citet{pen08}.

In galaxies fainter than $M_V \approx -18$, GCSs typically present only 
a few dozen members \citep[e.g.][]{har13}, meaning radial profiles 
cannot be obtained without significant scatter in their parameters. Hence, 
we decided to stack GCSs associated to galaxies with similar luminosities,
stellar masses \citep{pen08} and classification in order to fit a mean 
radial distribution. In each case, galaxies 
involved in the stacking process are indicated in the corresponding 
panel, and labelled in Table\,\ref{hubpar} with consecutive numbers, 
e.g. VS$\#$ for Virgo stacked samples. The first case, VS\,1 corresponds to 
galaxies VCC\,575, VCC\,1178 and VCC\,1261, presenting $V$ absolute magnitudes
between $-18.35$ and $-18.42$, and stellar masses ${\rm M_{\star} \approx 5-7 
\times 10^9\,M_{\odot}}$; in VS\,2 the stacked galaxies, VCC\,9, VCC\,437, 
VCC\,1087 and VCC\,1422, present $M_V$ in the range $-17.8$ and $-18.04$,
and ${\rm M_{\star} \approx 2.8-4 \times 10^9\,M_{\odot}}$; the last group
correspond to six galaxies with $M_V$ between $-17$ and $-17.5$ and 
${\rm M_{\star} \approx 1-2.3 \times 10^9\,M_{\odot}}$, these are
VCC\,200, VCC\,543, VCC\,1431, VCC\,1528, VCC\,1871 and VCC\,2019.

In total, 27 radial profiles were fitted (including the stacked 
galaxies), spanning absolute magnitudes from $M_B\approx-18.7$
to $M_B\approx-20.3$. The results are listed in Table\,\ref{hubpar}. The
columns ${\rm r_L}$ and ${\rm r_{eff,GCS}}$
correspond to the projected extension of the GCS calculated from our
profiles and its effective radius, respectively. This latter one 
depends on ${\rm r_L}$, $r_0$ and $b$. The extension ${\rm r_L}$ was 
assumed as the galactocentric distance for which the 
projected density falls to $30$ per cent of the contamination level, i.e. 
$\approx 0.3$\,arcmin$^{-2}$, according to the value for ${\rm n_b}$
previously indicated. This criterion has been used in past studies to 
define the GCS extension \citep[e.g.][]{bas17,cas17,cas19}.
We are aware that extrapolating radial profiles due to the limited FOV
might lead to uncertainties larger than those estimated for the most
extended GCS in our sample. The last three columns in Table\,\ref{hubpar}
correspond to the number of members of the GCS, the effective radius of
the host galaxy (${\rm r_{eff,gal}}$) and its central velocity dispersion.
This latter parameter was obtained from the HyperLeda web 
page\footnote{http://leda.univ-lyon1.fr} \citep{mak14}. For galaxies
belonging to the Fornax cluster, the number of members of the GCS was not 
available in literature for the full extension of the GCS. Then, we numerically
integrated the radial profiles up to the distance ${\rm r_L}$, resulting 
in the number of GCs brighter than 24\,mag in the z-band. From the set of 
parameters fitted by \citet{vil10} to the GCLF we calculated the fraction 
of GCs fainter than this magnitude limit, in order to obtain the total
population of GCs indicated in the Table.

In addition, we also compiled parameters from a large number of GCSs
from the literature, focusing on early type galaxies, because they are
thought to 
have undergone a rich merger history. These galaxies are listed in 
Table\,\ref{tab.otros}, together with their magnitudes in several bands 
and the parameters of their GCS radial profiles. For GCSs fitted by power-law 
profiles, the parameter $b$ in Table\,\ref{tab.otros} corresponds to 
half of the power law exponent indicated in those papers. When it was 
possible, projected densities were obtained and Hubble profiles were 
fitted for those GCSs represented by S\'ersic or de Vaucouleurs profiles. 
These cases are highlighted in Table\,\ref{tab.otros}, because data were 
not directly obtained and the uncertainties might be underestimated.

\subsection{Scaling relations for GCS}

\subsubsection{Scaling relations with the $b$ parameter of the Hubble profile}
\begin{figure*}    
\includegraphics[width=160mm]{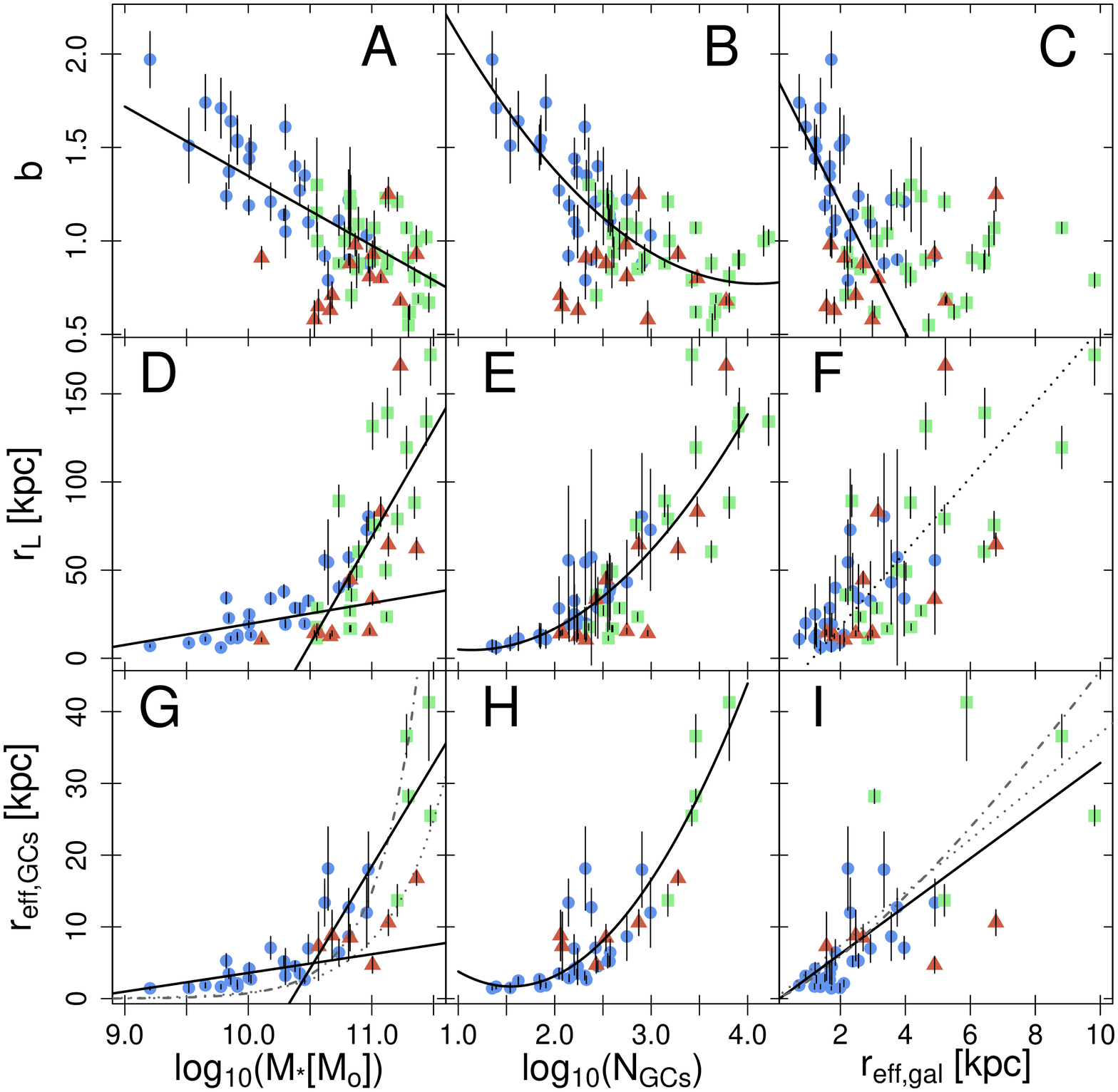}\\    
\caption{The exponent of the modified Hubble profile (b), the extension of 
the GCS (${\rm r_L}$) and its effective radius (${\rm r_{eff,GCS}}$), as 
functions of the logarithm of the stellar mass (${\rm M_{\star}}$), the 
logarithm of the number of GCs (${\rm N_{GCs}}$) and the effective radius 
of the host galaxy (${\rm r_{eff,gal}}$). Blue circles represent the galaxies
analysed in this paper (Table\,\ref{hubpar}), green squares and red triangles,
respectively, indicate ellipticals and lenticulars from the literature
(Table\,\ref{tab.otros}). Solid curves show relations fitted in
this paper, dotted and dashed-dotted curves correspond to literature
results \citep{kar14,for17,hud18}. See text for further details.}    
\label{reff2}
\end{figure*}

The first row in Figure\,\ref{reff2} shows the exponent of the Hubble
profile $b$ as a function of the logarithm of the stellar mass in units
of solar masses (${\rm M_{\star}}$, panel\,A), the logarithm of the total number 
of GCs (${\rm N_{GC}}$, panel\,B), and the effective radius of the galaxy, 
expressed in kpc (${\rm r_{eff,gal}}$, panel\,C). The stellar masses were 
obtained from the luminosities in $J$ and $K$ bands, applying the 
mass-to-light ratios ($M/L$) derived from \citet{bel03} and $(B-V)$ 
colours. Blue circles correspond to galaxies
from our sample (Table\,\ref{hubpar}), the other symbols represent
early-type galaxies from literature (Table\,\ref{tab.otros}), 
differentiated between ellipticals (green squares) and lenticulars (red
triangles).

The $b$ parameter seems to correlate with ${\rm M_{\star}}$ and ${\rm N_{GCs}}$,
pointing to steeper radial distributions for less massive galaxies
and poorer GCSs. On the other hand, galaxies with ${\rm M_{\star}} \gtrsim 
3\times 10^{10}\,{\rm M_{\odot}}$ and ${\rm N_{GCs}} \gtrsim 300$ present more 
extended systems, despite a large spread in the measurements.
Assuming linear relations, they can be described as:

\begin{equation}
b = 4.9\pm0.5 -0.36\pm0.05 \times {\rm X_A}
\end{equation}

\noindent represented with a solid line in panel\,A, with ${\rm X_A}$ 
being ${\rm log_{10}(M_{\star})}$, and

\begin{equation}
b = 3.1\pm0.4 -1.1\pm0.3 \times {\rm X_B} + 0.13\pm0.05 \times {\rm X_B^2}
\end{equation}

\noindent that corresponds to the solid line in panel\,B, with
${\rm X_B}$ being ${\rm log_{10}(N_{GCs})}$. This
latter relation could be obtained from the previous one and the 
correlation between the size of GCS and the luminosity of the host
galaxies. In Figure\,6 from \citet{har13} it is clear that this 
relation deviates from linearity when ${\rm L_K} \lesssim 3 \times 10^9
{\rm L_{\odot}}$, which corresponds to the faint end of our sample.
In panel\,C there seems to exist a correlation for galaxies 
with ${\rm r_{eff,gal}<4}$\,kpc, for whom a linear relation is
shown with a solid line, resulting in:

\begin{equation}
b = 1.9\pm0.14 - 0.34\pm0.06 \times {\rm X_C}
\end{equation}

\noindent with ${\rm X_C}$ being ${\rm r_{eff,gal}}$. Galaxies with 
${\rm r_{eff,gal}<4}$\,kpc typically

\begin{landscape}
\begin{table}   
\begin{minipage}{220mm}   
\begin{center}   
\caption{Galaxies analysed in this paper, listed in decreasing $B$-band  
luminosity. Magnitudes (col. 2 to 5) were obtained from NED and reddening
corrections from the recalibration by \citet{sch11}. Distance moduli 
correspond to SBF measurements listed in NED, typically from \citet{tul13} 
or \citet{bla09}. The parameters $a$, ${\rm r_0}$ and $b$ (col. 8 to 10) 
correspond to the modified Hubble profiles fitted to the GCS radial profiles. 
${\rm r_L}$ ${\rm r_{eff,GCS}}$ and ${\rm N_{GCs}}$ represent the 
projected extension of the GCS, its effective radius and the total GC population, 
respectively, obtained as indicated in the text (Section\,4.3). The
effective radius of the host galaxies ${\rm r_{eff,gal}}$ were obtained from the literature, and central velocity dispersions (${\rm \sigma_0}$) from 
HyperLeda database.}   
\label{hubpar}   
\begin{tabular}{@{}lcccccccccccccc@{}}   
\hline   
\multicolumn{1}{@{}c}{Name}&\multicolumn{1}{c}{$B$}&\multicolumn{1}{c}{$V$}&\multicolumn{1}{c}{$J$}&\multicolumn{1}{c}{$K$}&\multicolumn{1}{c}{$E_{(B-V)}$}&\multicolumn{1}{c}{$(m-M)$}&\multicolumn{1}{c}{$a$}&\multicolumn{1}{c}{${\rm r_0}$}&\multicolumn{1}{c}{$b$}&\multicolumn{1}{c}{${\rm r_L}$}&\multicolumn{1}{c}{${\rm r_{eff,GCS}}$}&\multicolumn{1}{c}{${\rm N_{GCs}}$}&\multicolumn{1}{c}{${\rm r_{eff,gal}}$}&\multicolumn{1}{c}{${\rm \sigma_0}$}\\   
 &\multicolumn{1}{c}{mag}&\multicolumn{1}{c}{mag}&\multicolumn{1}{c}{mag}&\multicolumn{1}{c}{mag}&\multicolumn{1}{c}{mag}&\multicolumn{1}{c}{mag}& &\multicolumn{1}{c}{arcmin}& &\multicolumn{1}{c}{arcmin}&\multicolumn{1}{c}{arcmin}& &\multicolumn{1}{c}{arcsec}&\multicolumn{1}{c}{km\,s$^{-1}$}\\   
\hline   
NGC\,4552 & $10.73$ & $9.75$ & $7.62$ & $6.73$ & $0.036$ & $31.00$ & $2.23\pm0.03$ & $0.53\pm0.08$ & $1.03\pm0.09$ & $15.8\pm4.7$ & $2.6\pm1.0$ & $984\pm198^1$ & $30.0^4$ & $250\pm2.9$\\
NGC\,3818 & $12.67$ & $11.71$ & $9.80$ & $8.87$ & $0.031$ & $32.81$ & $2.57\pm0.07$ & $0.14\pm0.02$ & $0.90\pm0.03$ & $5.4\pm1.2$ & $1.2\pm0.2$ & $240\pm--^2$ & $21.2^4$ & $193\pm3.7$\\
NGC\,1340 & $11.27$ & $10.39$ & $8.24$ & $7.40$ & $0.016$ & $31.35$ & $1.90\pm0.03$ & $0.71\pm0.12$ & $1.22\pm0.16$ & $8.0\pm2.5$ & $1.6\pm0.6$ & $560\pm--^3$ & $39.5^4$ & $163\pm3.4$\\
NGC\,4621 & $10.88$ & $9.63$ & $7.65$ & $6.75$ & $0.028$ & $30.86$ & $2.21\pm0.03$ & $0.32\pm0.04$ & $0.79\pm0.04$ & $25.9\pm5.7$ & $7.1\pm1.3$ & $803\pm355^1$ & $46.4^4$ & $228\pm3.8$\\
NGC\,4473 & $11.10$ & $10.20$ & $8.04$ & $7.16$ & $0.025$ & $30.92$ & $2.15\pm0.05$ & $0.42\pm0.07$ & $1.11\pm0.08$ & $9.0\pm2.9$ & $1.5\pm0.4$ & $376\pm97^1$ & $24.9^4$ & $179\pm2.5$\\
NGC\,1439 & $12.27$ & $11.39$ & $9.44$ & $8.57$ & $0.025$ & $32.05$ & $2.07\pm0.04$ & $0.29\pm0.04$ & $0.92\pm0.05$ & $7.4\pm1.6$ & $1.8\pm0.4$ & $139\pm--^2$ & $39.4^4$ & $146\pm5.4$\\
NGC\,1426 & $12.29$ & $11.39$ & $9.57$ & $8.67$ & $0.014$ & $31.82$ & $2.23\pm0.04$ & $0.35\pm0.06$ & $1.10\pm0.09$ & $4.9\pm1.5$ & $1.04\pm0.3$ & $159\pm--^2$ & $26.1^4$ & $147\pm1.9$\\
NGC\,7173 & $12.95$ & $12.03$ & $9.83$ & $8.96$ & $0.023$ & $32.48$ & $2.51\pm0.08$ & $0.25\pm0.08$ & $0.79\pm0.10$ & $6.0\pm1.7$ & $2.0\pm0.4$ & $208\pm--^2$ & $14.7^4$ & $201\pm4.6$\\
NGC\,4033 & $12.61$ & $11.70$ & $9.58$ & $8.70$ & $0.042$ & $31.66$ & $2.49\pm0.05$ & $0.26\pm0.03$ & $1.27\pm0.07$ & $4.5\pm1$ & $0.6\pm0.1$ & $111\pm--^2$ & $16.1^4$ & $120\pm7$\\
NGC\,1339 & $12.51$ & $11.58$ & $9.59$ & $8.69$ & $0.013$ & $31.55$ & $2.26\pm0.04$ & $0.43\pm0.04$ & $1.40\pm0.08$ & $4.8\pm0.8$ & $0.8\pm0.1$ & $280\pm--^3$ & $16.9^4$ & $157\pm3$\\
NGC\,4564 & $12.05$ & $11.12$ & $8.87$ & $7.94$ & $0.029$ & $31.01$ & $2.34\pm0.06$ & $0.29\pm0.04$ & $1.35\pm0.08$ & $4.2\pm1.0$ & $0.6\pm0.1$ & $213\pm31^1$ & $21.7^4$ & $156\pm2.2$\\
NGC\,1351 & $12.46$ & $11.58$ & $9.61$ & $8.79$ & $0.013$ & $31.42$ & $2.49\pm0.03$ & $0.28\pm0.02$ & $1.14\pm0.03$ & $6.8\pm0.8$ & $0.9\pm0.1$ & $370\pm--^3$ & $25.5^4$ & $137\pm2.9$\\
NGC\,1172 & $12.70$ & $11.86$ & $10.08$ & $9.22$ & $0.060$ & $31.63$ & $2.22\pm0.03$ & $0.51\pm0.06$ & $1.21\pm0.10$ & $5.5\pm1.4$ & $1.2\pm0.2$ & $265\pm--^2$ & $38.6^4$ & $111\pm4.1$\\
NGC\,3377 & $11.24$ & $10.38$ & $8.29$ & $7.44$ & $0.030$ & $30.13$ & $2.13\pm0.04$ & $0.44\pm0.08$ & $1.05\pm0.14$ & $6.20\pm1.9$ & $1.5\pm0.4$ & $173\pm--^2$ & $33.7^4$ & $136\pm2.3$\\
NGC\,4660 & $12.16$ & $11.24$ & $9.11$ & $8.21$ & $0.030$ & $30.88$ & $2.16\pm0.03$ & $0.55\pm0.06$ & $1.61\pm0.12$ & $4.6\pm1.0$ & $0.7\pm0.1$ & $205\pm28^1$ & $12.8^4$ & $192\pm3.2$\\
NGC\,1419 & $13.48$ & $12.59$ & $10.73$ & $9.89$ & $0.011$ & $31.82$ & $2.19\pm0.05$ & $0.38\pm0.05$ & $1.44\pm0.11$ & $3.7\pm0.9$ & $0.6\pm0.1$ & $160\pm--^3$ & $10.9^4$ & $117\pm3.1$\\
NGC\,1336 & $13.10$ & $12.26$ & $10.75$ & $9.81$ & $0.010$ & $31.36$ & $2.28\pm0.03$ & $0.41\pm0.04$ & $1.24\pm0.07$ & $6.3\pm1.2$ & $1.0\pm0.2$ & $355\pm--^3$ & $--$ & $97.7\pm4.2$\\
NGC\,4387 & $12.12$ & $13.01$ & $10.04$ & $9.15$ & $0.029$ & $31.14$ & $1.75\pm0.05$ & $0.38\pm0.05$ & $1.50\pm0.12$ & $2.7\pm0.6$ & $0.55\pm0.09$ & $70\pm10^1$ & $15.4^4$ & $100\pm2.8$\\
NGC\,4458 & $12.93$ & $12.07$ & $10.22$ & $9.32$ & $0.021$ & $31.06$ & $2.14\pm0.03$ & $0.31\pm0.04$ & $1.54\pm0.13$ & $2.8\pm0.7$ & $0.45\pm0.08$ & $72\pm12^1$ & $26.7^4$ & $97.4\pm2.0$\\
NGC\,4434 & $13.03$ & $12.15$ & $10.07$ & $9.21$ & $0.019$ & $31.15$ & $2.22\pm0.04$ & $0.21\pm0.02$ & $1.19\pm0.05$ & $3.9\pm0.6$ & $0.58\pm0.08$ & $141\pm34^1$ & $18.5^4$ & $116\pm2.8$\\
NGC\,4623 & $13.24$ & $12.36$ & $10.31$ & $9.47$ & $0.020$ & $31.17$ & $2.26\pm0.05$ & $0.23\pm0.03$ & $1.53\pm0.12$ & $2.3\pm0.5$ & $0.34\pm0.06$ & $71\pm14^1$ & $14.5^5$ & $77.0\pm2.8$\\
NGC\,1428 & $13.74$ & $--$ & $10.82$ & $10.03$ & $0.009$ & $31.56$ & $1.79\pm0.05$ & $0.34\pm0.05$ & $1.64\pm0.16$ & $1.9\pm0.5$ & $0.42\pm0.08$ & $40\pm--^3$ & $--$ & $80.0\pm10$\\
NGC\,4515 & $13.30$ & $12.54$ & $10.70$ & $9.89$ & $0.027$ & $31.01$ & $2.13\pm0.05$ & $0.34\pm0.04$ & $1.74\pm0.15$ & $2.4\pm0.4$ & $0.40\pm0.06$ & $81\pm10^1$ & $9.5^5$ & $82.2\pm2.6$\\
NGC\,1380B & $13.87$ & $12.92$ & $10.89$ & $10.04$ & $0.014$ & $31.52$ & $2.31\pm0.04$ & $0.32\pm0.03$ & $1.37\pm0.09$ & $3.9\pm0.8$ & $0.59\pm0.09$ & $170\pm--^3$ & $--$ & $76.6\pm2.7$\\
VS\,1$^{\rm a}$ & $13.64$ & $13.02$ & $11.06$ & $10.19$ & $0.060$ & $31.63$ & $2.37\pm0.04$ & $0.24\pm0.03$ & $1.71\pm0.16$ & $1.0\pm0.2$ & $0.27\pm0.04$ & $24.7\pm6^1$ & $13.5^5$ & $94\pm15^{\rm b}$\\
VS\,2$^{\rm a}$ & $14.05$ & $13.18$ & $12.21$ & $11.42$ & $0.030$ & $31.10$ & $2.62\pm0.09$ & $0.21\pm0.06$ & $1.51\pm0.20$ & $1.8\pm0.6$ & $0.30\pm0.08$ & $34.6\pm8^1$ & $24.7^5$ & $39\pm4^{\rm b}$\\
VS\,3$^{\rm a}$ & $14.38$ & $13.77$ & $12.37$ & $11.50$ & $0.032$ & $31.09$ & $2.66\pm0.03$ & $0.31\pm0.03$ & $1.97\pm0.15$ & $1.5\pm0.2$ & $0.30\pm0.04$ & $22.5\pm6^1$ & $21.5^5$ & $42\pm3^{\rm b}$\\
\hline
\end{tabular}
\end{center}
$^{\rm a}$ These correspond to the stacked galaxies from the Virgo cluster 
(see the text for further details).\\
$^{\rm b}$ These values correspond to the mean of the central velocity dispersions of individual galaxies.\\
{\bf References:} $^1$\citet{pen08}, $^2$\citet{cho12}, $^3$This paper, 
$^4$\citet[][we are aware that in this paper de Vaucouleurs profiles were used, instead of S\'ersic ones]{fab89}, $^5$\citet{fer06b}
\end{minipage}   
\end{table}   
\end{landscape}

\begin{figure*}   
\includegraphics[width=55mm]{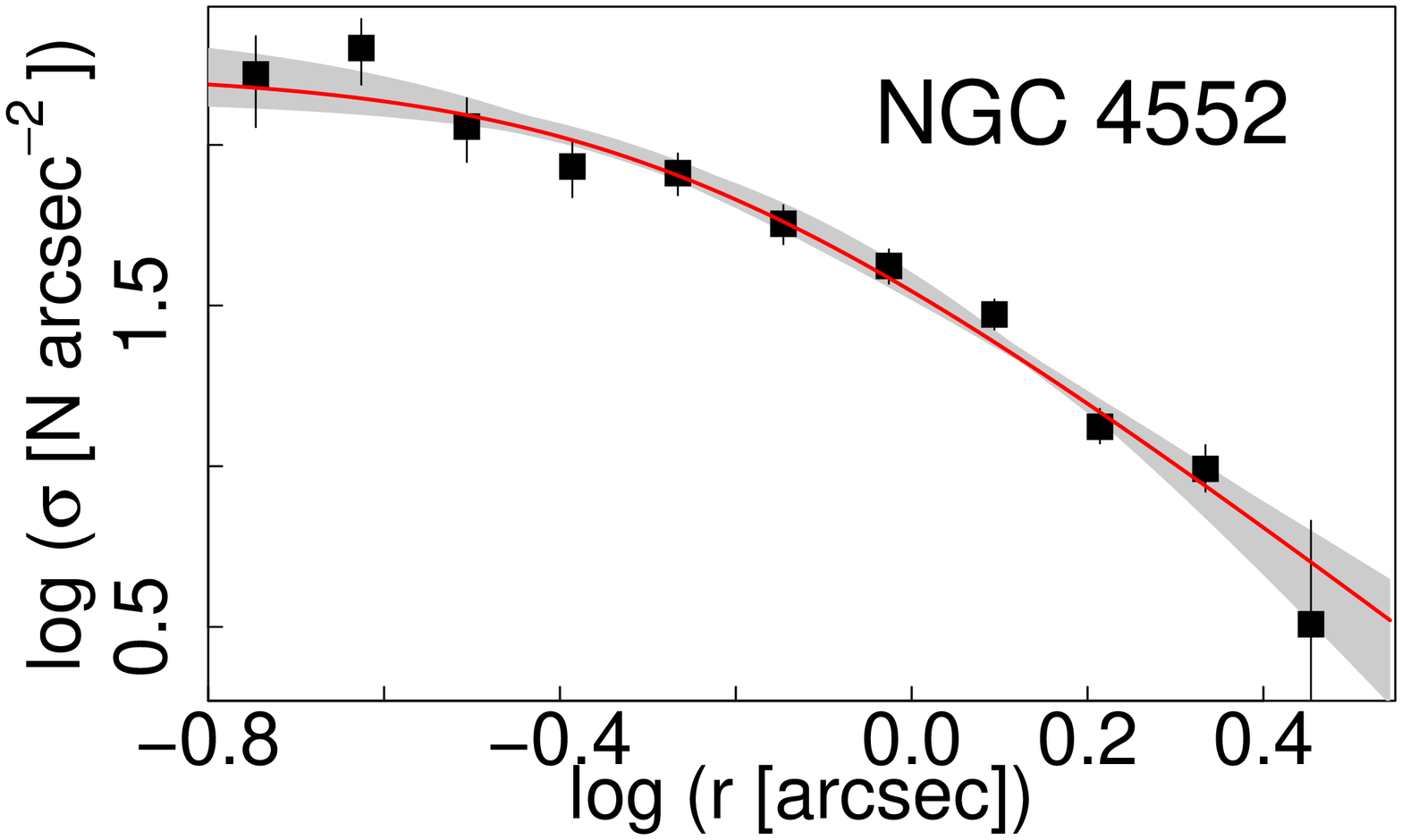}   
\includegraphics[width=55mm]{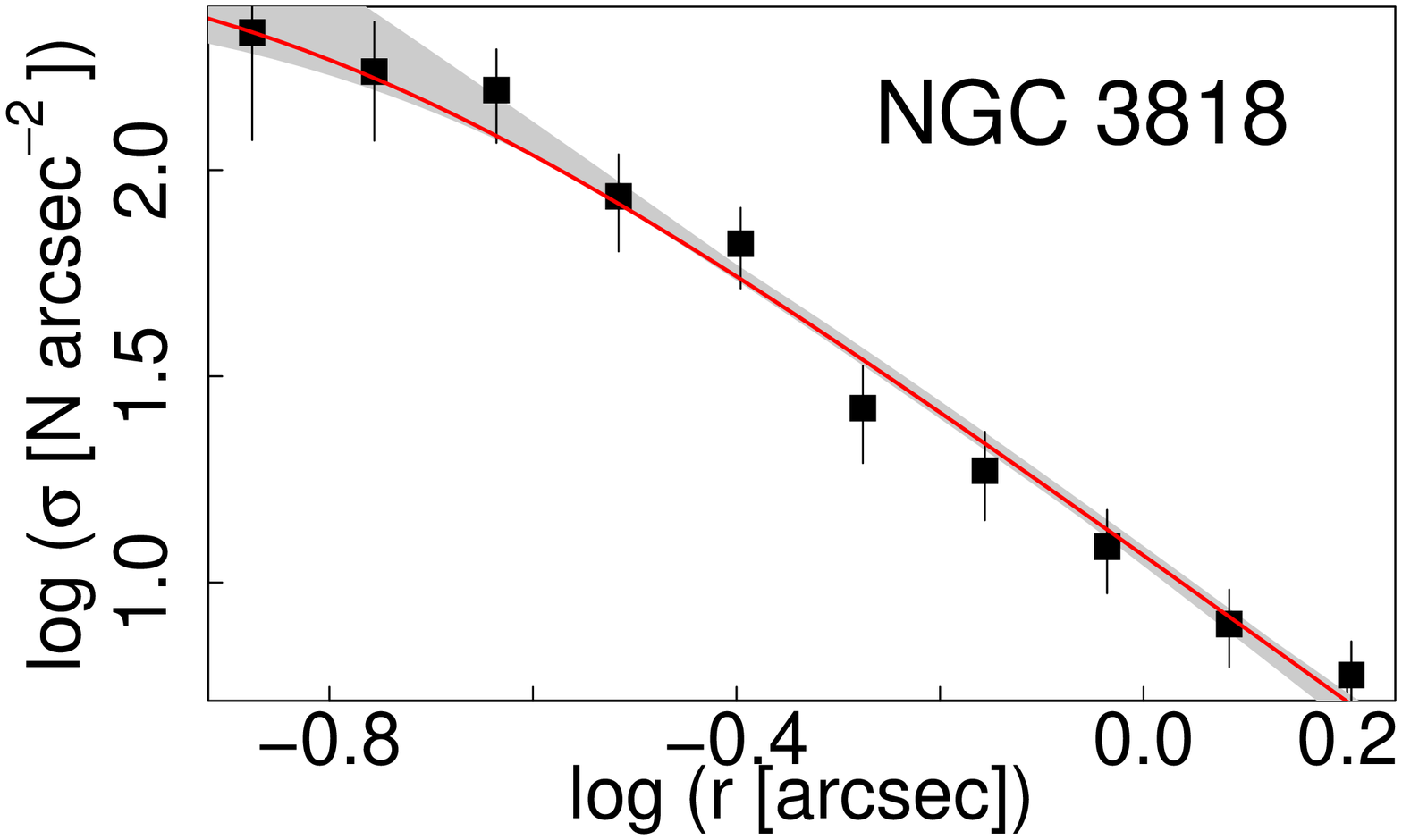}
\includegraphics[width=55mm]{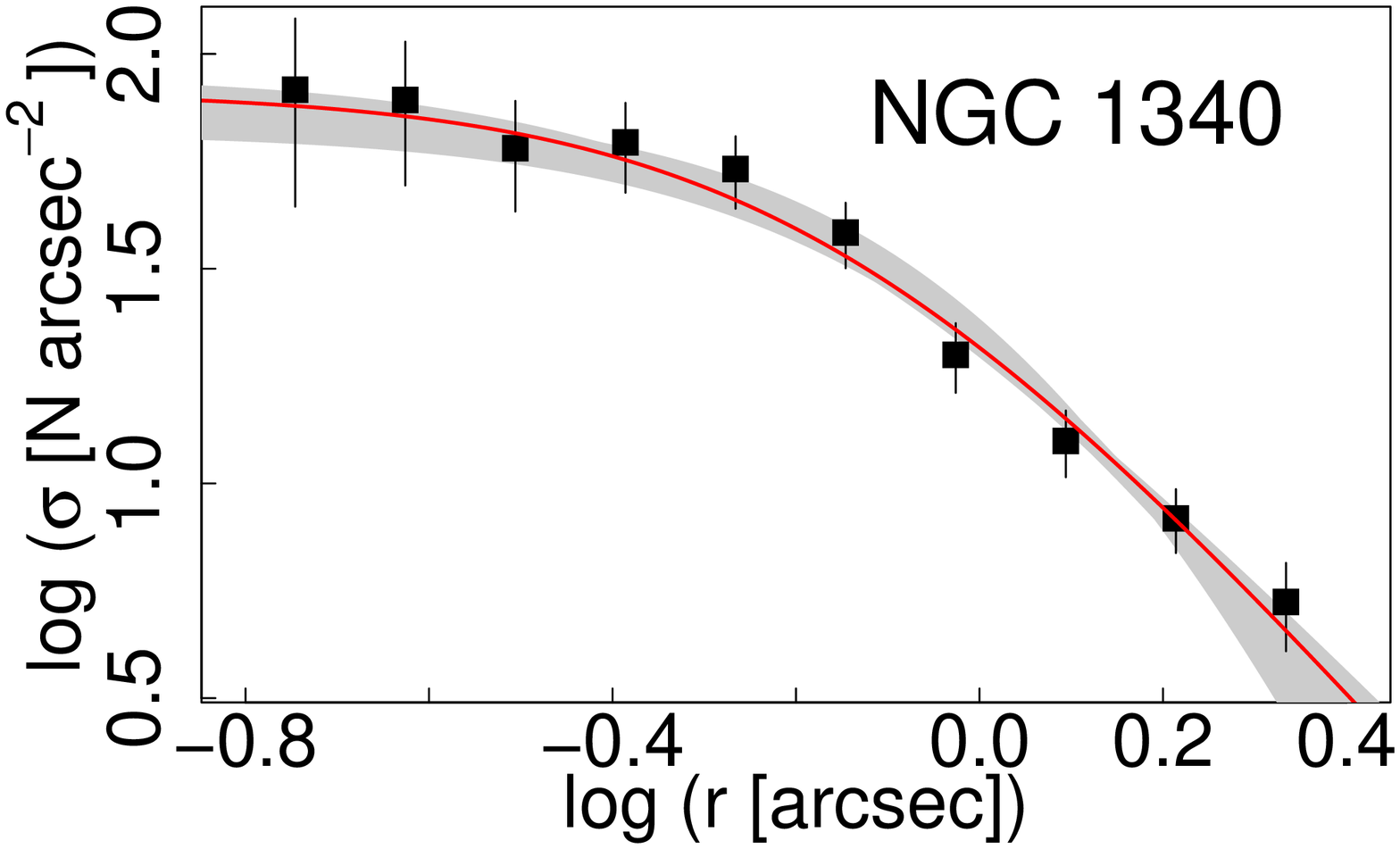}\\    
\includegraphics[width=55mm]{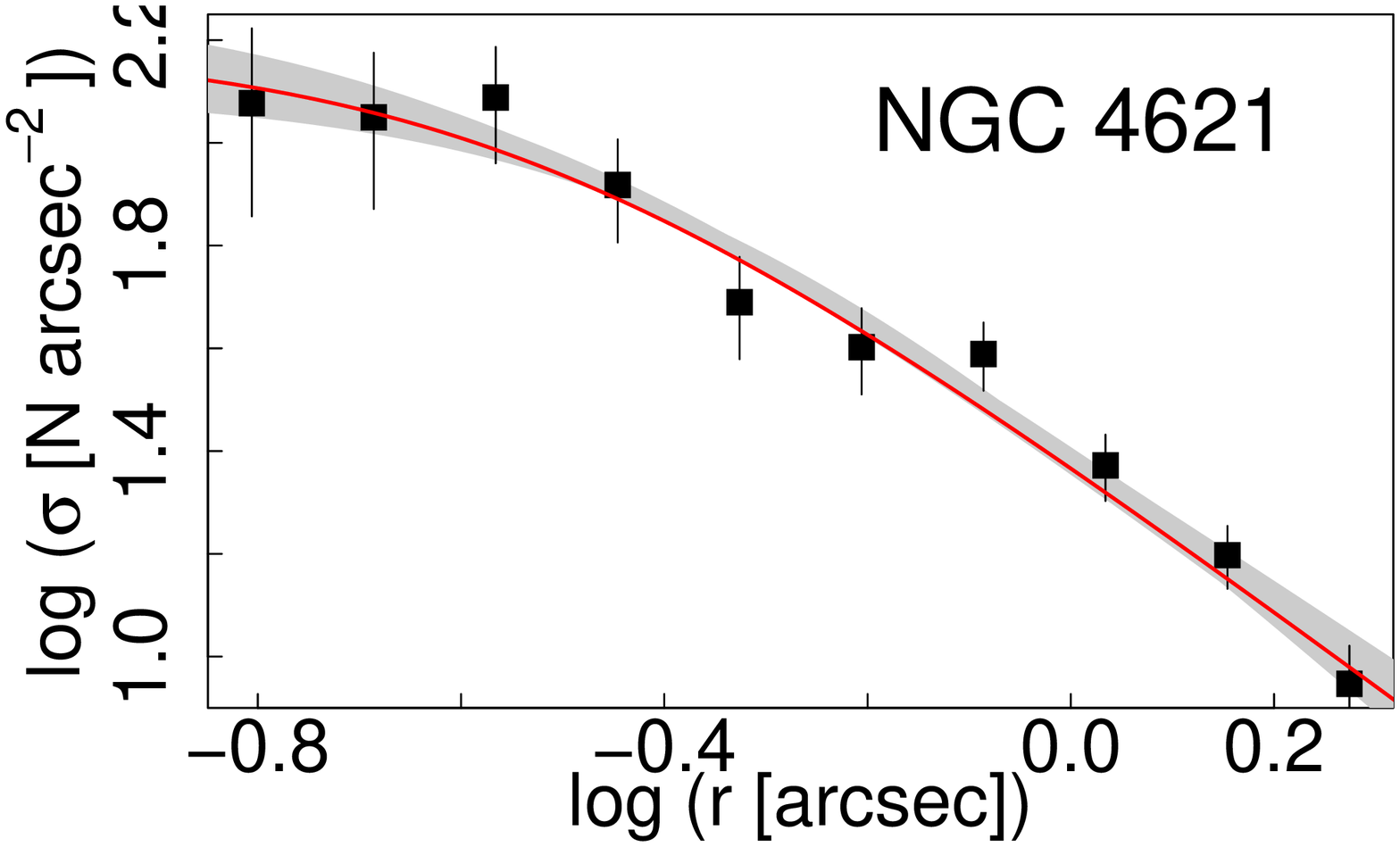}   
\includegraphics[width=55mm]{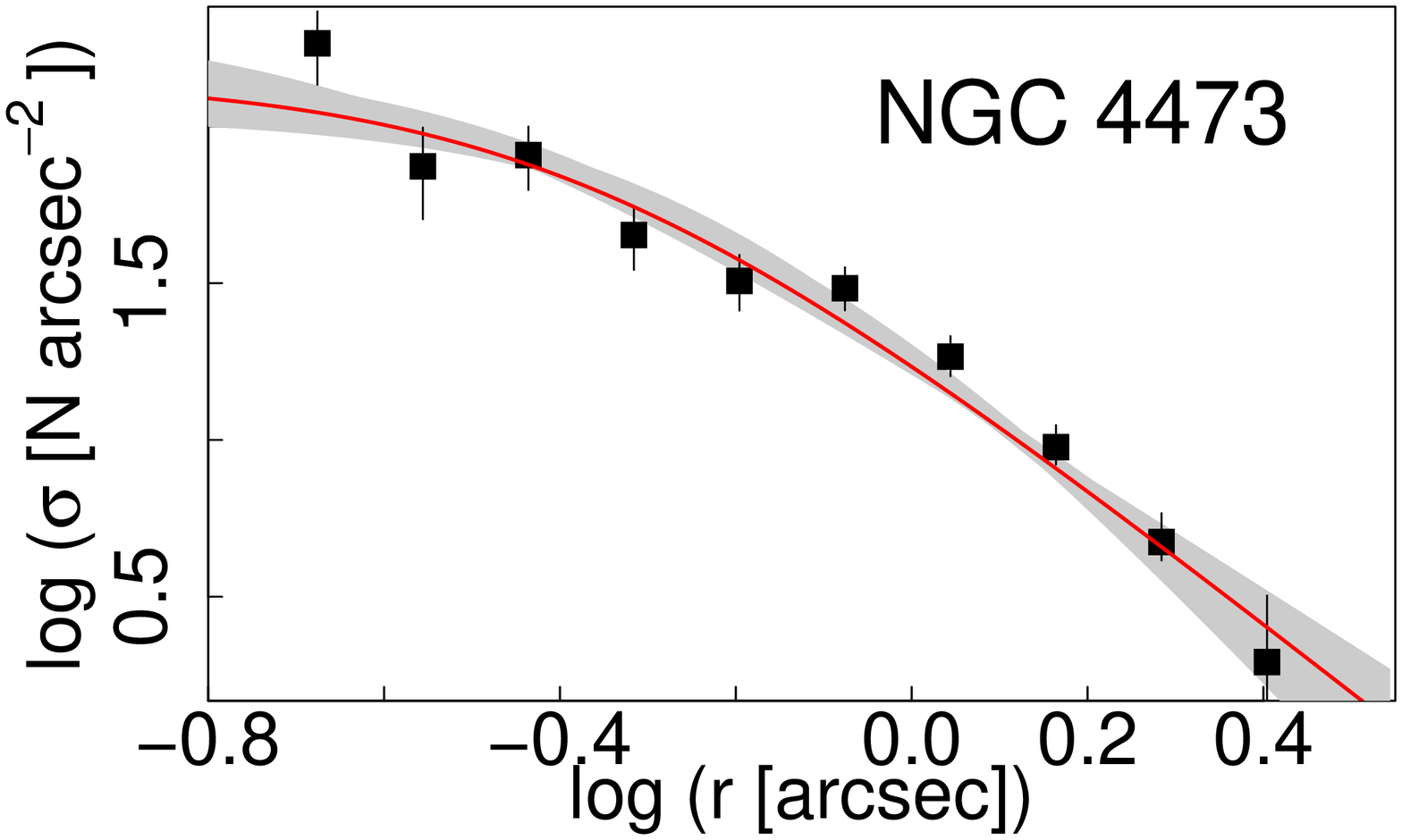}    
\includegraphics[width=55mm]{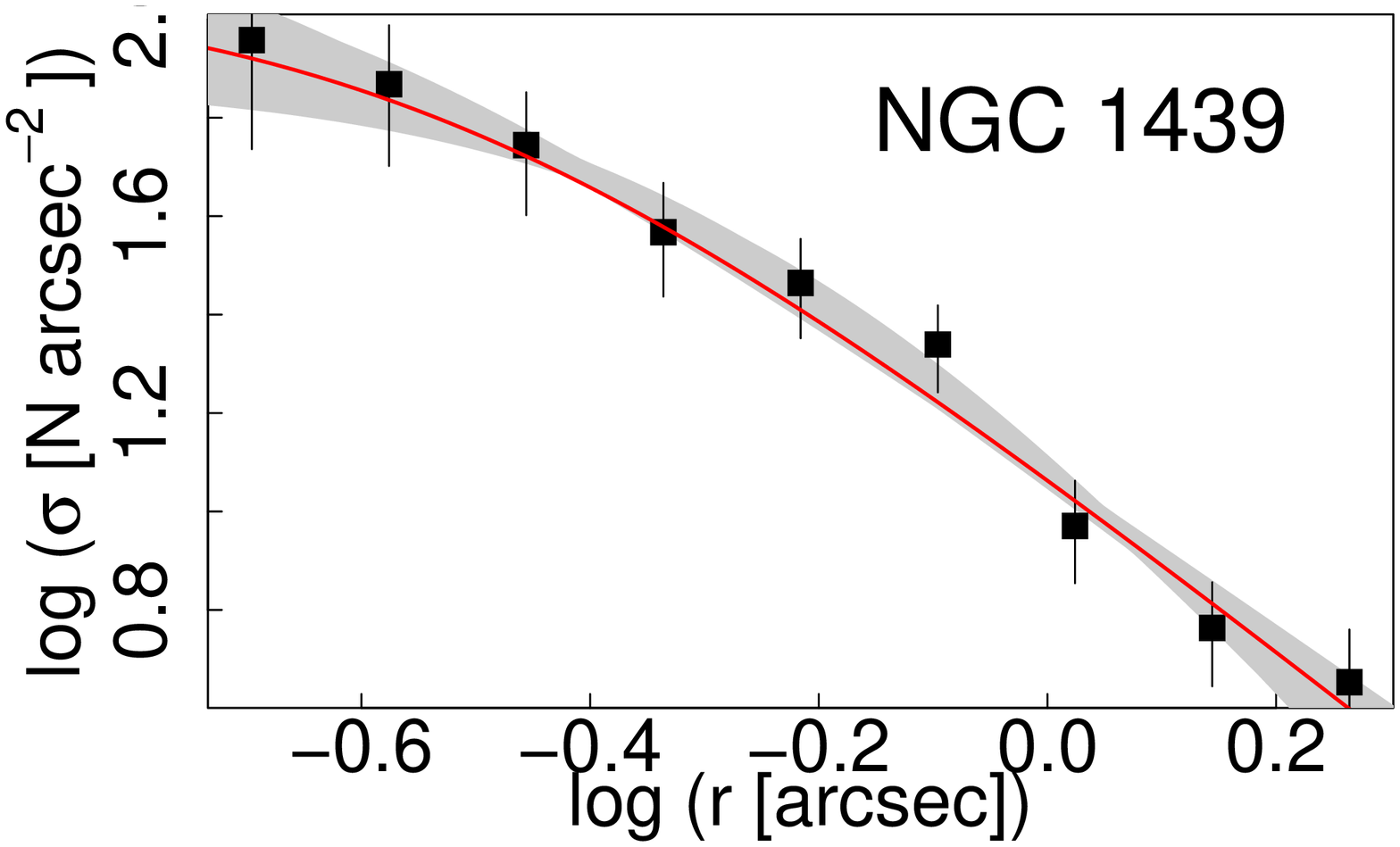}\\    
\includegraphics[width=55mm]{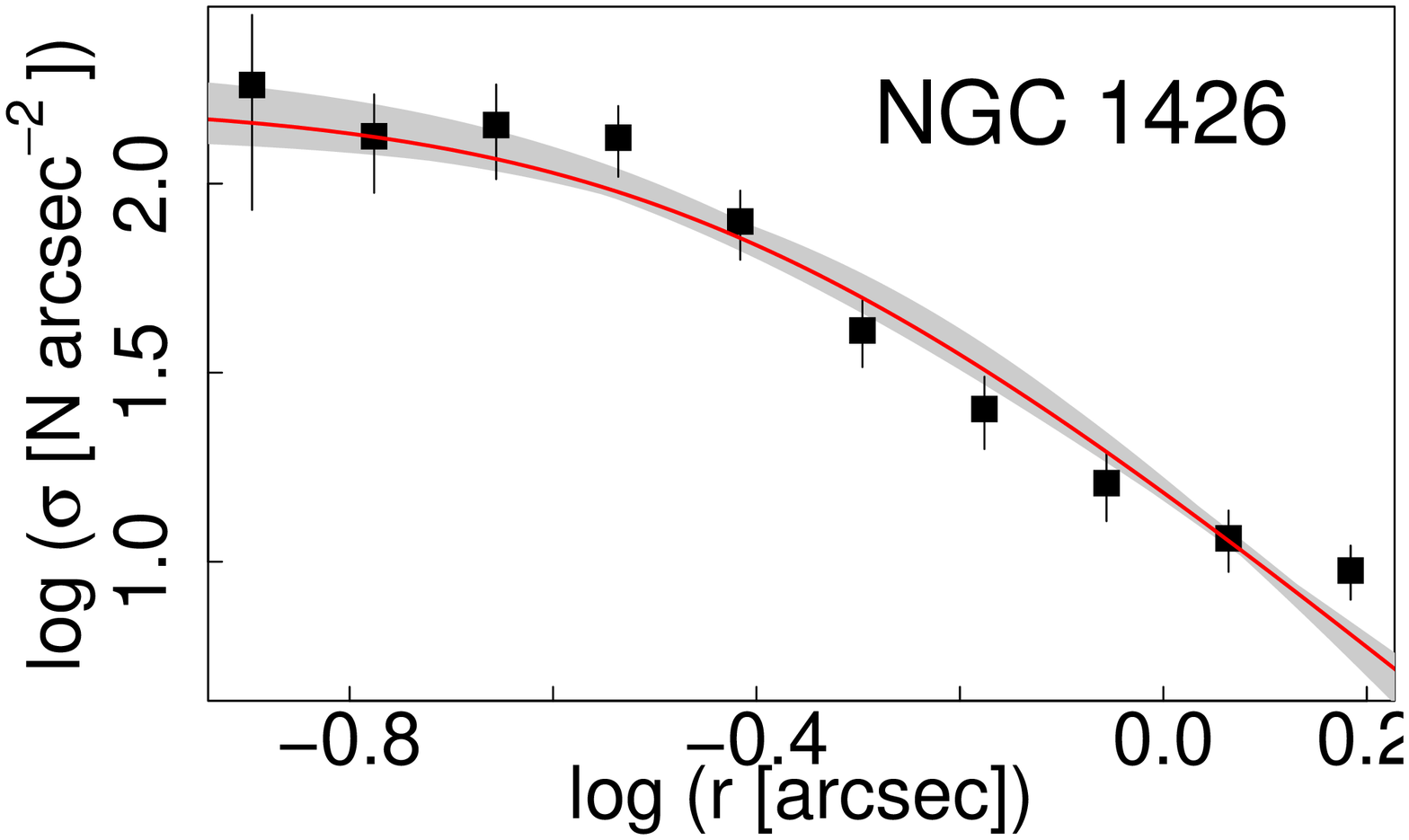}    
\includegraphics[width=55mm]{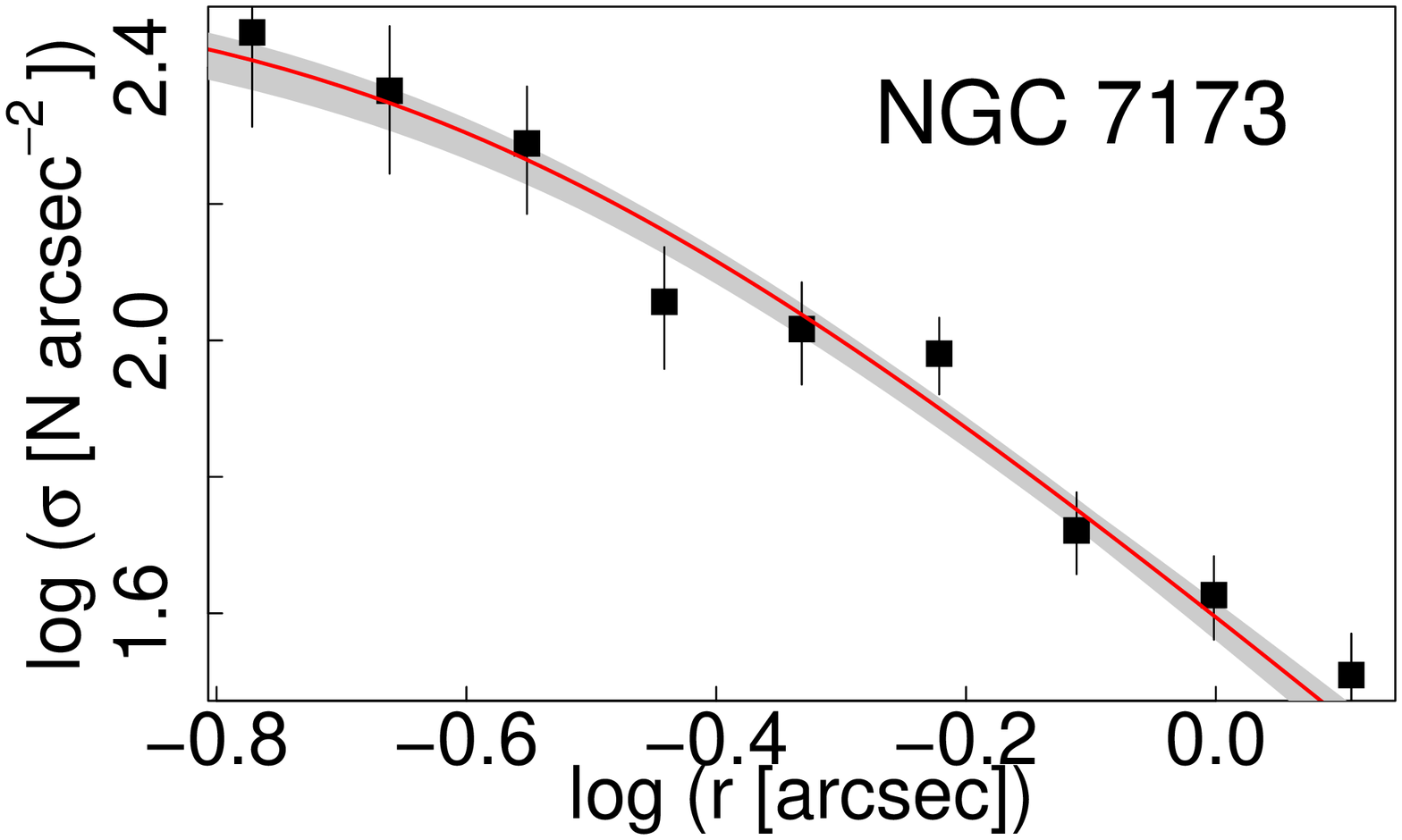}    
\includegraphics[width=55mm]{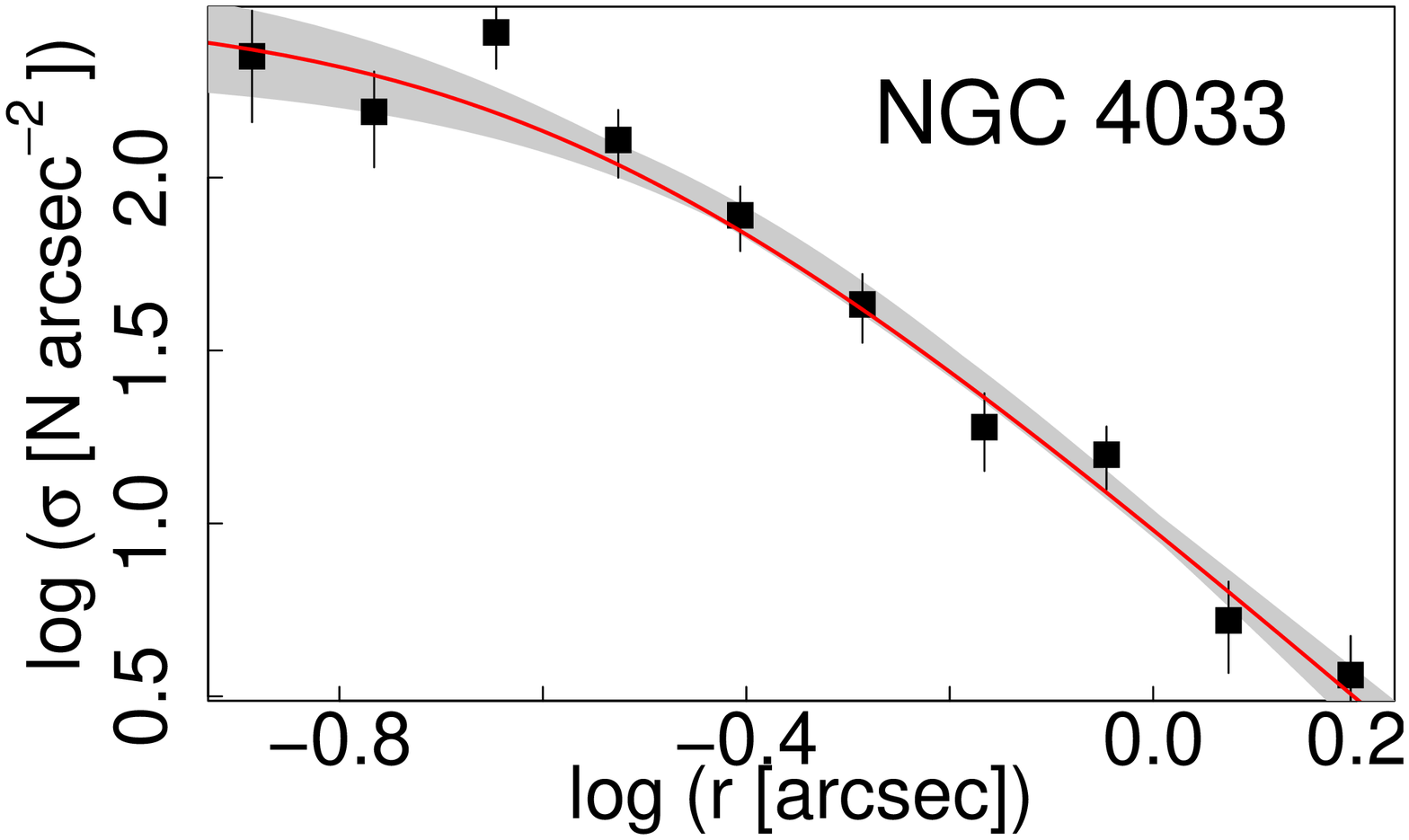}\\    
\includegraphics[width=55mm]{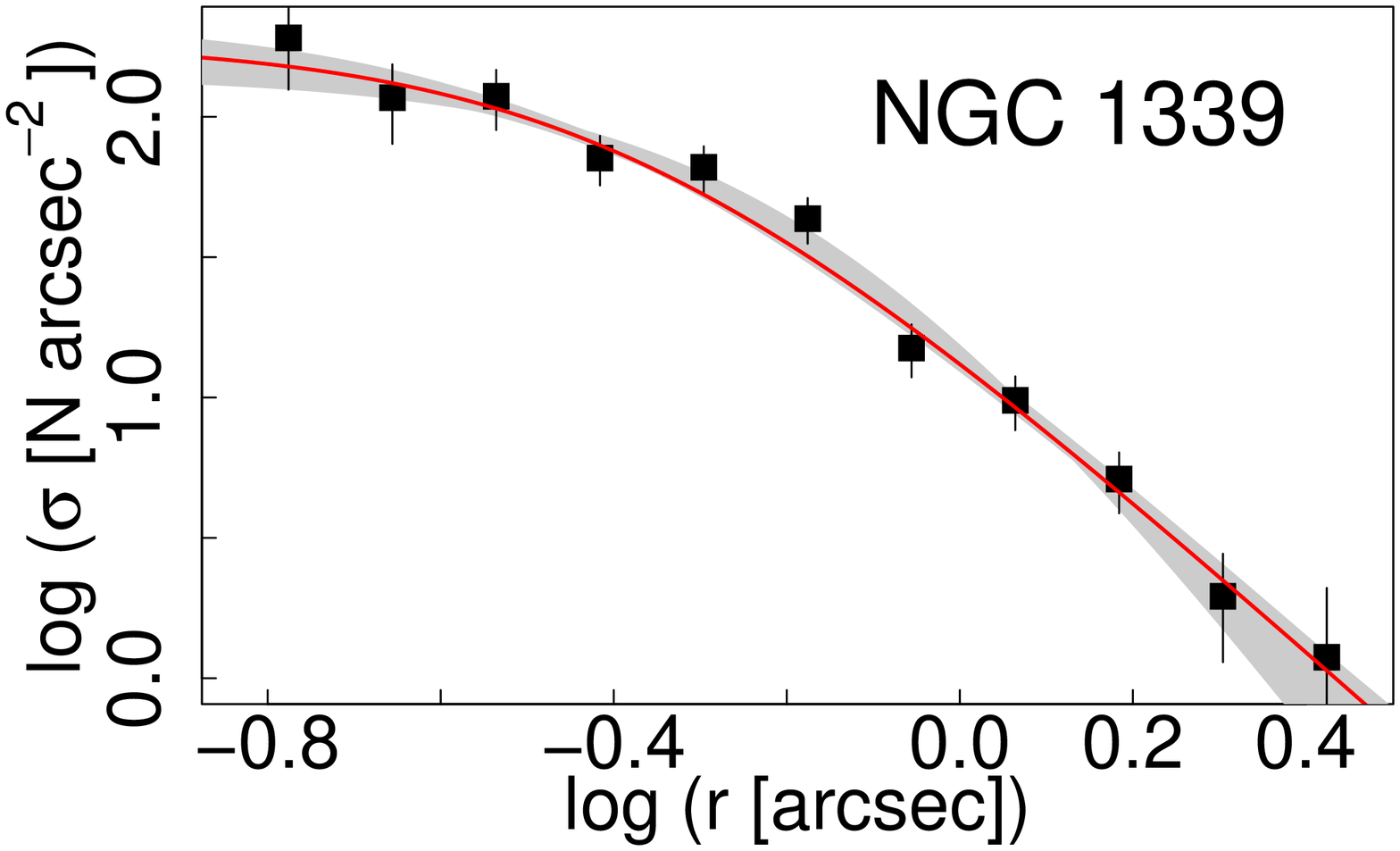}    
\includegraphics[width=55mm]{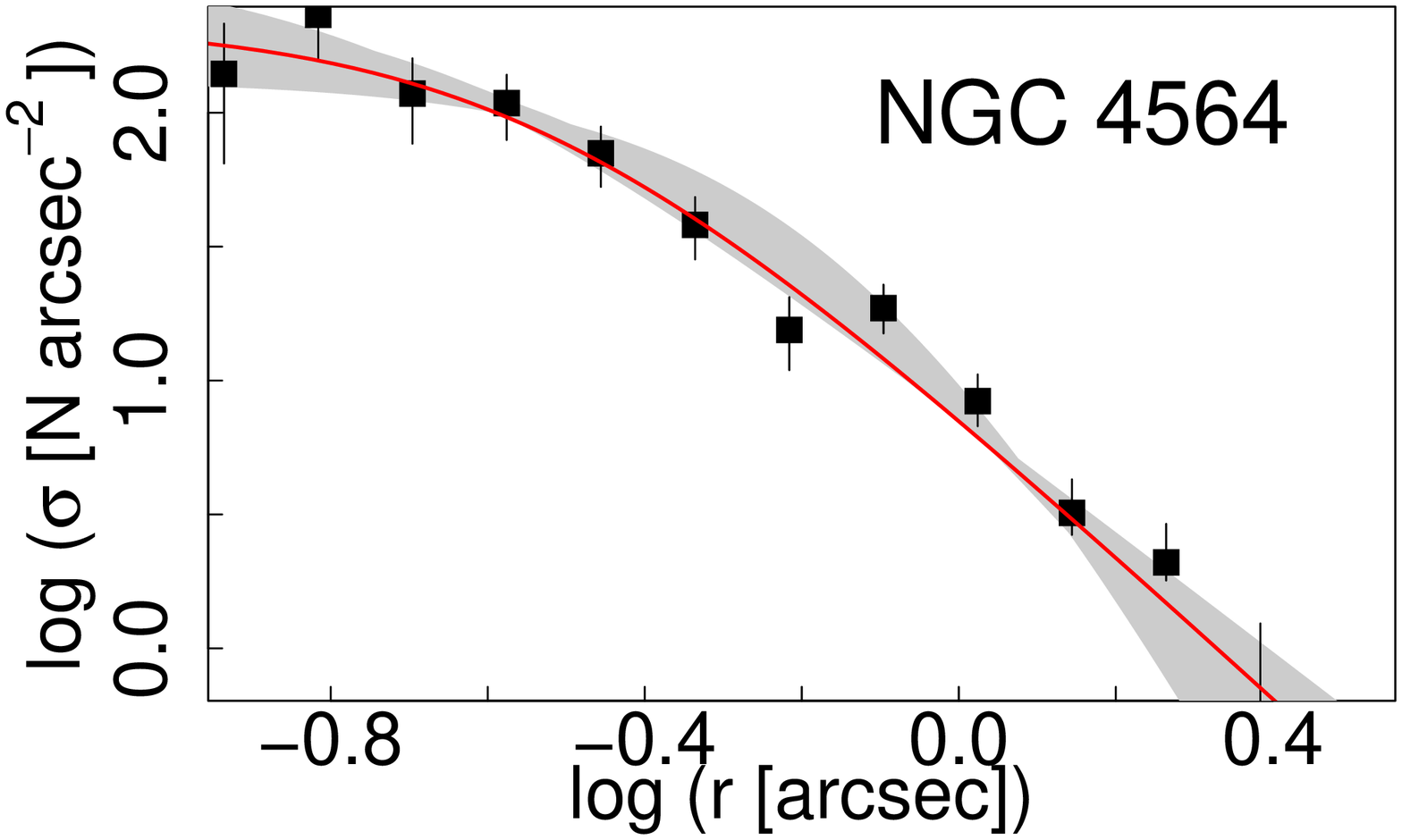}    
\includegraphics[width=55mm]{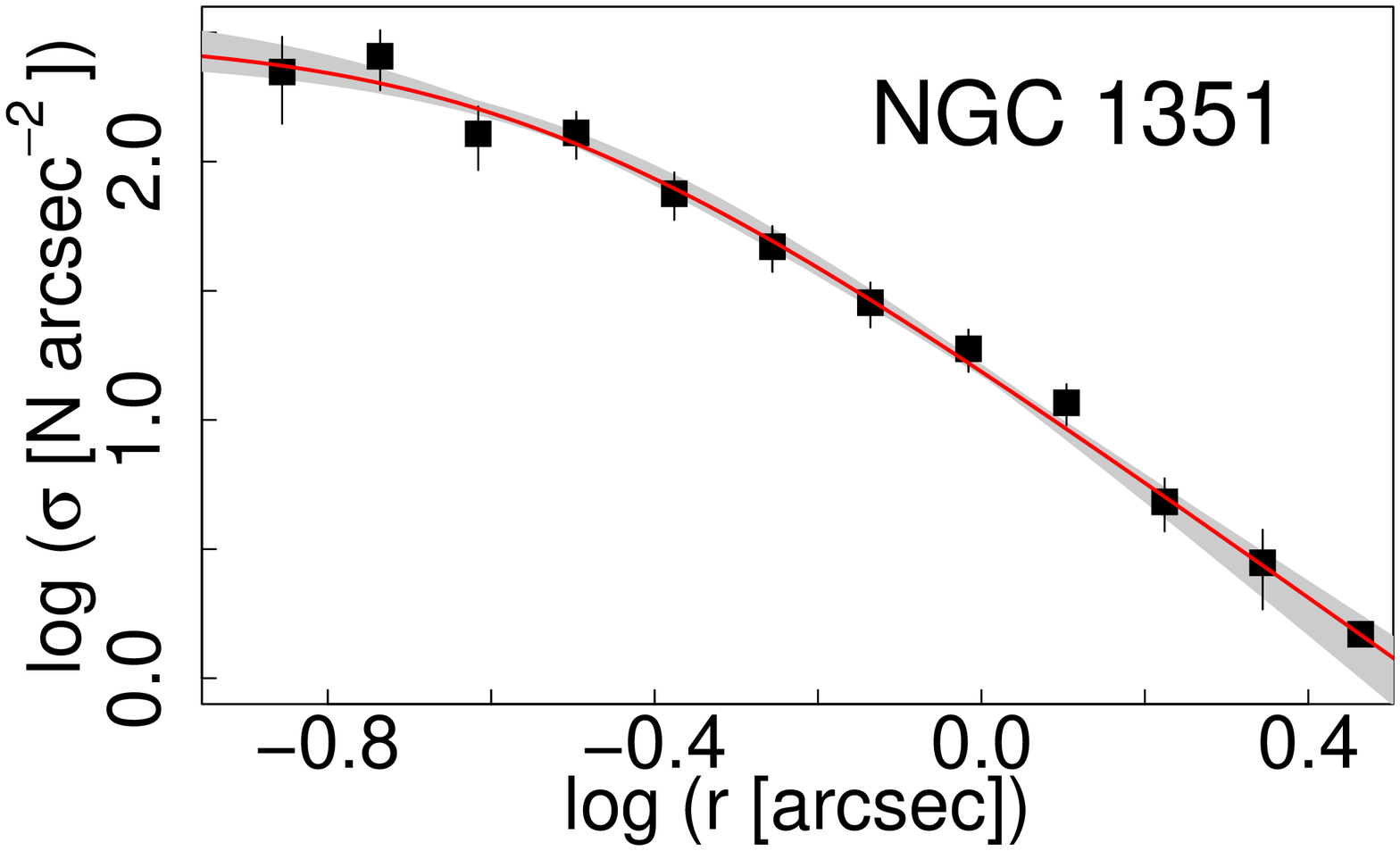}\\    
\includegraphics[width=55mm]{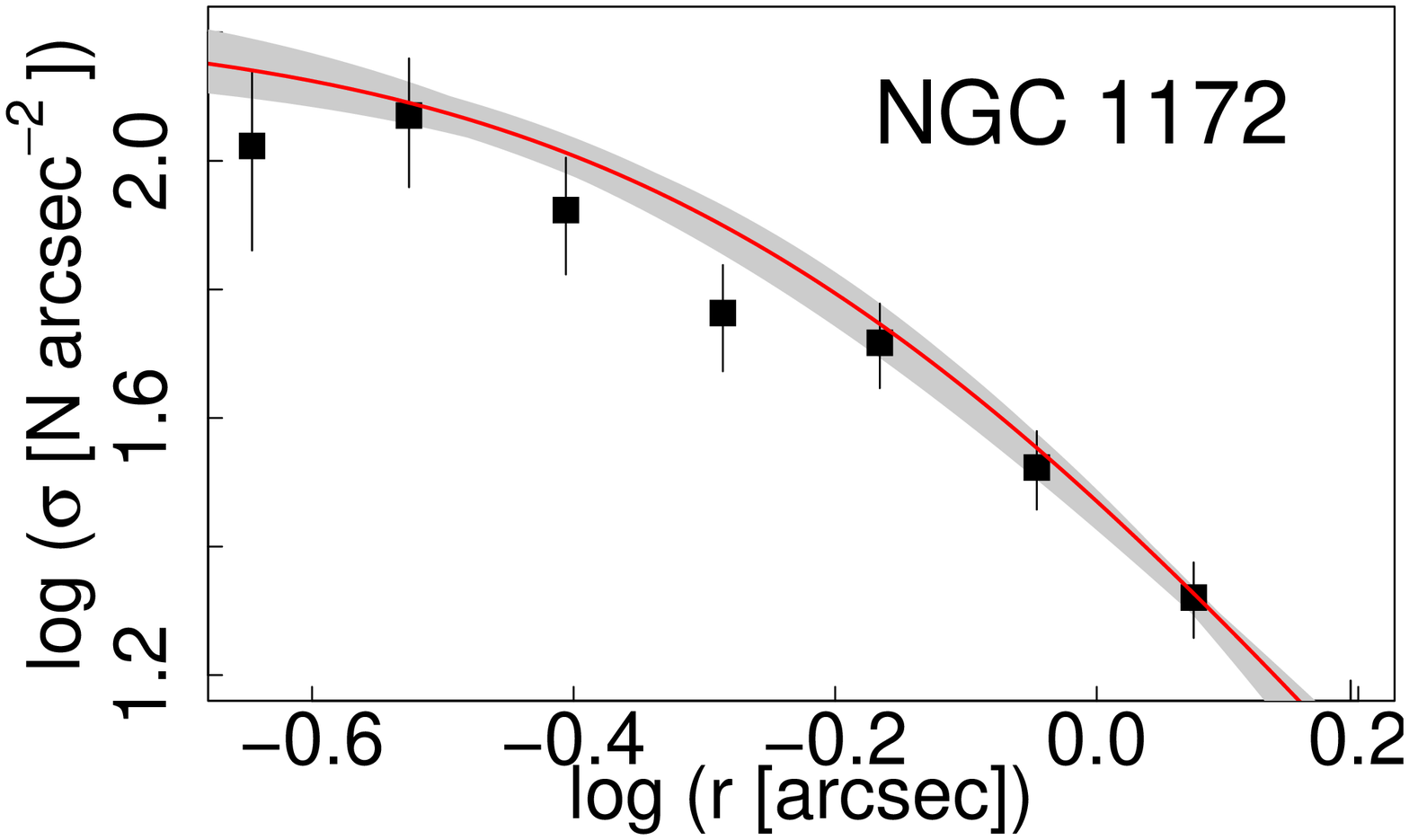}    
\includegraphics[width=55mm]{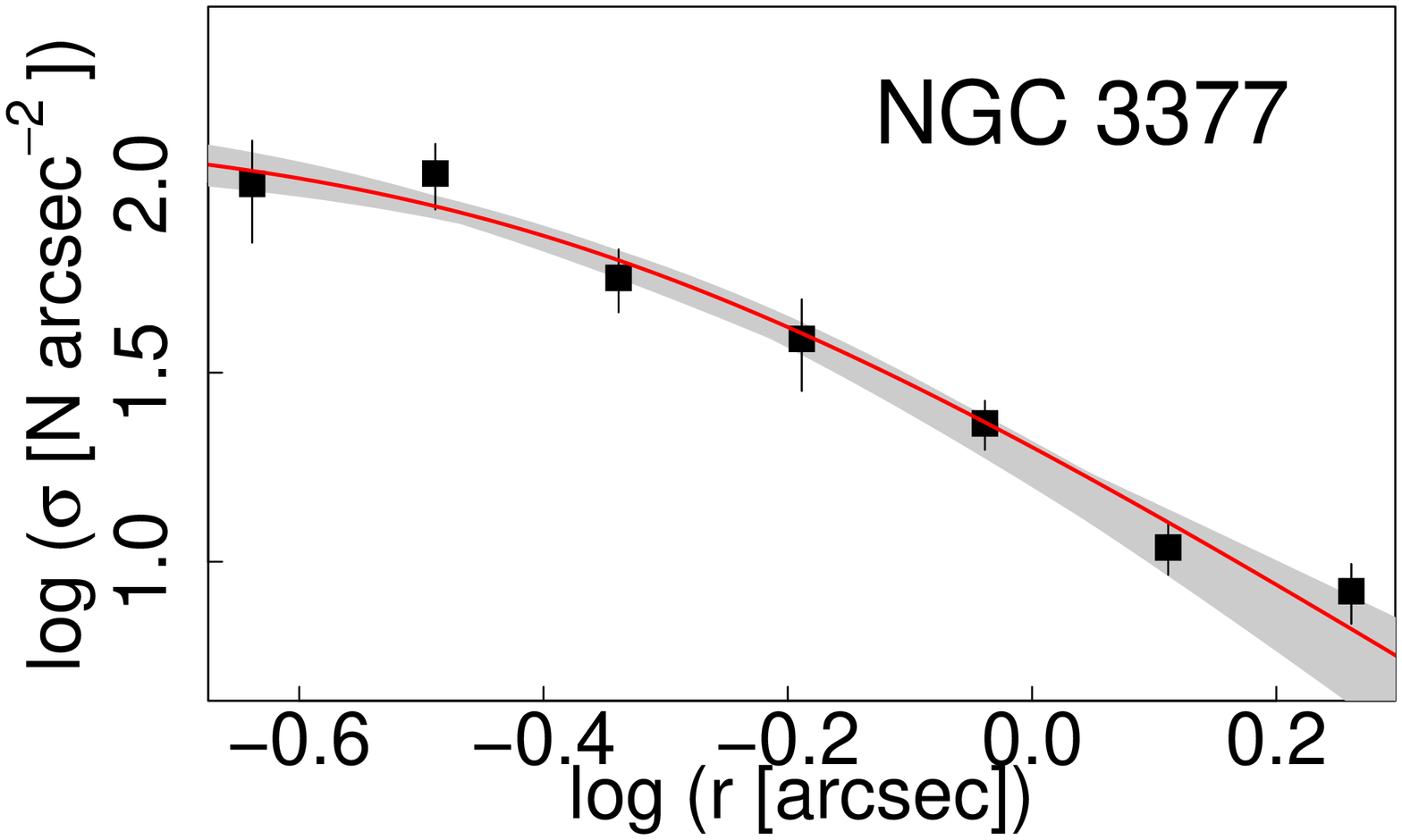}
\includegraphics[width=55mm]{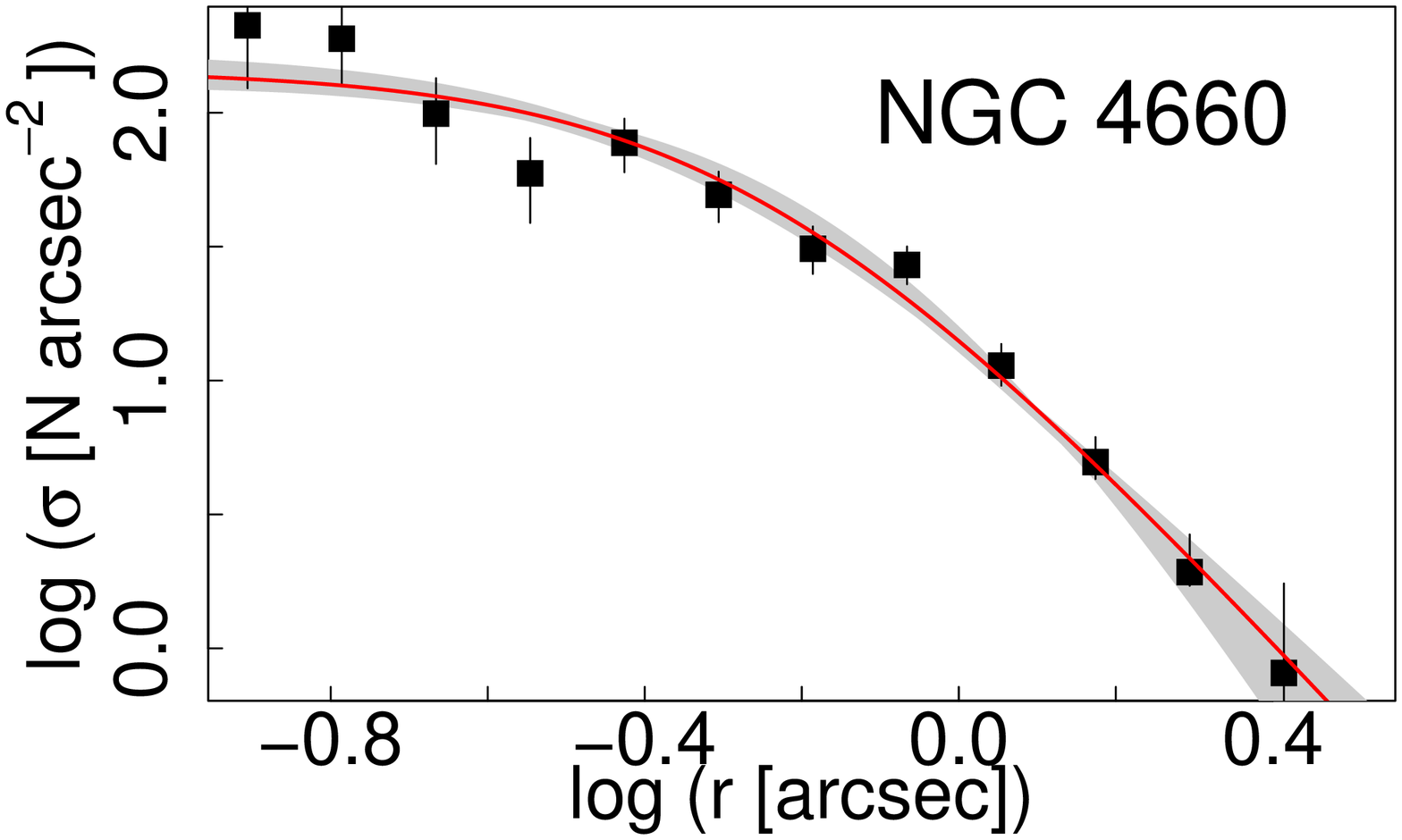}\\  
\includegraphics[width=55mm]{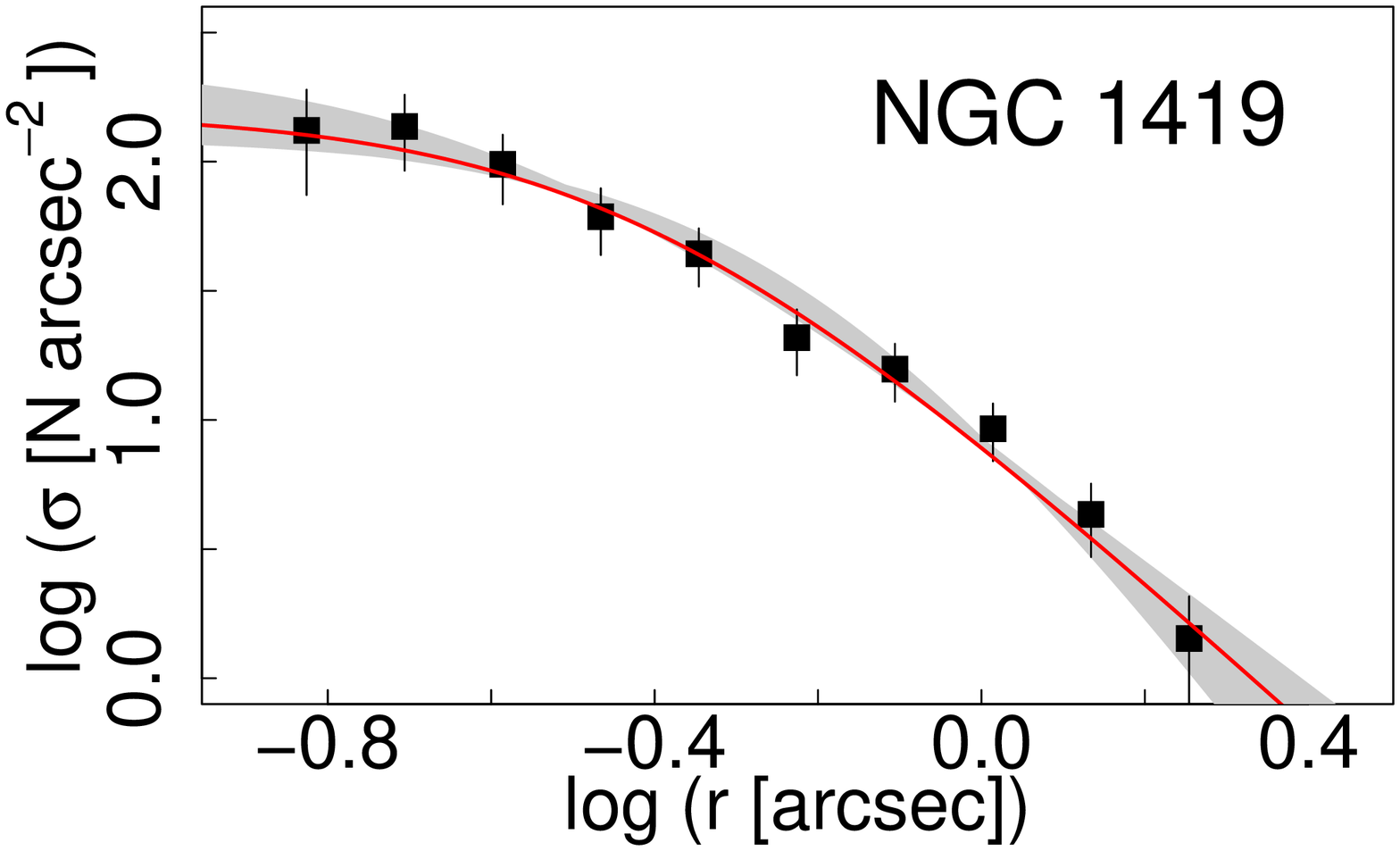}    
\includegraphics[width=55mm]{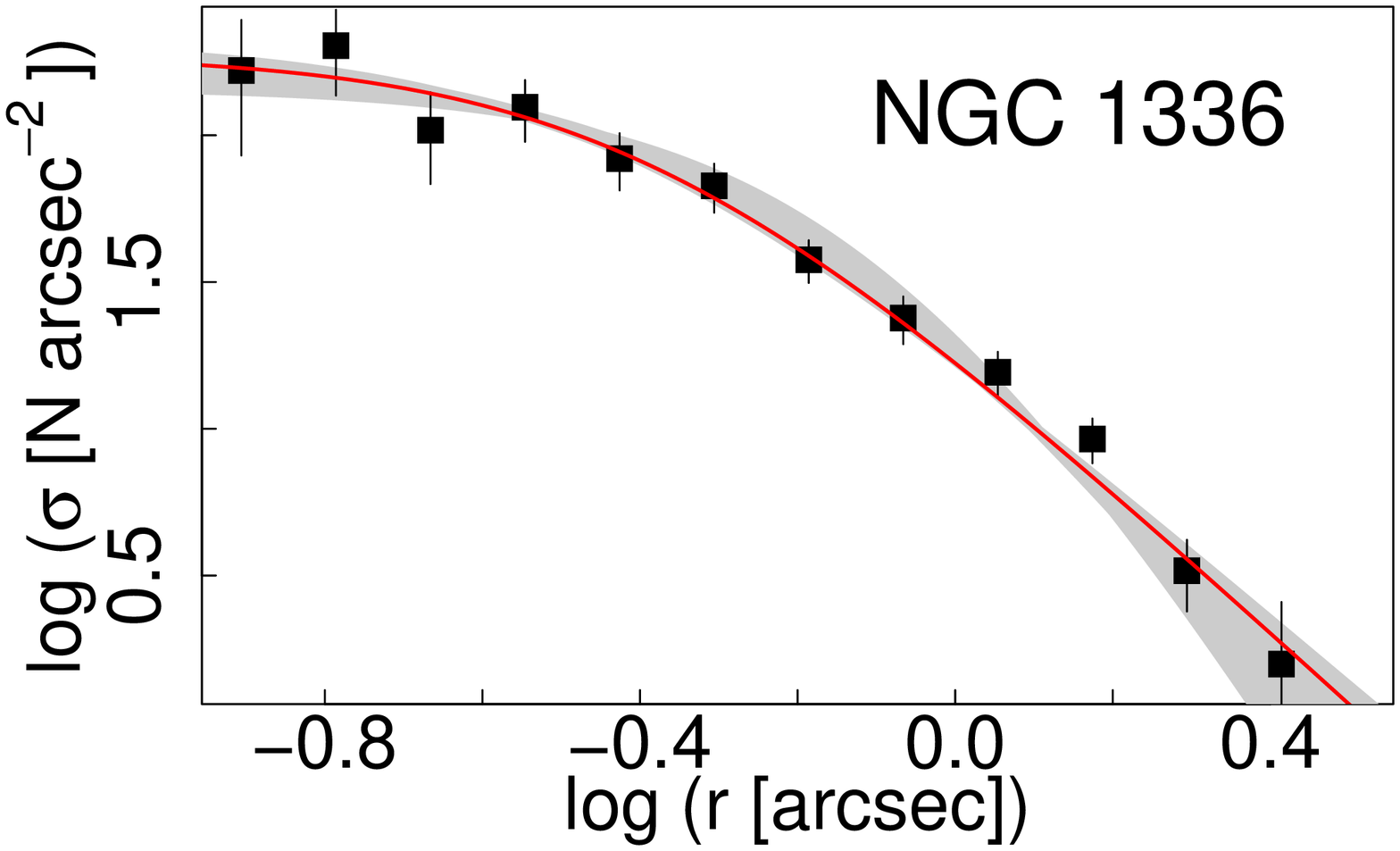}    
\includegraphics[width=55mm]{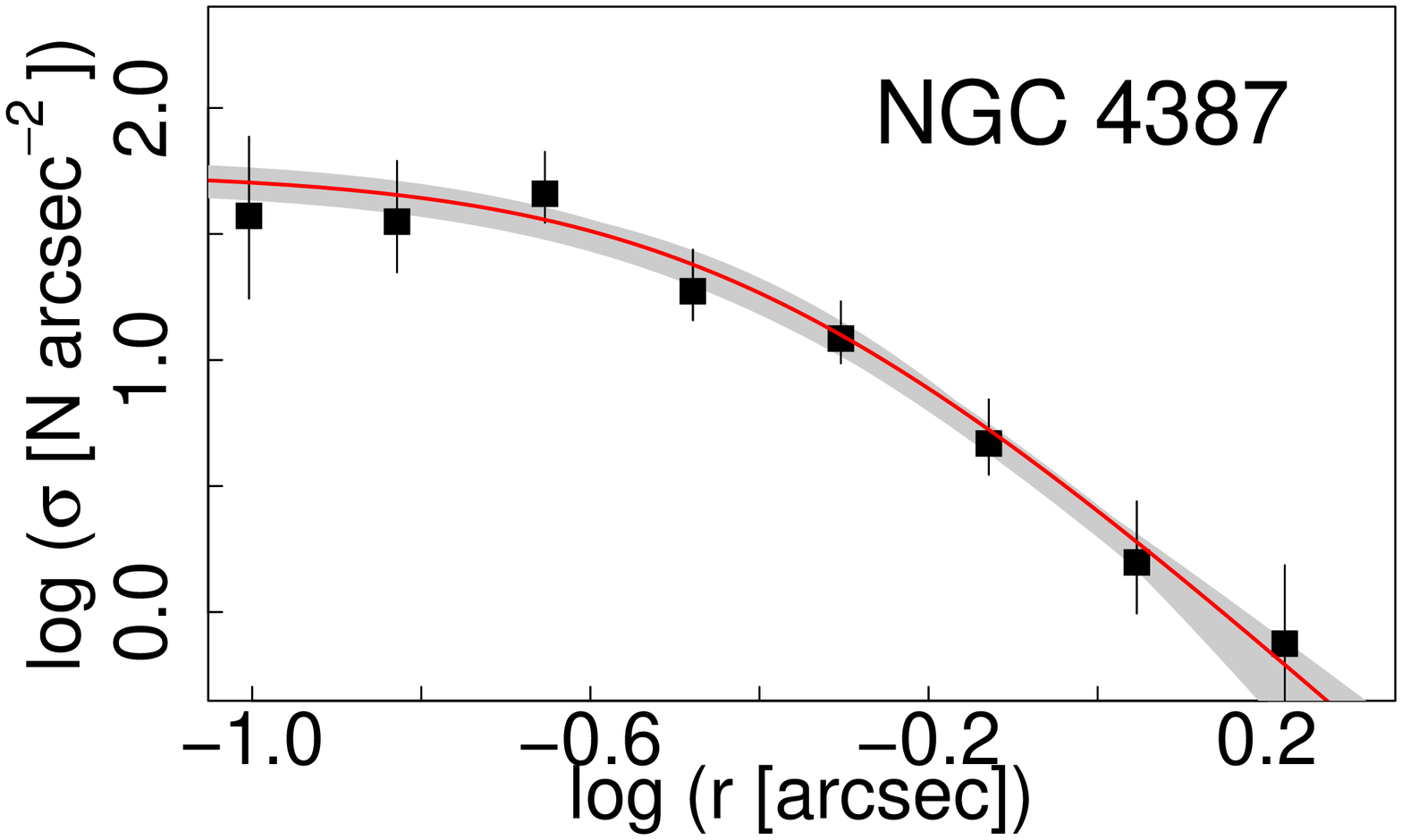} \\  
\includegraphics[width=55mm]{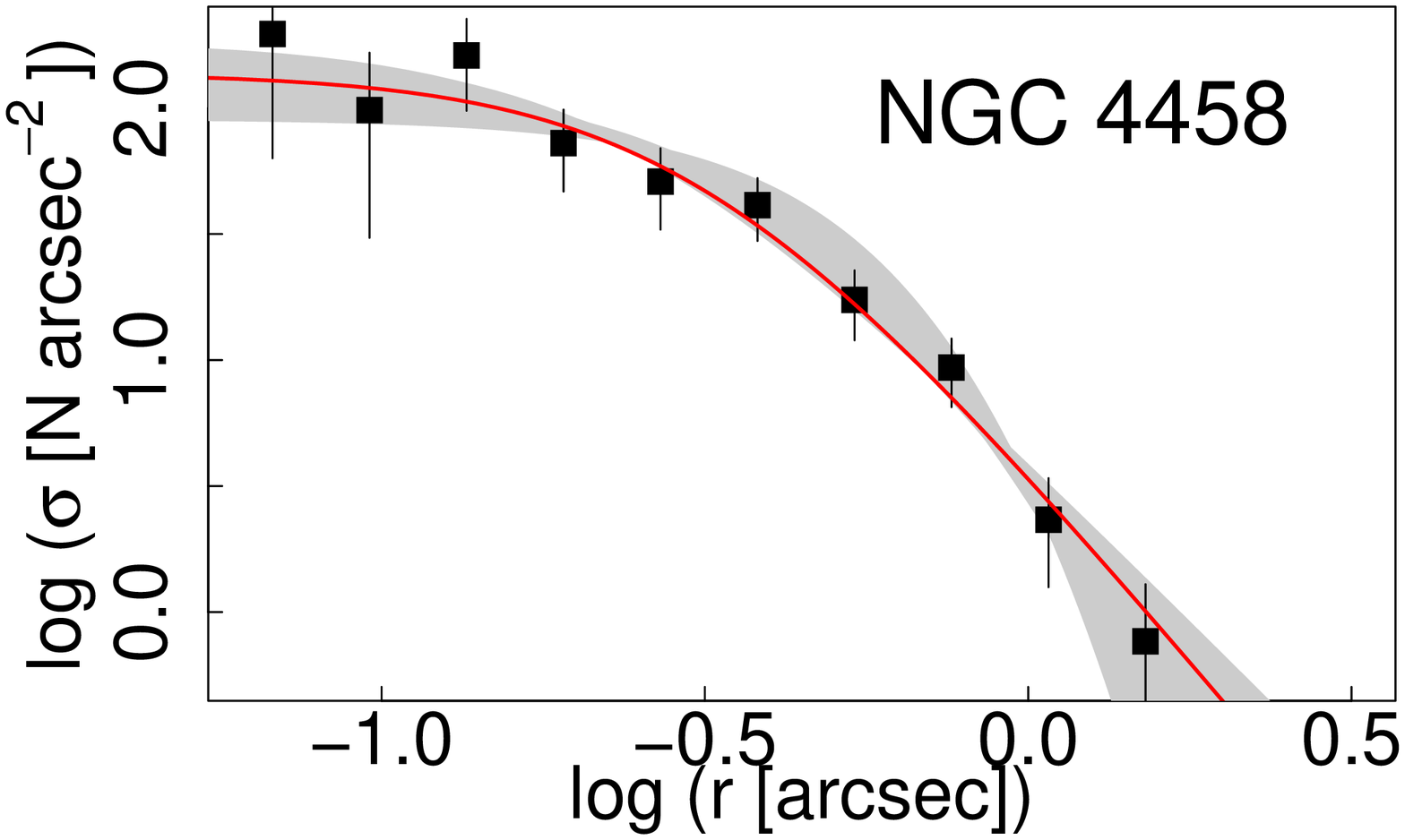} 
\includegraphics[width=55mm]{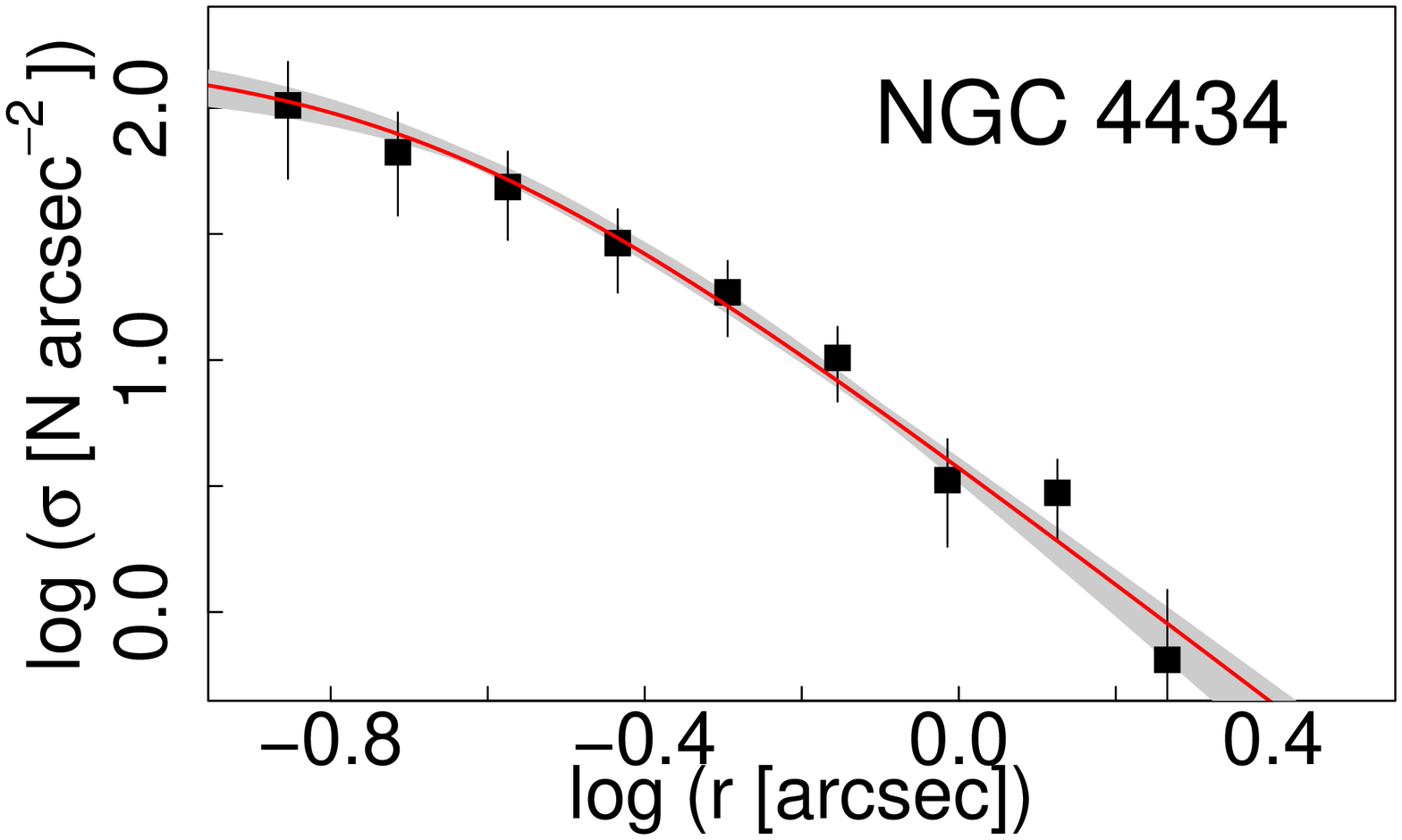}    
\includegraphics[width=55mm]{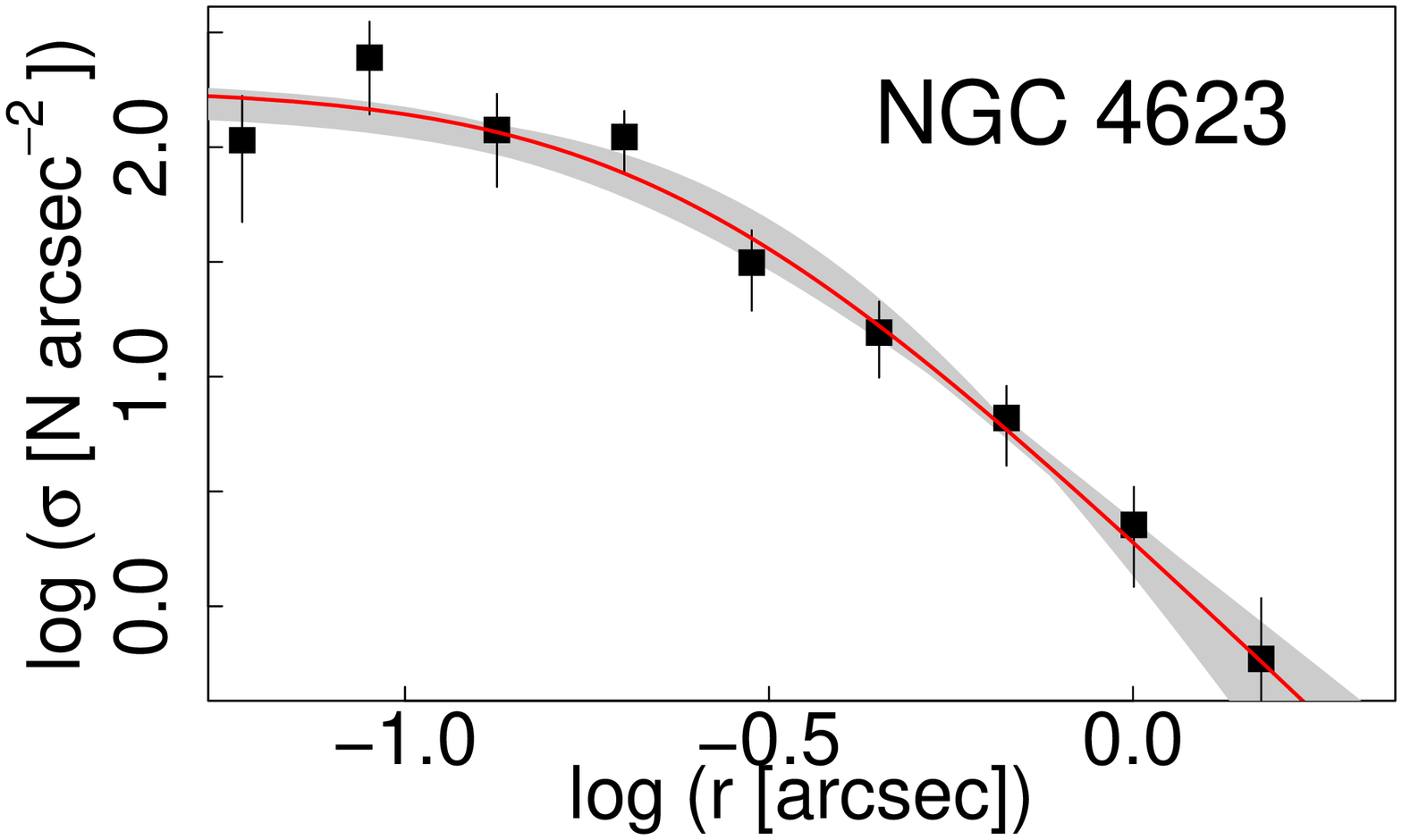}\\   
\caption{Projected radial distribution for GC candidates. The red solid 
line represents the modified Hubble profile fitted to the data. The grey 
region indicates the changes in the Hubble profile during individual 
iterations with different bin breaks (see the text for further details).}    
\label{hubprof}    
\end{figure*}   

\begin{figure*}    
\includegraphics[width=55mm]{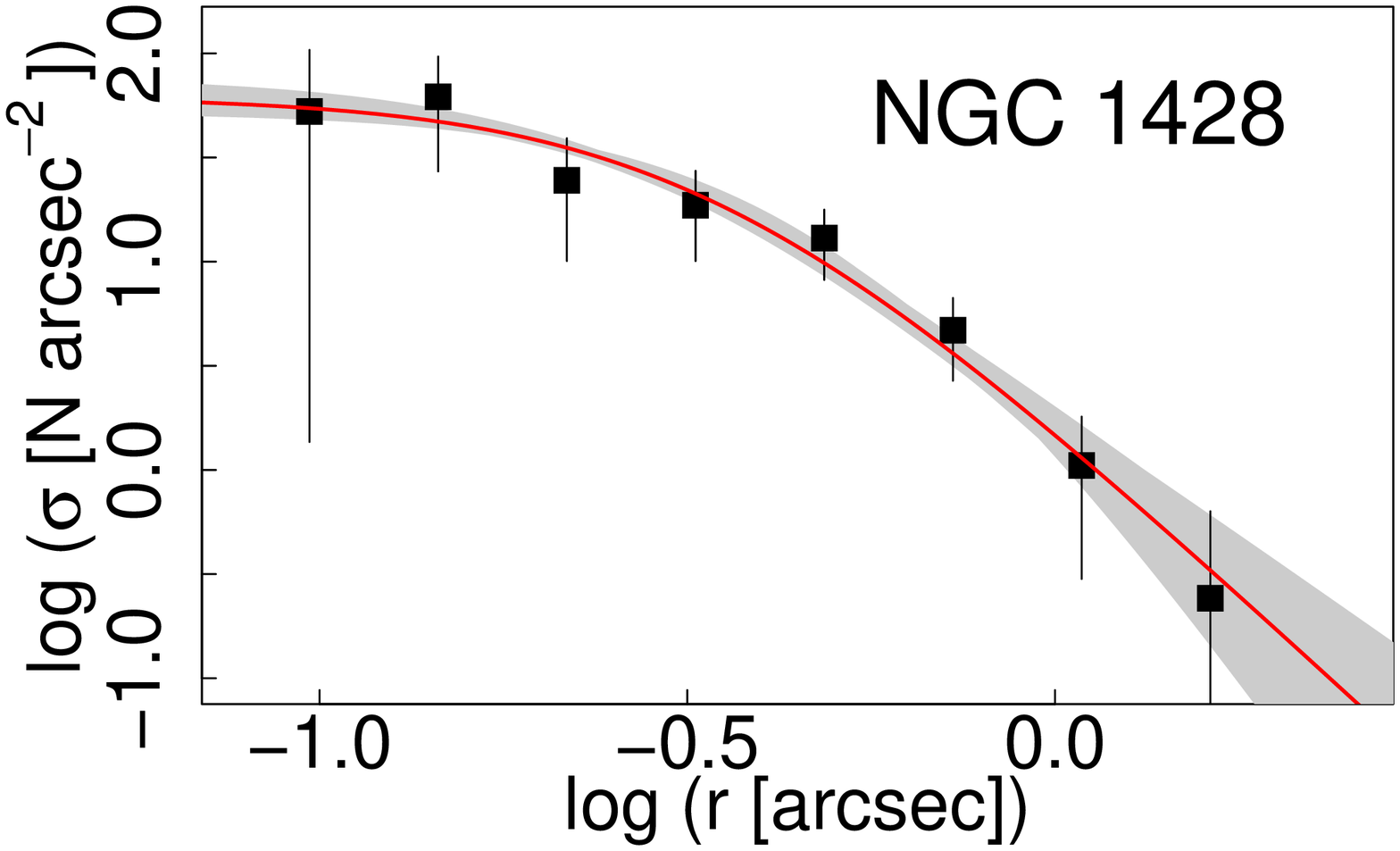}    
\includegraphics[width=55mm]{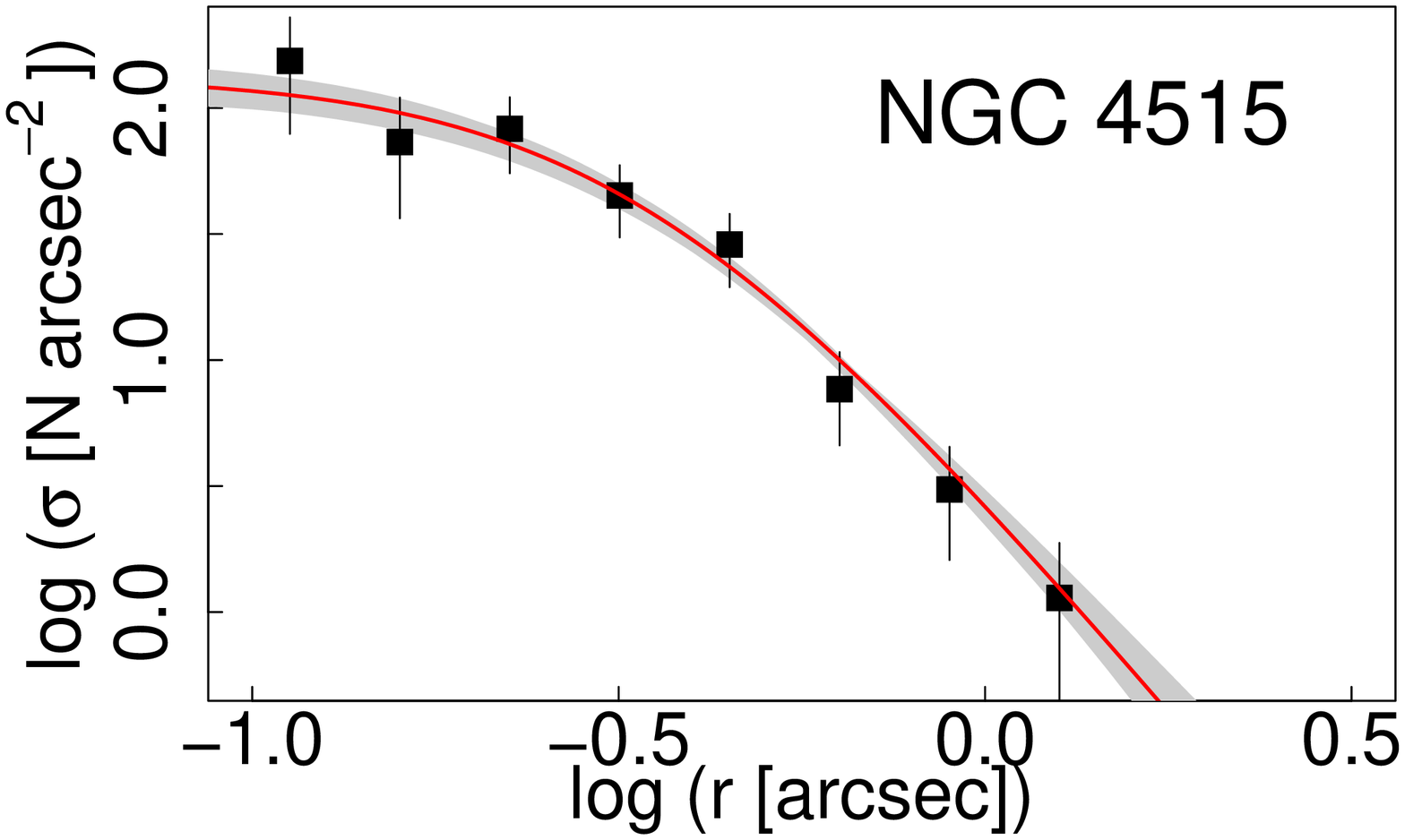}    
\includegraphics[width=55mm]{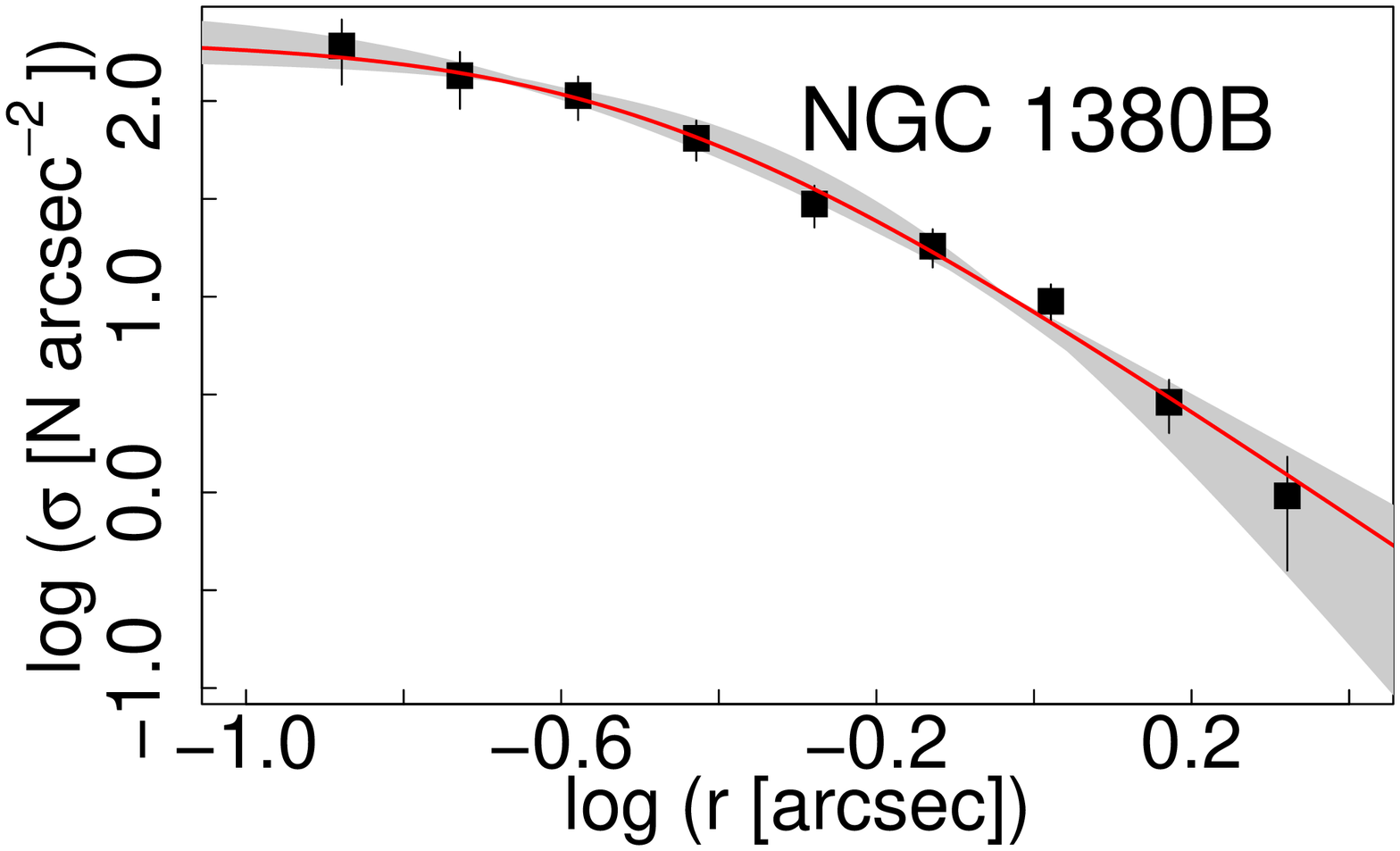}\\    
\includegraphics[width=55mm]{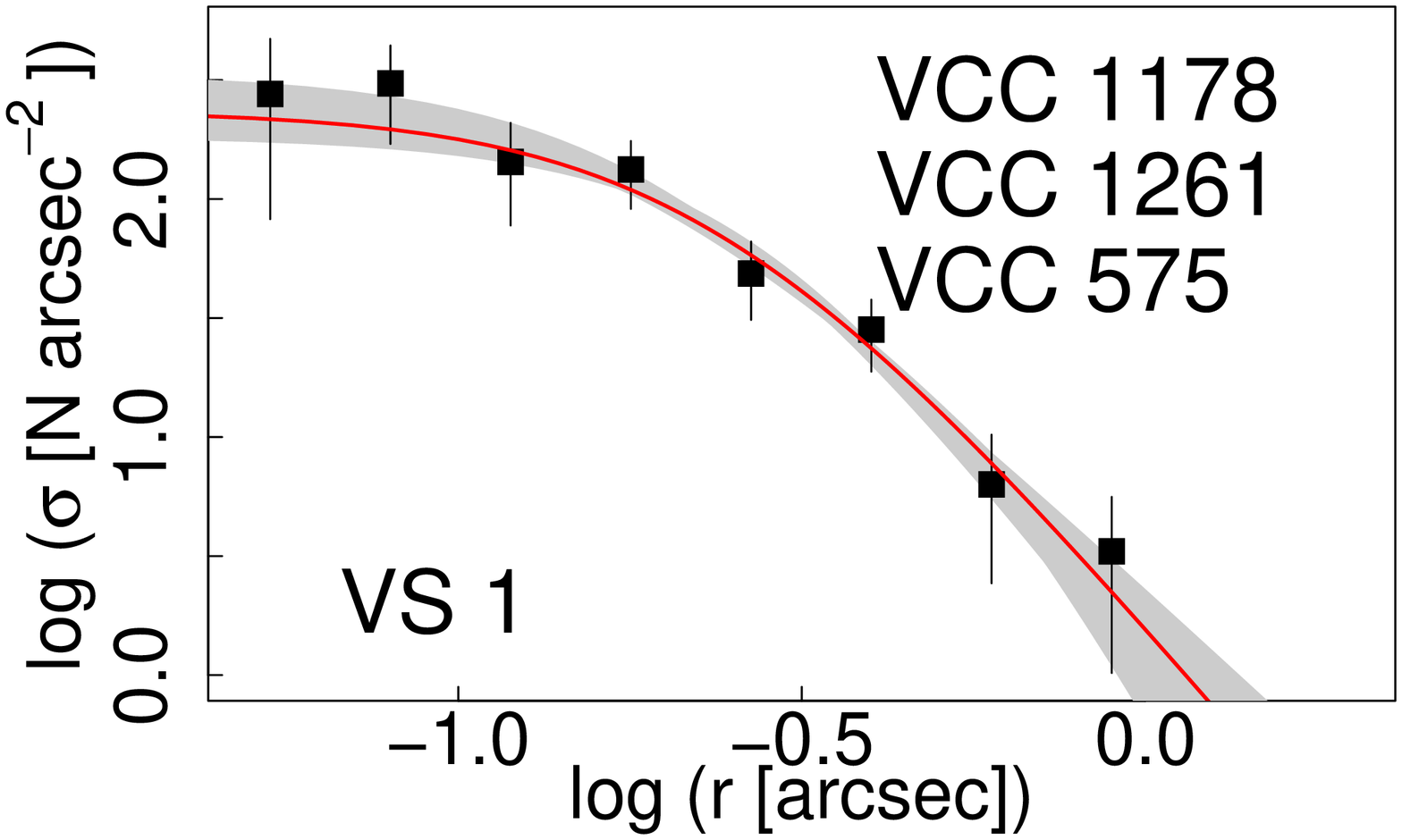}  
\includegraphics[width=55mm]{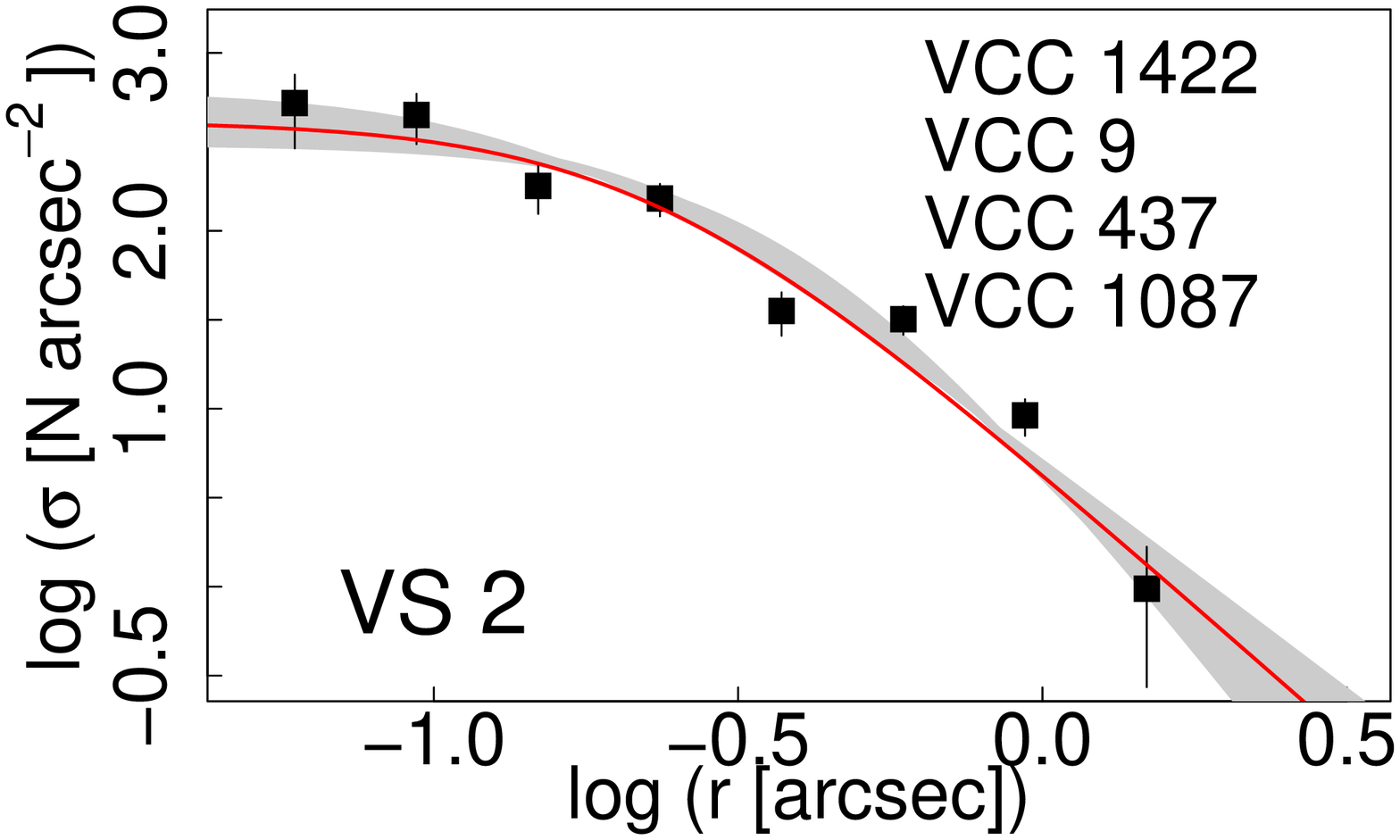}  
\includegraphics[width=55mm]{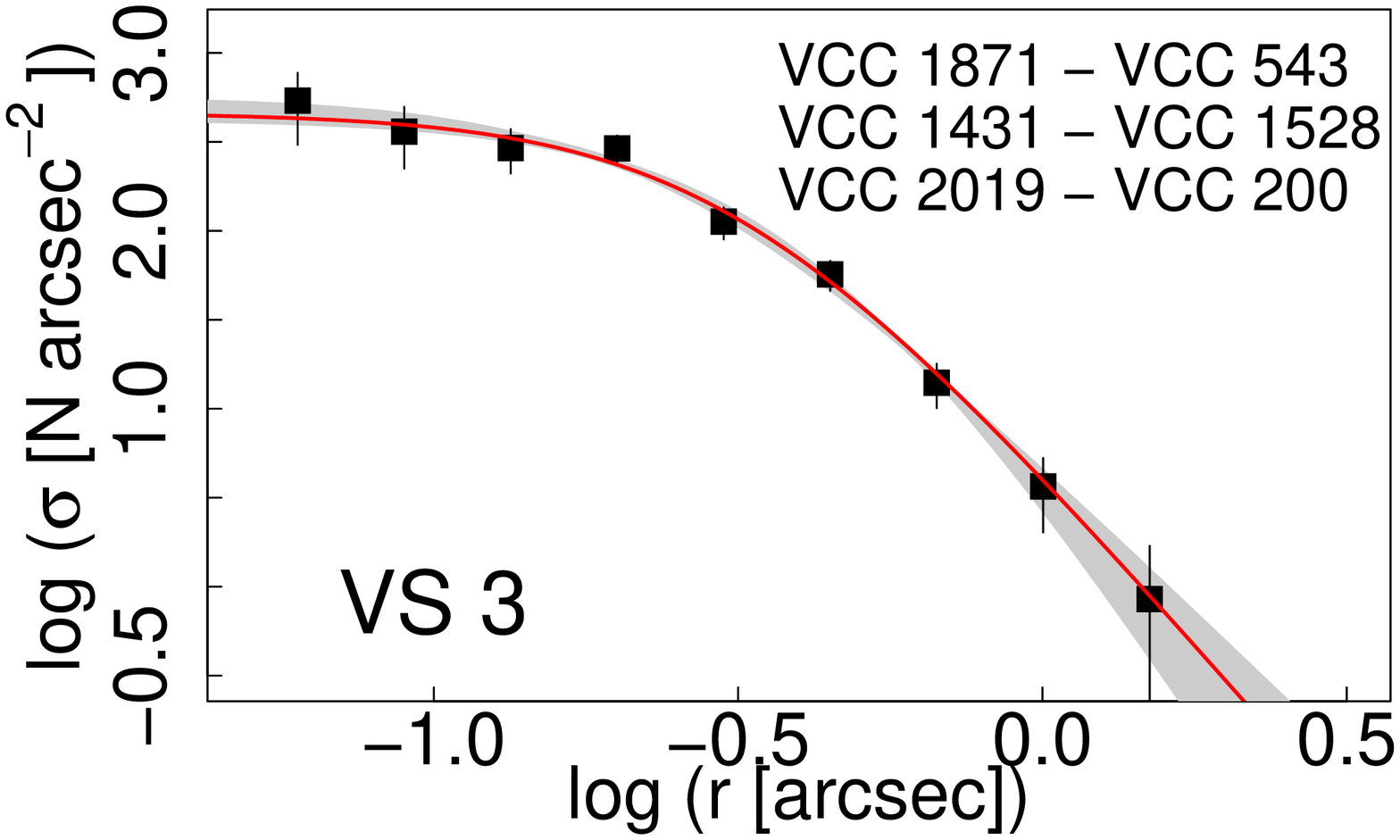}\\  
\contcaption{The stacked low-surface-brightness Virgo galaxies
are indicated with the acronym VS\# and the list of galaxies 
contributing to the sample.}    
\label{hubprof2}    
\end{figure*}

\begin{table*} 
\begin{minipage}{180mm}   
\begin{center}   
  \caption{Galaxies from the literature, listed in decreasing $B$-band luminosity.
Magnitudes (col. 2 to 5) were obtained from NED and reddening corrections from the 
recalibration by \citet{sch11}. Distance moduli correspond to SBF measurements listed in 
NED, typically from \citet{tul13}. The parameter $b$ corresponds to the exponent of the
Hubble modified profile (analogue to half of the power-law slope). ${\rm r_L}$, 
${\rm r_{eff,GCS}}$ and ${\rm N_{GCs}}$ represent the projected extension of the GCS, its 
effective radius and the total population of GCs, respectively. The ${\rm r_{eff,gal}}$ of 
the galaxies were taken from \citet[][ please note that in this paper de Vaucouleurs 
profiles were used, instead of S\'ersic ones]{fab89}. Central 
velocity dispersions (${\rm \sigma_0}$) were obtained from the HyperLeda database.}    
\label{tab.otros}   
\begin{tabular}{@{}l@{}c@{}c@{}c@{}c@{}c@{}c@{}c@{}c@{}c@{}c@{}c@{}c@{}}   
\hline   
\multicolumn{1}{@{}c}{Name}&\multicolumn{1}{c}{$B$}&\multicolumn{1}{c}{$V$}&\multicolumn{1}{c}{$J$}&\multicolumn{1}{c}{$K$}&\multicolumn{1}{c}{$E_{(B-V)}$}&\multicolumn{1}{c}{$(m-M)$}&\multicolumn{1}{c}{$b$}&\multicolumn{1}{c}{${\rm r_L}$}&\multicolumn{1}{c}{${\rm r_{eff,GCS}}$}&\multicolumn{1}{c}{${\rm N_{GCs}}$}&\multicolumn{1}{c}{${\rm r_{eff,gal}}$}&\multicolumn{1}{c}{${\rm \sigma_0}$}\\   
&\multicolumn{1}{c}{mag}&\multicolumn{1}{c}{mag}&\multicolumn{1}{c}{mag}&\multicolumn{1}{c}{mag}&\multicolumn{1}{c}{mag}&\multicolumn{1}{c}{mag}& &\multicolumn{1}{c}{arcmin}&\multicolumn{1}{c}{arcmin}& &\multicolumn{1}{c}{arcsec}&\multicolumn{1}{c}{km\,s$^{-1}$}\\   
\hline   
NGC\,1407 & $10.70$ & $9.67$ & $7.64$ & $6.70$ & $0.061$ & $32.25$ & $0.79\pm0.04^1$ & $21^1$ & $3.8\pm0.2^2$ & $6400\pm700^1$ & $71.9$ & $266\pm5.1$\\
NGC\,4486$^{\rm f}$ & $9.59$ & $8.63$ & $6.72$ & $5.81$ & $0.022$ & $31.11$ & $1.0\pm0.08^3$ & $--$ & $--$ & $14660\pm891^4$ & $81.3$ & $323\pm4.3$\\
NGC\,4406$^{\rm f}$ & $9.83$ & $8.9$ & $7.01$ & $6.10$ & $0.028$ & $31.26$ & $0.62\pm0.03^5$ & $--$ & $5.8\pm0.1^6$ & $2900\pm--^5$ & $35.2$ & $231\pm2.6$\\
NGC\,1395 & $10.55$ & $9.59$ & $7.83$ & $6.89$ & $0.021$ & $31.88$ & $0.68\pm0.02^7$ & $24^7$ & $--$ & $6000\pm1100^7$ & $45.4$ & $240\pm4.3$\\
NGC\,4649$^{\rm f}$ & $9.81$ & $8.84$ & $6.67$ & $5.74$ & $0.025$ & $31.08$ & $0.69\pm0.02^8$ & $--$ & $6.1\pm2.1^8$ & $4690\pm980^8$ & $66.1$ & $331\pm4.6$\\
NGC\,4594$^{\rm f}$ & $8.98$ & $8.00$ & $5.89$ & $4.96$ & $0.045$ & $30.26$ & $0.93\pm0.04^5$ & $19^5$ & $5.9\pm0.4^6$ & $1900\pm--^5$ & $--$ & $226\pm3.3$\\
NGC\,4374 & $10.09$ & $9.11$ & $7.12$ & $6.22$ & $0.036$ & $31.34$ & $0.55\pm0.06^9$ & $--$ & $--$ & $4301\pm1201^4$ & $52.5$ & $278\pm2.4$\\
NGC\,3962 & $11.62$ & $10.67$ & $8.56$ & $7.67$ & $0.039$ & $32.80$ & $0.91\pm0.07^{10}$ & $--$ & $--$ & $854\pm98^{10}$ & $34.4$ & $220\pm13$\\
NGC\,5813$^{\rm f}$ & $11.45$ & $10.46$ & $8.34$ & $7.41$ & $0.05$ & $32.50$ & $1.07\pm0.03^{11}$ & $13^{11}$ & $3.9\pm0.3^{11}$ & $2900\pm400^{11}$ & $57.5$ & $236\pm3.4$\\
NGC\,720$^{\rm f}$ & $11.16$ & $10.18$ & $8.18$ & $7.27$ & $0.014$ & $32.17$ & $1.21\pm0.05^2$ & $10^2$ & $2.0\pm0.3^2$ & $1489\pm96^2$ & $39.5$ & $239\pm4.6$\\
NGC\,1399 & $10.6$ & $9.59$ & $7.21$ & $6.31$ & $0.012$ & $31.53$ & $0.81\pm0.05^{12}$ & $45^{12}$ & $--$ & $6450\pm700^{13}$ & $42.4$ & $332\pm5.3$\\
NGC\,3311 & $12.65$ & $11.65$ & $8.97$ & $8.10$ & $0.076$ & $33.55$ & $1.02\pm0.04^{14}$ & $9^{14}$ & $--$ & $16500\pm2000^{14}$ & $--$ & $185\pm6.3$\\
NGC\,2768$^{\rm f}$ & $10.84$ & $9.87$ & $7.93$ & $6.99$ & $0.044$ & $31.73$ & $1.25\pm0.09^2$ & $10^2$ & $1.7\pm0.2^2$ & $744\pm68^2$ & $63.1$ & $185\pm2.8$\\
NGC\,4636$^{\rm f}$ & $10.0$ & $9.5$ & $7.31$ & $6.42$ & $0.027$ & $30.86$ & $0.88\pm0.05^{15}$ & $14^{15}$ & $--$ & $4200\pm120^{15}$ & $89.1$ & $199\pm2.7$\\
NGC\,3923$^{\rm f}$ & $10.80$ & $9.80$ & $7.42$ & $6.50$ & $0.071$ & $31.64$ & $0.62\pm0.04^8$ & $--$ & $0.6\pm0.2^8$ & $4580\pm820^8$ & $53.3$ & $246\pm4.9$\\
NGC\,4365 & $11.0$ & $9.6$ & $7.5$ & $6.6$ & $0.021$ & $31.82$ & $0.67\pm0.05^{16}$ & $--$ & $6.1\pm1.2^2$ & $6450\pm110^{16}$ & $52.5$ & $250\pm2.6$\\
NGC\,6411 & $12.79$ & $11.85$ & $10.02$ & $9.13$ & $0.048$ & $33.58$ & $1.07\pm0.09^{17}$ & $5^{17}$ & $--$ & $700\pm45^{17}$ & $26.7$ & $183\pm4.6$\\
NGC\,4762 & $11.1$ & $10.3$ & $8.2$ & $7.3$ & $0.021$ & $31.82$ & $0.93\pm0.0^{11}7$ & $5^{11}$ & $1.4\pm0.4^{11}$ & $270\pm30^{11}$ & $43.7$ & $141\pm4.1$\\
NGC\,7507 & $11.36$ & $10.38$ & $8.20$ & $7.29$ & $0.044$ & $31.95$ & $1.23\pm0.05^{18}$ & $7^{18}$ & $--$ & $350\pm50^{18}$ & $31.4$ & $217\pm2.7$\\
NGC\,1404 & $10.97$ & $10.00$ & $7.77$ & $6.82$ & $0.010$ & $31.53$ & $0.85\pm--^{19}$ & $4^{20}$ & $--$ & $725\pm145^{20}$ & $26.7$ & $230\pm3.8$\\
NGC\,4494 & $10.71$ & $9.83$ & $7.90$ & $6.99$ & $0.018$ & $31.14$ & $0.85\pm0.10^{21}$ & $10^{21}$ & $--$ & $392\pm49^{21}$ & $49.0$ & $148\pm2.6$\\
NGC\,2865 & $12.57$ & $11.66$ & $9.36$ & $8.46$ & $0.074$ & $32.95$ & $0.94\pm0.08^{10}$ & $--$ & $--$ & $410\pm8^{10}$ & $11.7$ & $171\pm2.8$\\
NGC\,1380 & $10.87$ & $9.93$ & $7.77$ & $6.86$ & $0.017$ & $31.23$ & $0.81\pm0.05^{22}$ & $3^{22}$ & $--$ & $560\pm30^{22}$ & $--$ & $215\pm4.6$\\
NGC\,3268 & $12.5$ & $11.45$ & $9.12$ & $8.15$ & $0.098$ & $32.83$ & $0.9\pm0.05^{23}$ & $13^{23}$ & $--$ & $8200\pm800^{23}$ & $36.1$ & $229\pm16$\\
NGC\,3258 & $12.5$ & $11.5$ & $9.25$ & $8.31$ & $0.077$ & $32.71$ & $0.9\pm0.05^{23}$ & $13^{23}$ & $--$ & $8000\pm800^{23}$ & $27.4$ & $261\pm9.8$\\
NGC\,5866 & $10.74$ & $9.89$ & $7.83$ & $6.87$ & $0.013$ & $30.93$ & $0.88\pm0.05^{11}$ & $10^{11}$ & $3.1\pm0.7^{11}$ & $340\pm80^{11}$ & $36.3$ & $162\pm4.7$\\
NGC\,6861 & $12.1$ & $11.1$ & $8.66$ & $7.71$ & $0.052$ & $32.28$ & $0.80\pm0.02^{24}$ & $10^{24}$ & $--$ & $3000\pm300^{24}$ & $22.8$ & $387\pm16$\\
NGC\,821$^{\rm f}$ & $11.67$ & $10.68$ & $8.80$ & $7.90$ & $0.097$ & $31.83$ & $1.24\pm0.26^{25}$ & $4^{25}$ & $--$ & $320\pm45^{25}$ & $39.8$ & $198\pm2.8$\\
NGC\,3115 & $9.87$ & $8.9$ & $6.78$ & $5.88$ & $0.044$ & $29.93$ & $0.98\pm0.06^8$ & $--$ & $--$ & $546\pm80^8$ & $36.1$ & $260\pm3$\\
NGC\,3379$^{\rm f}$ & $10.24$ & $9.28$ & $7.17$ & $6.27$ & $0.022$ & $30.25$ & $0.71\pm0.07^5$ & $11^5$ & $--$ & $270\pm--^5$ & $39.8$ & $202\pm1.8$\\
NGC\,1052$^{\rm f}$ & $11.41$ & $10.47$ & $8.37$ & $7.45$ & $0.023$ & $31.42$ & $1.04\pm0.07^{26}$ & $3^{26}$ & $--$ & $400\pm45^{26}$ & $36.9$ & $208\pm3.9$\\
NGC\,5128 & $7.84$ & $6.84$ & $4.98$ & $3.94$ & $0.101$ & $27.82$ & $1\pm0.1^{27}$ & $--$ & $--$ & $1550\pm--^{28}$ & $--$ & $103\pm6.2$\\
NGC\,4278$^{\rm f}$ & $11.09$ & $10.16$ & $8.09$ & $7.18$ & $0.026$ & $30.93$ & $0.88\pm0.02^{29}$ & $20^{29}$ & $2.8\pm0.5^{29}$ & $1378\pm200^{29}$ & $31.6$ & $237\pm4.5$\\
NGC\,1387 & $11.68$ & $10.69$ & $8.44$ & $7.43$ & $0.011$ & $31.43$ & $1.2\pm0.15^{30}$ & $3^{30}$ & $--$ & $390\pm27^{30}$ & $--$ & $167\pm12$\\
NGC\,1379 & $11.80$ & $10.91$ & $9.08$ & $8.24$ & $0.012$ & $31.54$ & $1.3\pm0.25^{30}$ & $3^{30}$ & $--$ & $225\pm23^{30}$ & $42.4$ & $117\pm2.2$\\
NGC\,1427 & $11.77$ & $10.86$ & $9.03$ & $8.14$ & $0.011$ & $31.46$ & $1\pm0.1^{31}$ & $5^{31}$ & $--$ & $470\pm40^{31}$ & $32.9$ & $155\pm2.8$\\
NGC\,7332 & $12.02$ & $11.11$ & $8.98$ & $8.07$ & $0.033$ & $31.66$ & $0.63\pm0.07^{32}$ & $2^{32}$ & $0.4\pm0.1^{33}$ & $175\pm15^{32}$ & $17.4$ & $128\pm3.3$\\
NGC\,4754 & $11.5$ & $10.6$ & $8.31$ & $7.41$ & $0.03$ & $31.04$ & $0.71\pm0.07^{11}$ & $3^{11}$ & $2.6\pm0.9^{11}$ & $115\pm15^{11}$ & $31.6$ & $177\pm3$\\
NGC\,1374 & $12.00$ & $11.08$ & $9.05$ & $8.16$ & $0.012$ & $31.46$ & $1.15\pm0.1^{30}$ & $2^{30}$ & $--$ & $360\pm17^{30}$ & $30.0$ & $179\pm3.3$\\
NGC\,2271 & $13.16$ & $--$ & $8.68$ & $9.69$ & $0.104$ & $32.53$ & $1.09\pm0.09^{10}$ & $--$ & $--$ & $562\pm9^{10}$ & $--$ & $148\pm20$\\
NGC\,1400 & $11.92$ & $10.96$ & $8.75$ & $7.81$ & $0.062$ & $31.06$ & $0.58\pm0.10^{34}$ & $3^{34}$ & $--$ & $922\pm280^{35}$ & $37.8$ & $246\pm3.4$\\
NGC\,3384 & $11.00$ & $9.9$ & $7.7$ & $6.8$ & $0.026$ & $30.01$ & $0.65\pm0.09^{11}$ & $5^{11}$ & $2.4\pm1.3^{11}$ & $120\pm30^{11}$ & $32.3$ & $144\pm2.5$\\
NGC\,7457 & $12.09$ & $11.20$ & $9.11$ & $8.19$ & $0.047$ & $30.41$ & $0.91\pm0.06^{36}$ & $3^{36}$ & $--$ & $210\pm30^{36}$ & $36.3$ & $68\pm3.5$\\
\hline
\end{tabular}
\end{center}
$^{\rm f}$ The power-law slopes indicated in the Table were obtained by fitting the density profiles published in the
corresponding paper.\\    
{\bf References:} $^1$\citet{for11}, $^2$\citet{kar14}, $^3$\citet{har09b}, $^4$\citet{pen08}, $^5$\citet{rho04}, $^6$\citet{kar16},
$^7$\citet{esc18}, $^8$\citet{fai11}, $^9$\citet{gom04}, $^{10}$\citet{sal15}, $^{11}$\citet{har12}, $^{12}$\citet{bas06a},
$^{13}$\citet{dir03a},
$^{14}$\citet{weh08}, $^{15}$\citet{dir05}, $^{16}$\citet{blo12}, $^{17}$\citet{cas19}, $^{18}$\citet{cas13b}, $^{19}$
\citet{cap09}, $^{20}$\citet{for98}, $^{21}$\citet{fos11}, $^{22}$\citet{kis97}, $^{23}$\citet{cas17}, $^{24}$\citet{esc15},
$^{25}$\citet{spi08}, $^{26}$\citet{for01a}, $^{27}$\citet{har04}, $^{28}$\citet{har06b}, $^{29}$\citet{ush13}, $^{30}$\citet{bas06b}, $^{31}$
\citet{for01b}, $^{32}$\citet{you12}, $^{33}$\citet{hud18}, $^{34}$\citet{for06}, $^{35}$\citet{per97}, $^{36}$\citet{har11}.    
\end{minipage}   
\end{table*}   

\noindent present ${\rm M_{\star} < 10^{11}\,M_{\odot}}$, 
which is in agreement with the range of masses for which the slopes of the scaling 
relations change in panels D and G. There is no clear evidence for more 
extended galaxies, that span a wide range of $b$. The fit of the entire 
sample with a quadratic polynomial produces large uncertainties.   

\subsubsection{Scaling relation with the extension of the GCS}
The panels in the second row show the extension of the GCS in kpc 
(${\rm r_L}$) as a function of the logarithm of ${\rm M_{\star}}$ 
(panel\,D), the logarithm of ${\rm N_{GC}}$ (panel\,E) and the 
${\rm r_{eff,gal}}$ of the host galaxy in kpc (panel\,F). The symbols 
follow the same prescription than in previous panels. As we previously 
indicated, for the galaxies
in our sample ${\rm r_L}$ was obtained as the galactocentric 
distance where the numerical density of GCs falls below the $30$ per cent
of the background level, and its uncertainty was calculated 
from the parameters of the Hubble profile fitted to the galaxy, 
as well as the distance estimator uncertainties, typically the 
$10$ per cent. For galaxies from literature, uncertainties of the GCSs 
extension are not always provided. Hence, for the ${\rm r_L}$ 
we assumed the mean of the uncertainties in arcmin of our sample, 
plus the uncertainties in the distance estimator for each case.
We find that ${\rm r_L}$ as a function of ${\rm log_{10}(M_{\star})}$
can be described by a bilinear relation of the form:

\begin{eqnarray}
\left.\begin{aligned}
{\rm r_L} =&-99\pm10 + 12\pm2 \times {\rm X_D}, & {\rm M_{\star} \lesssim 4\times 10^{10}\,M_{\odot}}\\
 &-1200\pm203 + 116\pm19 \times {\rm X_D}, & {\rm M_{\star} \gtrsim 4\times 10^{10}\,M_{\odot}}
\end{aligned}\right.
\end{eqnarray}

\noindent with ${\rm X_D}$ being ${\rm log_{10}(M_{\star})}$.
\citet{kar14} proposed a single linear relation between ${\rm r_L}$
and ${\rm log_{10}(M_{\star})}$ for early-type galaxies, but their 
sample spans stellar masses larger than the mass threshold where 
the slope changes. The slope of the linear relation fitted by \citet{kar14}
is $80.5\pm15.7$, but they obtained the stellar masses from the
${\rm M/L_V}$ ratios estimated by \citet{zep93}. The majority 
of the galaxies considered in this paper are ellipticals, hence the
slope from \citet{kar14} agrees with our fit for the massive galaxies,
considering that \citet{kar16} estimated that stellar masses from 
\citet{zep93} are $\approx 1.5$ times larger than those derived from
\citet{bel03} relations. As well as for the exponent of the Hubble 
profile, an order-two polynomial results in an accurate description 
of ${\rm r_L}$ as a function of ${\rm N_{GC}}$ in logarithmic scale,
resulting:

\begin{equation}
b = 25\pm4.1 -36\pm6.2 \times {\rm X_E} + 16.1\pm5.6 \times {\rm X_E^2}
\end{equation}

\noindent ${\rm X_E}$ representing the ${\rm log_{10}(N_{GCs})}$.
When the functions fitted in both panels are compared, the typical
residuals in panel\,D for galaxies more massive than $4\times 10^{10}
{\rm M_{\odot}}$ double those obtained for the latter ones. Avoiding 
an extensive discussion about the uncertainties involved in both
variables, this might imply that ${\rm N_{GC}}$ is more intrinsically
related to ${\rm r_L}$ than ${\rm M_{\star}}$ for massive galaxies.

In panel\,F there is a clear trend between the extension ${\rm r_L}$ and
the ${\rm r_{eff,gal}}$ of the galaxies, but the dispersions prevent any 
conclusion and further analysis on a larger sample is necessary. The
dotted line corresponds to the Equation\,11 from \citet{kar14},
scaled by a factor 1.5 due the differences in the stellar masses
previously indicated. The relation agrees with our data.

\subsubsection{Scaling relations with the effective radius of the GCS}
The third row shows the ${\rm r_{eff,GCS}}$ of the GCS as a function of 
the ${\rm M_{\star}}$ (panel\,G) and the ${\rm N_{GC}}$ (panel\,H), 
both in logarithmic scale, and the ${\rm r_{eff,gal}}$ of the host galaxy
(panel\,I). As in panel\,D, we fitted a bilinear relation:

\begin{eqnarray}
\left.\begin{aligned}
{\rm r_L} =&-22.5\pm6.8 + 2.6\pm0.7 \times {\rm X_G}, & {\rm M_{\star} \lesssim 4\times 10^{10}\,M_{\odot}}\\
 &-315\pm67 + 30.3\pm6.1 \times {\rm X_G}, & {\rm M_{\star} \gtrsim 4\times 10^{10}\,M_{\odot}}
\end{aligned}\right.
\end{eqnarray}

\noindent with ${\rm X_G}$ being ${\rm log_{10}(M_{\star})}$, 
plotted with black solid lines. The dotted grey curve represents the 
relation derived by \citet{for17} for early-type galaxies, while the 
dashed-dotted grey curve corresponds to a sample of early and late 
type galaxies from \citet{hud18}. Both relations seem to underestimate 
the ${\rm r_{eff,GCS}}$ of the GCS for the low stellar mass galaxies. Besides,
the relations deviate significantly for stellar masses above 
${\rm 10^{10}\,M_{\odot}}$, with observations showing a large spread
at fixed ${\rm M_{\star}}$.
In panel\,H the ${\rm r_{eff,GCS}}$ of the GCS is fitted by a quadratic
polynomial of the form:

\begin{equation}
b = 17\pm6.1 -20.6\pm5.7 \times {\rm X_H} + 6.8\pm1.2 \times {\rm X_H^2}
\end{equation}

\noindent with ${\rm X_H}$ being ${\rm log_{10}(N_{GC})}$. Although
there is a clear dependence in the calculus of both parameters, it is
worth to emphasize the tight correlation between them, pointing to
the richness of the GCS as the main factor to determine its extension.
On the other hand, panel\,I shows the ${\rm r_{eff,GCS}}$ of the GCS against
the ${\rm r_{eff,gal}}$ of the host galaxy. The solid curve corresponds
to a linear relation fitted to the data.

\begin{equation}
b = 0.4\pm2 + 3.3\pm0.55 \times {\rm X_I}
\end{equation}

The dashed-dotted grey curve corresponds to the relation derived by \citet{hud18}, while the dotted grey curve represents the mean ratio 
for both parameters from \citet{for17}. Both expressions are in agreement 
with our fit, considering the lack of ${\rm r_{eff,GCS}}$ measurements 
for many GCSs in massive galaxies, and the dispersion of the available
ones. These limitations prevent further conclusions.

\begin{figure}    
\includegraphics[width=80mm]{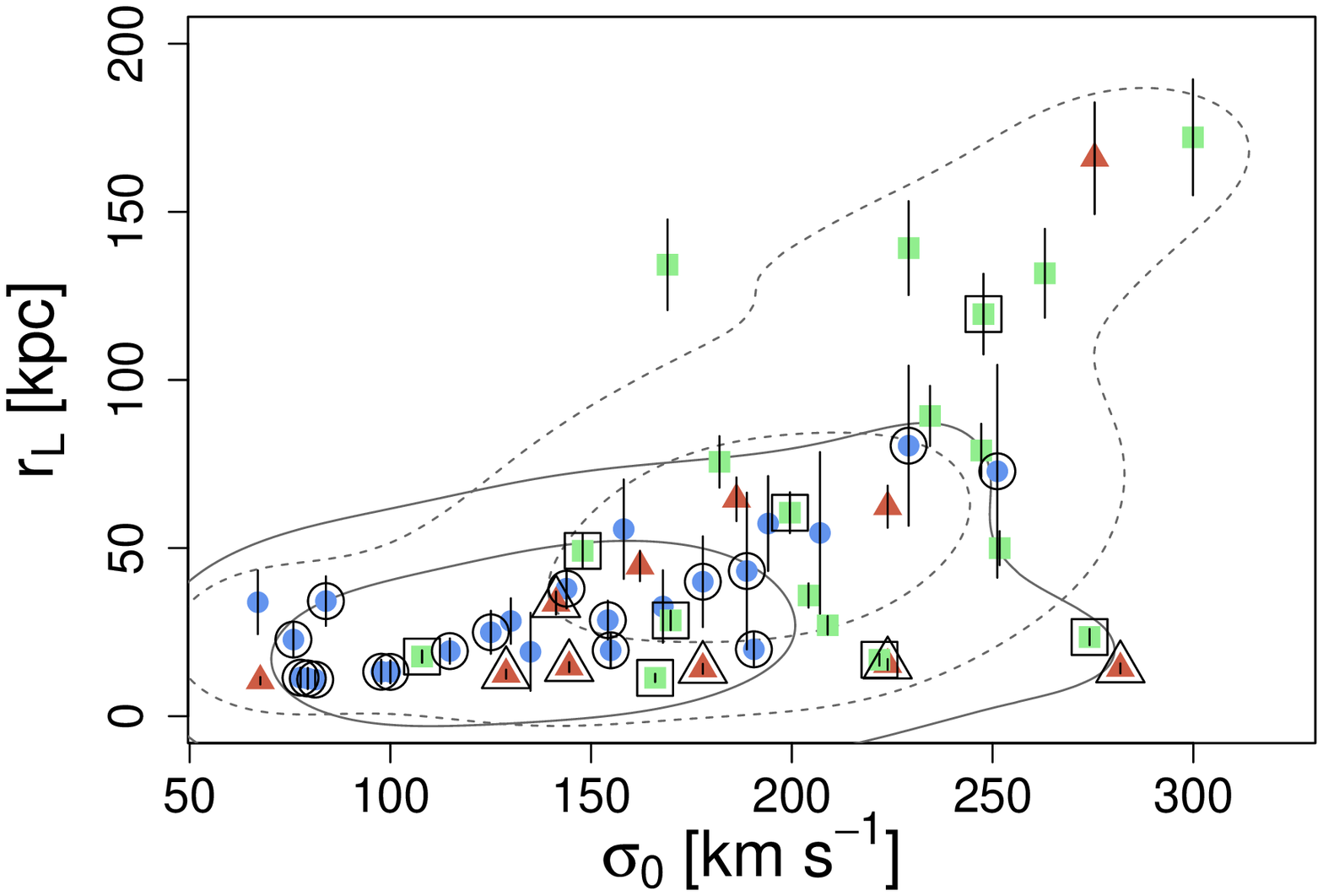}\\
\caption{The extension of the GCS ${\rm r_L}$ as a function of the central
velocity dispersion $\sigma_0$. The symbols follow the same prescription
than previous figures. Framed ones highlight the satellites of more massive 
galaxies. The solid contours are indicative of the locus of the satellites,
while the dashed ones correspond to central galaxies.}  
\label{disp}    
\end{figure}    

\subsubsection{Comparison with other parameters}
In Figure\,\ref{disp} we explore the differences in the 
extension of the GCS ${\rm r_L}$ as a function of the central velocity 
dispersion $\sigma_0$ for the galaxies listed in Tables\,\ref{hubpar} and 
\ref{tab.otros}. The symbols follow the same prescription that previous
figures, with framed ones representing satellite galaxies. This classification
was based on the information indicated in the papers that analysed their
corresponding GCSs. We are aware of the effect of possible misclassifications. 
The contours are only indicative of the locus that
satellites (solid curves) and central galaxies (dashed curves) occupy.
In the latter ones ${\rm r_L}$ shows a correlation with $\sigma_0$,
but it is nearly invariant for satellites. There are four GCSs labelled as
satellites that follow the central galaxies correlation. Two of them
come from the literature sample and correspond to NGC\,4636 and NGC\,4649,
giant ellipticals from the Virgo cluster that dominate respective
cluster subgroups, present very populated GCSs (see Table\,\ref{tab.otros}) and 
extended dark matter haloes \citep[e.g][]{das11,sch12}. The other two galaxies
are VC1903 and VCC1632, which also belong to the Virgo cluster.
Their GCSs contain around a thousand members and, as we previously indicated,
the derived value of ${\rm r_L}$ is only indicative, because it largely
exceeds our FOV.

\begin{figure}    
\includegraphics[width=80mm]{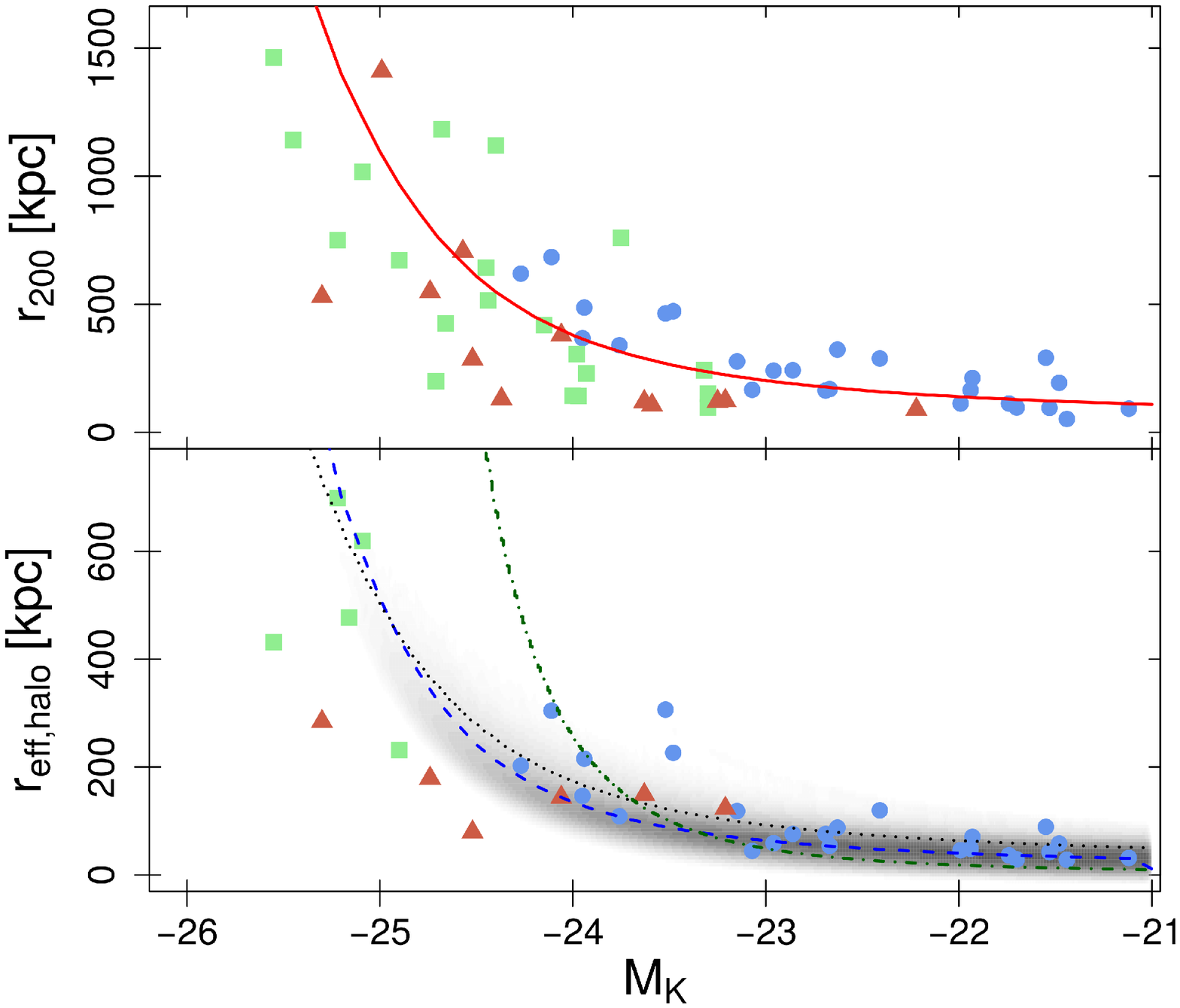}\\
\caption{{\bf Upper panel:} The red solid line corresponds to the 
distribution of ${\rm r_{200}}$ for the haloes from SMDPL as a function 
of the $K$ absolute magnitude assigned by the HOD method, and 
the symbols are the ${\rm r_L}$ of the GCS, scaled to fit this 
latter one by a factor $f_{200}$. {\bf Lower panel:} Smoothed distribution 
of the ${\rm r_{eff,halo}}$ fitted with a S\'ersic profile to the projected 
haloes from the SMDPL simulation. The dashed curve indicates the mean 
values of ${\rm r_{eff,halo}}$ as a function of $M_K$. Filled symbols represent
the ${\rm r_{eff,GCS}}$ of the GCS, scaled to the ${\rm r_{eff,halo}}$
distribution by a factor $f_{\rm eff}$. The dotted curve shows the relation 
between the size of galaxies and the virial radius derived by \citet{kra13}, 
as represented in \citet{for17}, but scaled by the factor $f_{\rm eff}$ to 
follow the ${\rm r_{eff,GCS}}$ in our plot. The dashed-dotted curve 
corresponds to the relation fitted by \citet{hud18} between the 
${\rm r_{eff,GCS}}$ and the ${\rm r_{200}}$, applying the same scaling factor
$f_{\rm eff}$.}   
\label{sim1}    
\end{figure}

We lack of characterisations of the dark matter haloes for the galaxies in 
our sample, hence direct comparison between the halo mass or virial radius
and the parameters of the GCS is not possible. Instead we applied a 
statistical point of view. We selected from the SMDPL simulation those haloes 
with $K$ luminosities in the same range as the galaxies analysed in this 
paper. Then we projected their density distribution, described by a Navarro, 
Frenk \& White profile \citep[hereafter, NFW profile][]{nav96}, on the Cartesian
xy plane. S\'ersic profiles provided an accurate fit to the resulting projected 
distributions. The red solid line in the upper panel of Figure\,\ref{sim1} 
corresponds to the $r_{200}$ radius, defined as the galactocentric distance 
where the volumetric density equals 200 times the critic density at $z=0$. The 
 symbols indicate the extent of the GCS (${\rm r_L}$), scaled to the 
distribution of $r_{200}$ of the haloes for comparison purposes. There seems 
to be an agreement in the behaviour of both parameters as a function of $M_K$. 
The scaling factor results ${\rm r_{200}} = f_{200} \times {\rm r_L}$ with 
$f_{200}=8.5\pm0.5$. The symbols follow the same prescription that in previous 
figures.
The smoothed distribution of ${\rm r_{eff,halo}}$ of the haloes in terms of $M_K$ 
is shown in the lower panel of Figure\,\ref{sim1}, and its mean values are
represented by the blue dashed curve.
The filled symbols indicate the ${\rm r_{eff,GCS}}$ of the systems with 
available measurements, scaled to the ${\rm r_{eff,halo}}$ of the haloes.  
In this case the scaling factor was fitted to the dashed line, on the basis of a 
possible correlation between the parameters, resulting that ${\rm r_{eff,halo}}
= f_{\rm eff} \times {\rm r_{eff,GCS}}$, with $f_{\rm eff}=16.7\pm2.3$. 
The distribution of scaled ${\rm r_{eff,GCS}}$ of the GCS seems to follow the 
distribution of ${\rm r_{eff,halo}}$ of the haloes, despite a larger sample of 
bright galaxies would provide a more accurate result. This gives confidence 
to the assumption that the ${\rm r_{eff}}$ of haloes and GCSs are correlated. 

The dotted curve shows the relation derived by \citet{kra13} for the galaxy
size as function of the virial radius, scaled by $3.7$ to consider the
mean ratio between galaxies and GCS sizes \citet{for17}, and by $f_{\rm eff}$. 
The relation is in agreement with the ${\rm r_{eff,GCS}}$ of the GCS. The 
dashed-dotted curve corresponds to the relation between the $r_{200}$ 
radius and the ${\rm r_{eff,GCS}}$ of the GCS derived by \citet{hud18}, once 
again scaled by $f_{\rm eff}$, but it seems to overestimate ${\rm r_{eff}}$ for
galaxies brighter than $M_K=-23.5$\,mag.

\begin{figure}    
\includegraphics[width=80mm]{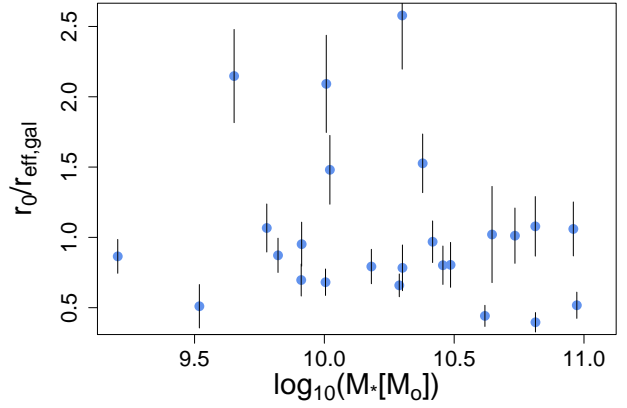}\\
\caption{Ratio between the core radius from the Hubble profile 
(${\rm r_0}$) and the ${\rm r_{eff,gal}}$ of the host galaxy, as
a function of ${\rm M_{\star}}$. Points with large ratio
might be overestimated due to small measurements of ${\rm r_{eff,gal}}$.}    
\label{core}    
\end{figure}

GCS in elliptical galaxies usually present a flattened radial 
distribution, less peaked in the inner arcsecs than the galaxy light 
profile \citep[e.g.][]{har79,cap09,cas17}, even when the variation 
in the completeness with the galactocentric distance, is taken into 
account \citep[e.g.][]{bas17}. In Figure\,\ref{core} the ratio between
the core radius ${\rm r_0}$ from the Hubble profile and the ${\rm
r_{eff,gal}}$ of the galaxy is plotted as a function of the logarithm
of the ${\rm M_{\star}}$ for the galaxies in our sample. In this
case we avoid the comparison with literature data based in two
reasons, (i) the treatment of the completeness as a function of
the galactocentric radius results in more accurate measurements
of GCs projected densities, particularly for bright ellipticals,
and the lack of this analysis might lead to significant differences
in the inner region of the radial profile, (ii) \citet{broc14}
studied the dissolution processes ruling the GC erosion with
numerical simulations, pointing that the core size depends on 
the threshold mass for the GCs, because low-mass GCs
are more affected by disruption processes than the most
massive ones. Then, the faint limit in magnitude achieved in the
observations is important, and the inclusion of
results from 
different instruments and photometric depth would introduce
unnecessary noise.

The majority of GCSs in our sample present ${\rm r_0 /
r_{eff,gal}} \approx 1$, which resembles the results from
\citet{broc14} for the models MOD2 and MOD3, with ${\rm M_{\star}
\approx 1-3 \times 10^{11}\,M_{\odot}}$. There are few galaxies 
presenting ratios above 1.5, but in all cases they present
lower ${\rm r_{eff,gal}}$ than galaxies with similar stellar
masses, pointing that large ratios are due to the underestimation 
of this latter parameter.
These galaxies are NGC\,4660, NGC\,4515 and NGC\,1419. \citet{broc14}
indicate that the fraction might be lower for the most massive
and extended galaxies, but our sample does not allow us to test it.

\section{Discussion}

\citet{rod16} analysed the population of dark matter haloes from the 
cosmological simulations Bolshoi-Planck and Multidark-Planck \citep{kly16}. 
They found that the distribution of concentration index at $z=0$ depends 
on the virial mass, becoming more extended for more massive haloes.
Moreover, in the surface-brightness profiles in early-type galaxies
the S\'ersic index $n$ and the ${\rm r_{eff,gal}}$ correlates with luminosities
\citep[e.g.][]{cal15}, implying that galaxies with moderate luminosity
are more compact than the brightest ones. Under the assumption that
GCSs are related with the mass distribution of the host galaxy, it is 
expected to obtain steeper radial distributions when the galaxies
become less massive.

\citet{kar14} compared the properties of GCSs for a sample of early and 
late-type massive galaxies, with the host galaxy stellar mass, resulting
in a linear relation with ${\rm log_{10}(M_{\star})}$, that is in agreement
with our results for galaxies with ${\rm M_{\star} \gtrsim 4\times 10^{10}\,
M_{\odot}}$ when a scaling relation due to the different sources of the $M/L$ 
relations \citep{kar16} is considered. The change in slope for lower masses
might be related with the stellar mass-size relation for galaxies, 
that flattens for central galaxies \citep[e.g.][]{she03,sha14}.
\citet{sha09} indicated that the late evolution of the most massive galaxies
in rich environments, mainly driven by minor mergers, might explain the 
gradual steepening of the size-mass relation for larger luminosities.
We are aware that most of the galaxies in our sample are satellites in 
dense environments like the Virgo and Fornax clusters, but \citet{hue13}
showed that central and satellite early-type galaxies follow a similar 
stellar mass-size relation. Similar results were found by \citet{spi17},
who claimed that it cannot be ruled out that environmental processes may
modify the size and mass for a given galaxy. In fact, they propose to
the central velocity dispersion ($\sigma_0$) as invariant variable to 
changes due to environmental processes. They found that, at fixed $\sigma_0$, 
quiescent central galaxies are larger and more massive than their satellite 
counterparts. This is expected considering that satellite galaxies, moving 
through a high density environment like the intracluster medium, should
experienced a ram pressure that might strip its gas, leading to the
reduction of the star formation and a subsequent reduced size 
\citep[e.g.][]{kap09}. 
Taking into account that GCSs are typically more extended
than the field population of the galaxy, it is expected for environmental
processes to also affect them, particularly their extension (${\rm r_L}$).
Central velocity dispersions tend to be invariant to growth by minor mergers,
\citet{bez12} found that the internal dynamics of quiescent galaxies,
in the high central velocity dispersion regime, remains roughly unchanged
with time. The authors point to a rapid quench, becoming more efficient
with the increase of the velocity dispersion. Minor mergers should have
played a relevant role in the mass increase in later stages, that in 
central ellipticals might represent an important fraction of their
mass at $z=0$ \citep{vdo10}. The mergers that increase the stellar mass
of the central galaxy also provide GCs that enlarge the preexisting 
population, in detriment of satellites which hardly experienced
merging episodes. 

The halo mass-size relation for galaxies has also been studied by 
\citet{kra13} and \citet{cha17}. The latter ones estimated halo mass from
weak lensing analysis and found a differential measurement of the halo 
mass-size relation at fixed stellar mass, in the form of a power-law.
Although the fitted exponents vary with stellar mass, the average
values differ between blue and red galaxies, described in their paper
as primarily star-forming discs and quiescent ellipticals, respectively.
\citet{hud18} derived a correlation between the ${\rm r_{eff,GCS}}$ of the GCS
and the halo extension and mass for a sample of early and late-type galaxies.
Although the previously mentioned evolutionary differences between central
and satellite galaxies might play a role, they found that GCSs with larger 
${\rm r_{eff,GCS}}$ occupy larger and more massive haloes.

The stellar mass at which the slope changes 
in our Equation\,8 matches with that corresponding to the maximum of the 
ratio ${\rm M_{\star}/M_{halo}}$ from numerical simulations \citep{beh10,mos10}. 
\citet{cor18} differentiated satellite and main haloes, for halo masses below
$\approx 10^{12}\,{\rm M_{\odot}}$ (i.e. ${\rm M_{\star} \approx 
3\times 10^{10}\,M_{\odot}}$) they found that central galaxies inhabit 
more massive haloes than satellites at fixed stellar mass. This might
be understood in terms of the mass loss in subhaloes, mainly due to
dynamical friction, tidal stripping and tidal heating \citep[e.g.][]{gan10}.
Moreover, the calculus of the tidal radius ${\rm r_t}$ in satellite galaxies 
after the accretion epoch has to reflect the fact that the satellite galaxy 
is bound to more massive halo, instead of ${\rm r_{200}}$. In a simplified 
approach, the ${\rm r_t}$ is reached when the gravitational acceleration 
towards the satellite centre equals the tidal acceleration from the host 
potential. Although a more accurate treatment should involve the phase 
space distribution of the satellite particles \citep[e.g.][]{kam07}, the
qualitative idea that the virial radius of a halo is shorten afterwards
it is accreted by a more massive one remains valid. The majority of the 
galaxies in our sample are indeed satellites in density environments 
like the Virgo and Fornax clusters, and hence their haloes should have
experienced this environmental effects.

It is scarcely a novelty that the concentration of the radial distribution
of GCS (represented by the parameter $b$), as well as its ${\rm r_L}$ and
${\rm r_{eff,GCS}}$, are related with the richness of the GCS. This latter 
property is closely connected to the merger history of the host galaxy,
responsible for the mass accretion but also for the built up of the GCS
through major starburst driven by merging episodes \citep[e.g.][]{mur10,li14} and accretion of GCs
\citep{for11,amo19}. Moreover, \citet{kru15} pointed that the environmental 
conditions that favour GCs formation also lead to their tidal disruptions 
at early stages in their evolution, and that subsequent mergers are
needed to eject them to the host galaxy halo, improving their survival
ratio. Although there is a large dispersion, reviews on the subject seem 
to confirm this connection with the stellar 
\citep[e.g.][]{har13} and virial masses \citep[e.g.][]{hud14} for early
type galaxies. The dispersion in the relations might be ruled by the environmental 
conditions that affect the formation and evolution of the GCS in 
cluster-like environments \citep[e.g.][]{pen08} as well as in the field 
\citep[e.g.][]{sal15}. 

Environmental conditions increase the disruption rate of GCs in
the inner regions of galaxies, leading to the flattening in 
their radial profiles. \citet{cap09} proposed dynamical friction as 
the mechanism behind this GC erosion, but it has been ruled out in 
more recent papers \citep{broc14}. These authors tested the efficiency of 
different dissolution processes involved in the GC erosion through 
numerical simulations. They found that GCs density profiles are 
typically flattened in less than a Hubble time. The resulting cores 
depend on the mass and effective radius of the galaxy, but radial 
anisotropies of the GCS might also play a main role. Other studies 
focused on the evolution of Galactic GCs also point to the relevance 
of the mass-loss rate when GCs are subject to strong tidal fields 
close to the centre of the Galaxy \citep[e.g.][]{web14,mad17}. The 
accurate analysis of the inner density distribution of GCS might 
provide relevant information about the mechanisms ruling the 
kinematical behaviour of the field population in the inner region of 
the galaxies.

\section{Summary}
We performed the photometry
of HST/ACS archive observations of several intermediate 
luminosity galaxies located in low density environments. It was 
supplemented with available photometries of GCSs from the Virgo and 
Fornax clusters, resulting in a sample of almost 30 GCS for whom we 
fitted their radial profiles. Additional literature studies were
compiled to enlarge the sample. We summarize our conclusions in the 
following.

\begin{itemize}
\item For the galaxies in low density environments, we obtained
the effective radii of their GCs. Blue GCs are more extended than 
red ones, and mean values are in agreement with previously
published results.
The reduced luminosity range spanned by these galaxies does not 
allow us to observe any trend between the mean effective radii
and the stellar mass.

\item Hubble modified profiles provide an accurate fit for the 
entire sample of GCSs. The exponent of the power law correlates
with the stellar mass of the host galaxy and the number of GCs,
being steeper for low mass galaxies. The relation with the 
effective radius of the galaxy is not clear at the luminous end. 
This suggests that the concentration of the GCS depends on the
general properties directly related with the mass growth of 
the galaxy.

\item The extension of the GCS also correlates with the stellar
mass, the number of GCs and the effective radius of the galaxy.
In the first case, the relation flattens for galaxies with
stellar masses below $4\times10^{10}\,{\rm M_{\odot}}$. Due to the
commonly known non-linear relation between the stellar mass
and the number of GCs, the correlation with this later property
is soften and a quadratic curve is an accurate description.

\item The effective radius of the GCS correlates with the
effective radius of the host galaxy, as indicated in previous
studies, but with a large dispersion.
The comparison with the stellar mass and the number of GCs
shows a similar behaviour than that described for the GCS 
extension.

\item The extension of the GCS of central galaxies seems to 
correlate with the central velocity dispersion, but it presents 
a distinctive behaviour for satellites. We interpret this in
the context of the different mass accretion history of the
two groups of galaxies.

\item From the statistical comparison with numerical simulations,
the effective radius of the GCS scales with the projected effective 
radius of the haloes, and the extension of the GCS scales with their
virial radius.

\item The size of the core of the Hubble modified profile for GCS
correlates with the effective radius of the galaxy, in agreement
with results from numerical simulations for low and intermediate-mass
ellipticals.
\end{itemize}

\section*{Acknowledgments}
This work was funded with grants from Consejo Nacional de Investigaciones   
Cient\'{\i}ficas y T\'ecnicas de la Rep\'ublica Argentina, Agencia Nacional de   
Promoci\'on Cient\'{\i}fica y Tecnol\'ogica, and Universidad Nacional de La Plata  
(Argentina).
This research has made use of the NASA/IPAC Extragalactic Database (NED) which 
is operated by the Jet Propulsion Laboratory, California Institute of Technology, 
under contract with the National Aeronautics and Space Administration.

\bibliographystyle{mnras}
\bibliography{biblio}

\end{document}